 \documentclass[twocolumn,showpacs,nofootinbib]{revtex4}

\usepackage[latin1]{inputenc}
\usepackage{graphicx}%
\usepackage{dcolumn}
\usepackage{amsmath}

\makeatletter
\def\btt#1{\texttt{\@backslashchar#1}}%
\DeclareRobustCommand\bblash{\btt{\@backslashchar}}%
\makeatother

\begin{document}

 \title{Microscopic features of the effect of vehicle overacceleration  on   traffic flow}

\author{Boris S. Kerner$^1$ and Sergey L. Klenov$^2$}

 \affiliation{$^1$
Physics of Transport and Traffic, University of Duisburg-Essen,
47048 Duisburg, Germany}

    \affiliation{$^2$
Moscow Institute of Physics and Technology,
Department of Physics, 141700 Dolgoprudny, Moscow Region, Russia}

\pacs{89.40.-a, 47.54.-r, 64.60.Cn, 05.65.+b}

\begin{abstract} 
Through the development of a microscopic deterministic   model in the framework of three-phase traffic theory,
 microscopic  features of  vehicle overacceleration, which determines the occurrence of the  metastability
of free traffic flow  at a bottleneck,
have been revealed:  (i) The greater the impact of vehicle overacceleration 
on   free traffic flow   at a bottleneck, the higher the maximum  flow rate
 at which free flow can   
persist   at the bottleneck, i.e., the better traffic breakdown can be avoided.
(ii) There can be  at least two   mechanisms of overacceleration  in   road lane caused by safety acceleration at the  bottleneck.
(iii) Through a microscopic analysis of  
spatiotemporal competition between speed adaptation and vehicle acceleration behaviors, traffic conditions
have been found at which safety acceleration in   road lane or/and   
vehicle acceleration due to lane-changing on  multi-lane road become overacceleration.  
 (iv) There is   spatiotemporal cooperation of  different   overacceleration mechanisms.
(v)  The stronger  the    overacceleration cooperation, the stronger the maintenance of free flow
at the bottleneck due to overacceleration.  (vi)
On two-lane road, 
both speed adaptation  
and    overacceleration in   road lane can effect qualitatively on  
 the  overacceleration mechanism
caused by lane-changing. These    microscopic features of the effect of vehicle overacceleration on traffic flow
  are related to traffic flow consisting of    human-driving or/and
automated-driving vehicles.
 \end{abstract}

 \maketitle

\section{Introduction  \label{Int}}

One of the most well-known and widely accepted theoretical results in vehicular traffic science is the assumption that
traffic breakdown results from driver {\it overreaction} on an unexpected braking  of the preceding vehicle:
Due to the existence of driver reaction time, a driver begins to decelerate with a time delay;
as a result, the driver decelerates stronger than it needs to avoid the collision with the preceding vehicle.
Because of this driver {\it overdeceleration} caused by   overreaction,
the vehicle speed becomes less than the speed of the preceding vehicle.
If such overdeceleration effect is realized for the following drivers, traffic flow instability is realized:
Due to overdeceleration of the following vehicles, a growing wave of a   speed decrease occurs that propagates upstream. 
This classical  traffic flow instability that should explain traffic breakdown was  discovered theoretically
in 1950s in     
car-following models by Herman, Gazis,  Montroll, Potts, and  Rothery~\cite{GM_Com1,GM_Com2,GM_Com3} as well as
Kometani and  Sasaki~\cite{KS,KS1,KS2,KS4}. Through different mathematical approaches,
the classical traffic flow instability  
has later been incorporated in a diverse variety of mathematical
 traffic flow 
models, in particular, well-known and widely used models by Newell~\cite{Newell1961,Newell}, Gipps~\cite{Gipps1981,Gipps1986},
Wiedemann~\cite{Wiedemann},
macroscopic model of Payne~\cite{Payne_1,Payne_2}, cellular automation model of Nagel and Schreckenberg~\cite{Nagel_S},
 optimal velocity model of Bando et al.~\cite{Bando_1,Bando,Bando_2,Bando_3},
lattice traffic flow model of Nagatani~\cite{Nagatani_1,Nagatani_2},
intelligent driver model   of Treiber~\cite{Treiber}, and
stochastic microscopic   model of Krau{\ss}~\cite{Krauss}. The above-mentioned   models
 as well as many other standard traffic flow models incorporating the classical traffic flow instability
 (e.g.,~\cite{Jiang2001,Barlovic,Aw-Raschle})  are currently one of the main fundamentals  
of standard traffic science and the basis of the most microscopic traffic simulation
 tools widely used in traffic 
engineering~\cite{Chen2012A,Chen2012B,Chen2014,Ashton,Drew,Gerlough,Gazis,Gartner,Barcelo,Elefteriadou,DaihengNi,Kessels,Treiber-Kesting,Schadschneider,Chowdhury,Helbing,Nagatani_R,Nagel}. It has been found~\cite{KK1994} that the development
of this classical traffic flow  instability lead to the formation of moving traffic jam (J) in free flow (F)
(F$\rightarrow$J transition)~\footnote{It should be emphasized that in models of traffic flow consisting of 100$\%$  
automated-driving vehicles (e.g.,~\cite{Ioannou,Ioannou1993,Levine,Liang,Liang2,Swaroop,Swaroop2,Shladover,Shladover2,Rajamani,Davis-1,Davis-2}) so-called {\it string instability} can occur (e.g.,~\cite{Ioannou1993,Levine,Liang,Liang2,Swaroop,Swaroop2}).   
As in traffic flow  consisting of  human-driving vehicles,
 the cause of string instability is   overdeceleration of an automated-driving vehicle, which
occurs under particular dynamic characteristics of the automated-driving vehicle, at which
 the vehicle decelerates stronger than it needs to avoid the collision with a
 slower moving  preceding vehicle.
Contrary to   human-driving vehicles for which
    overdeceleration is caused by   driver reaction time that cannot be zero,
			overdeceleration of the automated-driving vehicle and, therefore, string instability
			of a platoon of such automated-driving vehicles can be avoided through 
		 an appropriated choice of  dynamic characteristics of the automated-driving vehicles~\cite{Liang}.
		Despite this physical difference,    string instability in traffic flow consisting 
		of automated-driving vehicles is qualitatively similar
to the classical traffic flow instability in traffic flow consisting of human-driving vehicles. This is because in both cases
vehicle overdeceleration is the cause of these traffic flow instabilities. For this reason, 
below we will use the term {\it vehicle overdeceleration}		as common one for human-driving and automated-driving vehicles.}.

\begin{figure}
\begin{center}
\includegraphics[width = 8 cm]{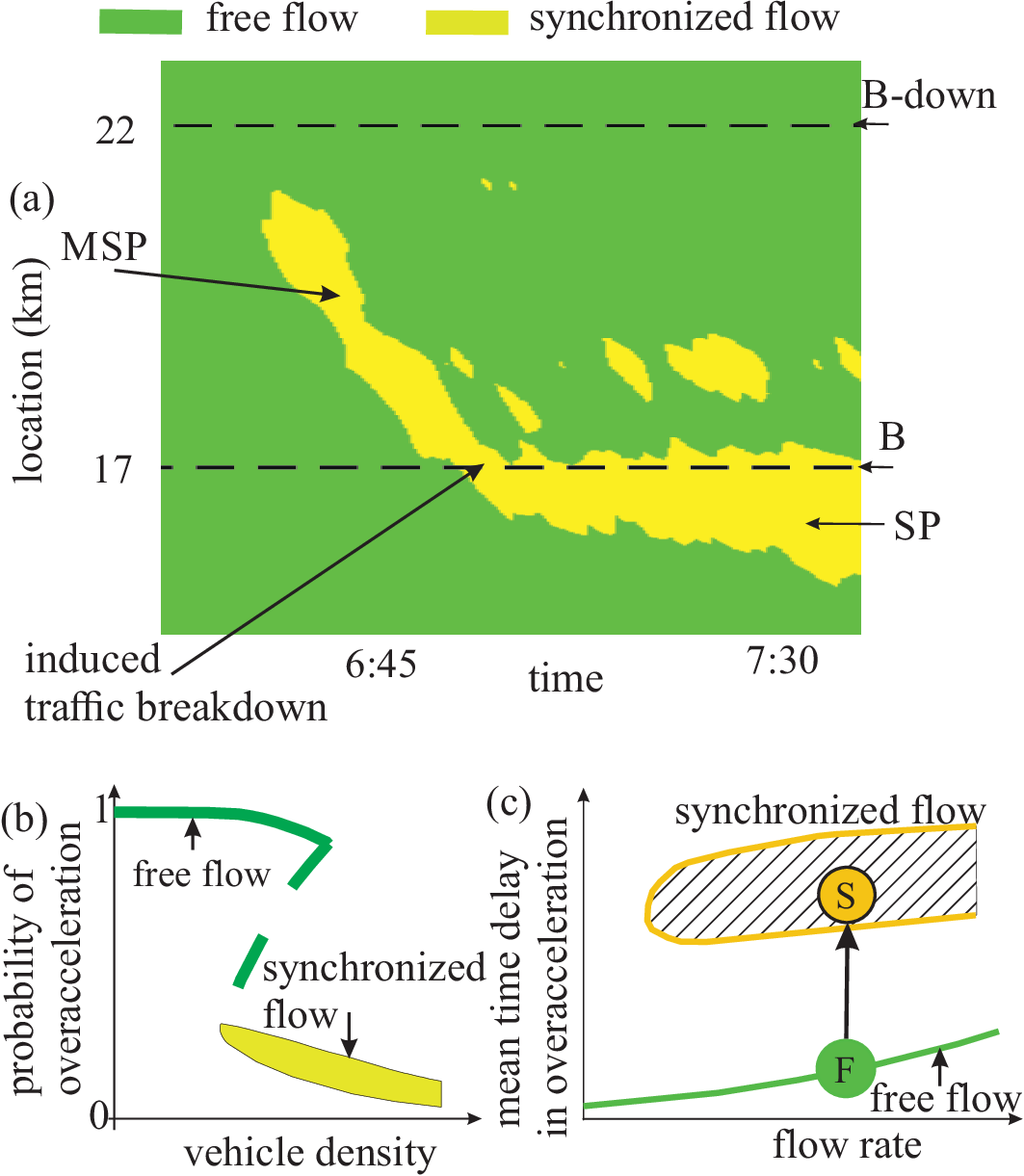}
\end{center}
\caption[]{Hypothesis of three-phase traffic theory about the 
discontinuous character of driver overacceleration~\cite{OA}. (a)
Empirical nucleation nature of  traffic breakdown (F$\rightarrow$S transition) at bottlenecks.   Speed data    presented  
in space and time with averaging method described in Sec. C.2 of Ref.~\cite{TomTom} were measured
with road detectors installed along   road: 
 A moving synchronized flow pattern (MSP) that has emerged   at   downstream bottleneck (B-down) while propagating upstream induces
F$\rightarrow$S transition (induced traffic breakdown) leading to
emergence of a synchronized flow pattern (SP) at upstream  bottleneck (B); adapted 
from~\cite{KernerBook1,KernerBook2,KernerBook3,KernerBook4}. 
(b), (c) Qualitative vehicle density-dependence of driver overacceleration
probability per a   time interval (b) and equivalent presentation of (b)
as  a
 discontinuous flow-rate dependence of the mean time-delay in
 overacceleration  (c): At given initial traffic parameters
(e.g., at given flow rates at traffic network boundaries), 
during F$\rightarrow$S phase transition the probability of overacceleration drops  abruptly whereas the mean time delay in overacceleration jumps abruptly;  F and S are states of free flow and
synchronized flow, respectively.
}
\label{Hyp-OA}
\end{figure}

However,
 at the end of 1990s it was found that rather than the F$\rightarrow$J
transition of standard traffic flow models, real empirical  traffic breakdown
is a phase transition from free flow (F) to synchronized flow (S)
 (F$\rightarrow$S
transition)~\cite{KR1997,Kerner1998}.
Empirical studies  have shown that the F$\rightarrow$S
transition at a bottleneck exhibits the nucleation nature.
None of the standard traffic flow models can explain the empirical 
nucleation nature of traffic breakdown (F$\rightarrow$S
transition) at a bottleneck.
To explain the empirical nucleation nature of the F$\rightarrow$S
transition  [Fig.~\ref{Hyp-OA}(a)],
 Kerner introduced three-phase traffic theory~\cite{OA,Kerner-Three,KernerBook1,KernerBook2,KernerBook3,KernerBook4,KernerBook5}.

As mentioned, in standard traffic theory vehicle overdeceleration  caused by driver overreaction  in traffic flow consisting of
 human-driving vehicles 
should be responsible for traffic breakdown. Contrary to standard traffic flow theory,
in   three-phase traffic theory it is assumed
that
the empirical nucleation nature of traffic breakdown (F$\rightarrow$S
transition) is caused by a discontinuity in the probability (per time interval) of 
driver {\it acceleration} that shows up
when a driver tries to accelerate  from a local speed decrease
in traffic flow:  The probability to accelerate   from a small   local speed decrease
 occurring in free flow   drops abruptly when the same  local speed decrease occurs in synchronized flow. 
It is assumed that in free flow drivers can accelerate
from car-following at a lower speed to a higher speed with
a larger probability than it occurs in synchronized flow [Fig.~\ref{Hyp-OA}(b)]~\cite{OA}. 
Driver acceleration behavior that leads  to such a discontinuity  in 
  the probability to accelerate to free flow has been called driver {\it overacceleration}.
		The discontinuity  in 
  the  overacceleration probability  
  [Fig.~\ref{Hyp-OA}(b)]
can be considered the discontinuous character of    overacceleration.  
In   three-phase traffic theory, the discontinuous character of   overacceleration is the cause for
	the nucleation nature of the F$\rightarrow$S   transition at a bottleneck.
	 The behavioral origin of driver overacceleration is related
to the wish of drivers to move in free flow.

The discontinuous character of   overacceleration is not solely the feature of   
 human-driving vehicles. Indeed,
 in traffic flow consisting of 100$\%$ of
  automated-driving vehicles there is also a discontinuity in the probability of
  overacceleration of  the automated-driving vehicles that
	is the cause for
	the nucleation nature of the F$\rightarrow$S   transition at the bottleneck~\cite{Kerner2023C}.
   Therefore,
for both
human-driving and automated-driving vehicles we can generalize the definition of the term {\it  overacceleration}
as follows:
\begin{itemize}
\item [--] 
Vehicle
acceleration behavior  that  
causes the free flow metastability with respect to the F$\rightarrow$S transition at the
bottleneck is called vehicle
  overacceleration. 
\end{itemize}
 The discontinuous character of overacceleration is a basic feature of   overacceleration
[Fig.~\ref{Hyp-OA}(b)]~\footnote{An equivalent presentation of the discontinuous character of    overacceleration
is a discontinuous flow-rate dependence of the mean time-delay in
vehicle  overacceleration  [Fig.~\ref{Hyp-OA}(c)].}. In general, for both human-driving and automated-driving vehicles moving in
traffic flow,
 the behavioral origin of vehicle overacceleration is related
to   the vehicle's desire to move in free flow.

The term {\it  overacceleration} should 
 distinguish   vehicle overacceleration behaviors that do
result in the free flow metastability with respect to the F$\rightarrow$S transition at the
bottleneck from usual vehicle acceleration behaviors that do not cause the free flow
metastability.

 The overacceleration effect is in a spatiotemporal competition with 
the opposite effect of speed adaptation related to  deceleration
of the vehicle that approaches a slower moving preceding vehicle. Contrary to overacceleration,
speed adaptation
 describes the tendency   from free flow to congested traffic
at the bottleneck: Only if the overacceleration effect is on average stronger than   speed adaptation,
free flow remains at the bottleneck; otherwise, speed adaptation leads to the F$\rightarrow$S transition at the bottleneck.

	In stochastic three-phase  traffic flow models~\cite{KKl,OA_Stoch,Kerner2015_SF,TPACC},  overacceleration
	has been described {\it on average} through model fluctuations. Although in~\cite{KKl,OA_Stoch,Kerner2015_SF,TPACC} many empirical spatiotemporal features of traffic breakdown and resulting congested patterns have been explained~\footnote{These
	three-phase stochastic traffic flow models~\cite{KKl,OA_Stoch,Kerner2015_SF,TPACC} are currently widely used  for
	 simulations of a diverse variety of traffic scenarios 
	(e.g.,~\cite{Hu2012A,Qian2017,Yang2018A,Hu2021A,Lyu2022A,Hu2019A,Zeng2019A,Hu2020A,Zhao2020A,Yang2024A,Hu2022A,Zeng2021A,Chen2024A,Hu2025A}).},
	however,   microscopic  characteristics of overacceleration is   difficult 
	to disclose through  a study of the stochastic models.   	In~\cite{Kerner2023B},   a simple
	deterministic model of overacceleration in a road lane has been introduced that exhibits the discontinuous character. In this model,
	overacceleration   exceeds zero only when the vehicle speed is equal or larger than
	a threshold synchronized vehicle speed. This overacceleration model 
explains the empirical nucleation nature of traffic breakdown. 
However, this   simple overacceleration model~\cite{Kerner2023B} cannot alone explain 
a possible diversity of microscopic  features of the effect of vehicle overacceleration 
on  traffic flow.

 Indeed, a microscopic theory of the effect of vehicle overacceleration on   traffic flow
developed  in this paper   will show
that there are other microscopic mechanisms of overacceleration.
Moreover, we will find that there is a cooperation between the different 
mechanisms of overacceleration, which appear to be very essential for the occurrence of the critical nucleus for  
  the F$\rightarrow$S transition as well as for an   
S$\rightarrow$F instability.

The paper is organized as follows. 
A deterministic
three-phase traffic flow model  used for   simulations has been developed 
 in Sec.~\ref{Model-Sec},
 overacceleration mechanisms caused by safety acceleration on single-lane road are studied
in Sec.~\ref{Safety_Acc_S}, a
cooperation of different mechanisms of overacceleration  is the subject of Sec.~\ref{Cooperation_Sec},
the effect of overacceleration in a road lane on overacceleration through lane-changing
is studied in Sec.~\ref{2-lane_Coop_Sec2}, a theory of the
critical nucleus for spontaneous traffic breakdown is presented in Sec.~\ref{Physics_v_cr},
a proof of  the effect of vehicle overacceleration
 on   traffic flow is given in Sec.~\ref{Suppression_S}, an
 S$\rightarrow$F instability on two-lane road caused by overacceleration cooperation  
is analyzed in Sec.~\ref{Competition_SF_Sec}.
In  Sec.~\ref{Dis}, we explain what microscopic behaviors of vehicle motion are the cause of
 overacceleration     (Sec.~\ref{OA_behavior_Sec}),
generalize the model  for moving jam simulations (Sec.~\ref{Gener_Sec}),
discuss  vehicle overacceleration versus vehicle overdeceleration   
(Secs.~\ref{OA-OV_Sec} and~\ref{Confusion-S}) as well as formulate   conclusions (Sec.~\ref{Con-S}).

\section{Microscopic  model of   vehicular   traffic     \label{Model-Sec}}

\subsection{Basic three-phase traffic flow model   \label{Basic_Model}}

As a basic model for  vehicle motion, we apply  Kerner's model of Ref.~\cite{Kerner2023B}. In the     microscopic
deterministic traffic flow model,    vehicle acceleration/deceleration $a$   in a road lane
is described by
a system of equations:  
 \begin{equation}
a = a_{\rm OA} +
K_{\Delta v}\Delta v   \ \textrm{at $g_{\rm safe} \leq g \leq G$}, 
\label{g_v_g_min1}  
\end{equation}
 \begin{equation}
a =  a_{\rm G} \ \textrm{at $ g > G$},
 \label{g_v_g_min2}
\end{equation}
 \begin{equation}  
 a =  a_{\rm safety}(g, v, v_{\ell}) \ \textrm{at $ g < g_{\rm safe}$},
 \label{g_v_g_min3}
\end{equation}
where   overacceleration $a_{\rm OA}$ in a road lane is given by equation:
\begin{equation}
a_{\rm OA}=\alpha   \Theta (v - v_{\rm syn}),
\label{a_OA}
\end{equation} 
  $v$ is the vehicle speed,   $0\leq v \leq v_{\rm free}$, $v_{\rm free}$ is a maximum vehicle speed;
 $\alpha$ is a   coefficient of    overacceleration;  $\Theta (z) =0$ at $z<0$ and $\Theta (z) =1$ at $z\geq 0$;
 $v_{\rm syn}$ is a given synchronized flow speed ($v_{\rm syn}<v_{\rm free}$);
$g$ is  a space gap to the preceding vehicle, $g= x_{\ell} - x - d$, $x$ and $x_{\ell}$
	are, respectively, the coordinates of the vehicle and the
	 preceding   vehicle, $d$ is the vehicle length;
	\begin{equation} 
\Delta v=v_{\ell}-v, \nonumber
	\end{equation}
  $v_{\ell}$ is   the speed of the  preceding vehicle;
$K_{\Delta v}$ is a positive  dynamic coefficient; $G$ is a synchronization space gap;
  $g_{\rm safe}$ is a safe space gap; $a_{\rm safety}(g, v, v_{\ell})$ is a safety vehicle 
	acceleration~\footnote{In~\cite{Kerner2023B}, 
			safety  vehicle acceleration	$a_{\rm safety}(g, v, v_{\ell})$   in (\ref{g_v_g_min3}) has been called
			$\lq\lq$safety vehicle deceleration". The choice of this term in~\cite{Kerner2023B}
			should emphasize that when at  small space gaps $g<g_{\rm safe}$ a vehicle should decelerate, then
			the value of this vehicle deceleration must prevent  collisions between vehicles.
			However, under condition $g<g_{\rm safe}$ {\it and} at a large enough positive value $\Delta v>0$  
			 it can also occur that 	$a_{\rm safety}(g, v, v_{\ell})>0$, i.e., the
			vehicle accelerates. Therefore, the term $\lq\lq$safety vehicle deceleration" for $a_{\rm safety}(g, v, v_{\ell})$
			  in (\ref{g_v_g_min3}) used in~\cite{Kerner2023B}  might lead to confusions. For this reason, here and furthermore for $a_{\rm safety}(g, v, v_{\ell})$ in Eq.~(\ref{g_v_g_min3})
			we use the term $\lq\lq$safety vehicle acceleration": As usual in the science, the term $\lq\lq$acceleration" has a general meaning, i.e., it can get either a positive value, or negative value, or else null.};
				  $a_{\rm G}$ is vehicle acceleration at large space gaps; 
	vehicle acceleration and speed are limited 
by maximum acceleration $a=a_{\rm max}$ and $v=v_{\rm free}$, respectively.

\subsection{Vehicle acceleration at large space gaps \label{Model_Gap_S}}

In~\cite{Kerner2023B},  to prove the
metastability of free flow at a bottleneck through  the effect of overacceleration
without vehicle overdeceleration,  for  simplification 
of Eq.~(\ref{g_v_g_min2}) we have set
$a_{\rm G}$ as constant value: $a_{\rm G}=a_{\rm max}$~\footnote{To avoid 
possible problems with a considerable  drop in acceleration at the boundary $g=G$ of the indifferent zone that  can occur when   space gap  decreases continuously
beginning from large gaps $g>G$,  for simulations of  the occurrence of the
metastability of free flow at the bottleneck through  the effect of overacceleration in~\cite{Kerner2023B}
we have chosen a long enough synchronization time-headway $\tau_{\rm G}$ at which no
 drop-effect in acceleration at the boundary $g=G$
is realized at model parameters used in~\cite{Kerner2023B}.}. However, to develop a microscopic theory
of the effect of vehicle overacceleration on   traffic flow, the model
should be able to capture a variety of real traffic scenarios.   Therefore, 
 we get:
	\begin{equation} 
a_{\rm G}(g, v, v_{\ell})= a_{\rm OA} +  K_{1}(g-v\tau_{\rm G})+ K_{2}\Delta v,
  \label{a_G-f}  
	\end{equation}
where $a_{\rm OA}$ is given by Eq.~(\ref{a_OA});
$K_{1}$  and $K_{2}$  are    positive dynamic coefficients. Note that the two last terms
	in the right hand of Eq.~(\ref{a_G-f}) is well-known Helly's 
	formula~\cite{Helly}~\footnote{Helly's formula~\cite{Helly} is often used in
				adaptive cruise control (ACC), which is one of the basic models for 
				automated-driving vehicles 
				(e.g.,~\cite{Ioannou,Ioannou1993,Levine,Liang,Liang2,Swaroop,Swaroop2,Shladover,Shladover2,Rajamani,Davis-1,Davis-2}). }.
	In particular, through the use of the term $K_{2}\Delta v$ of Helly's formula
	we take into account situations at which a vehicle that initial gap $g>G$ approaches a slower moving preceding vehicle.  

\subsection{Space-gap dependence  of   overacceleration}

In numerical simulations of   (\ref{g_v_g_min1})--(\ref{a_OA}) made in~\cite{Kerner2023B}, for
 simplification the coefficient of   overacceleration
$\alpha$ in (\ref{a_OA}) has been chosen a constant. 
However, it is physically more realistic to assume
 that $\alpha$ in (\ref{a_OA}) should  depend on the space-gap $g$. In particular,
within the space-gap range
$g_{\rm safe} \leq g \leq G$   overacceleration should be the larger, the larger  the space-gap $g$.
 Indeed, the less the  difference $g-g_{\rm safe}$ between the  space-gap and the safe space-gap, the smaller should be the probability
of   overacceleration. Therefore, in (\ref{a_OA}) we get
\begin{equation}
\alpha =\left\{\begin{array}{ll}
\alpha_{0}    \ \textrm{at $g > G$}, \\
(\alpha_{0}-\alpha_{1})\Biggl(\frac{g-g_{\rm safe}}{G-g_{\rm safe}}\Biggr)^{k} +  \alpha_{1} 
  \  \textrm{at $g_{\rm safe} \leq g \leq G$}, \\
\end{array} \right.
\label{a_OA_gap}
\end{equation}
where $k$, $\alpha_{0}$, and $\alpha_{1}$  are positive model parameters, $\alpha_{0}>\alpha_{1}$. 
	
		 \subsection{Safety vehicle acceleration}

	Eq.~(\ref{g_v_g_min3}) describes safety vehicle acceleration that should
    prevent collisions between vehicles at  small space gaps $g<g_{\rm safe}$. 
There are many   concepts developed in   standard  traffic flow  models~\cite{GM_Com1,GM_Com2,GM_Com3,KS,KS1,KS2,KS4,Newell1961,Newell,Gipps1981,Gipps1986,Wiedemann,Nagel_S,Bando_1,Bando,Bando_2,Bando_3,Nagatani_1,Nagatani_2,Treiber,Jiang2001,KK1994,Krauss,Barlovic,Gartner,Barcelo,Elefteriadou,DaihengNi,Chowdhury,Helbing,Nagatani_R,Nagel,Treiber-Kesting,Schadschneider,Helly}
		that can be used for safety vehicle acceleration
		$a_{\rm safety}(g, v, v_{\ell})$. 
 In~\cite{Kerner2023B},  Helly's formula~\cite{Helly} has been used  for safety vehicle 
				acceleration:  
$a_{\rm safety}(g, v, v_{\ell})=   K_{g}(g-g_{\rm safe})+ K_{v}\Delta v$,
  where
  $K_{g}$  and $K_{v}$  are    positive dynamic coefficients. However,
 from the analysis of Helly's formula  made in~\cite{Kerner2023C}, we know that 
in some traffic situations at a large enough negative speed difference $\Delta v$,
specifically by   lane changing   at low enough vehicle speeds, vehicle collisions can be 
possible~\footnote{See footnote 19 of~\cite{Kerner2023C}.}. 

 To avoid vehicle collisions,   
   for safety  
				acceleration $a_{\rm safety}(g, v, v_{\ell})$  in (\ref{g_v_g_min3}) we   apply ideas of
				  dynamic breaking strategies of General Motors car-following model
					 of Herman, Gazis, Montroll, Potts,
Rothery, and Chandler~\cite{GM_Com1,GM_Com2,GM_Com3,Gazis}, in which at $\Delta v<0$ vehicle deceleration is proportional to the term
	\begin{eqnarray}
\frac{v\tau_{\rm safe}}{g}\Delta v, \nonumber
\end{eqnarray}
where $\tau_{\rm safe}$ is a safe time headway.				
					This idea of the GM-model~\cite{GM_Com1,GM_Com2,GM_Com3,Gazis} insures that
						the dynamic vehicle breaking should be the stronger, the lower the space gap $g$ between vehicles is. 
						Accordingly, instead of Helly's formula we   get 
 \begin{equation}
a_{\rm safety}(g, v, v_{\ell})=   K_{3}(g-g_{\rm safe})+ K_{4}(g, v, \Delta v)\Delta v,
\label{Helly_st2}
\end{equation}
where
\begin{equation}
K_{4}=\left\{\begin{array}{ll}
K^{(1)}_{4}   \ \textrm{at $\Delta v > 0$}, \\
K^{(2)}_{4} \frac{v\tau_{\rm safe}}{g} \ \textrm{at $ \Delta v \leq 0$}. \\
\end{array} \right.
\label{Helly_st3}
\end{equation}
  $K_{3}$, $K^{(1)}_{4}$,  and $K^{(2)}_{4}$  are    positive dynamic coefficients. To make   vehicle deceleration without jumps at $\Delta v <0$,
when the space gap $g$ intersects boundaries $G$ and $g_{\rm safe}$ of the indifferent zone,
 in (\ref{g_v_g_min1}) we apply~\footnote{Contrary to~\cite{Lee2004}, 
	in model (\ref{g_v_g_min1})--(\ref{K_Deltav2}) no different states (like optimistic
state or defensive state of~\cite{Lee2004}) are assumed in collision-free
traffic dynamics governed by safety vehicle braking.}
\begin{equation}
K_{\Delta v}=\left\{\begin{array}{ll}
K_{2}   \ \textrm{at $\Delta v > 0$}, \\
K^{(1)}_{\Delta v}  \ \textrm{at $ \Delta v \leq 0$}, \\
\end{array} \right.
\label{K_Deltav}
\end{equation}
where
\begin{equation} 
K^{(1)}_{\Delta v}=\Biggl(K_{2}-K^{(2)}_{4} \frac{v\tau_{\rm safe}}{g}\Biggr)\Biggl(\frac{g-g_{\rm safe}}{G-g_{\rm safe}}\Biggr) +  K^{(2)}_{4} \frac{v\tau_{\rm safe}}{g}.
\label{K_Deltav2}
\end{equation}
 
In addition to traffic flow consisting of human-driving vehicles, the   model (\ref{g_v_g_min1})--(\ref{K_Deltav2})
is also applicable for traffic flow  consisting of automated-driving vehicles as well as for mixed traffic flow in which  
automated-driving vehicles move  with the use of three-phase adaptive
cruise control (TPACC)~\cite{TPACC,TPACC2}~\footnote{The basic feature of TPACC-model is the existence of 
the indifferent zone for car-following applied
in Eq.~(\ref{g_v_g_min1}). In comparison with the TPACC-model of~\cite{TPACC,TPACC2}, in the   TPACC-model 
(\ref{g_v_g_min1})--(\ref{K_Deltav2})   vehicle overacceleration $a_{\rm OA}$ (\ref{a_OA}) of~\cite{Kerner2023B} as well as
 further developments (\ref{a_G-f})--(\ref{K_Deltav2}) of the basic model  
(\ref{g_v_g_min1})--(\ref{a_OA}) are added.
}.

\subsection{Models of lane-changing and on-ramp bottleneck \label{Mode_2-Lane_S}}

We use well-known incentive lane changing rules from the right to left lane R$\rightarrow$L
(\ref{RL}) and from the left to right lane L$\rightarrow$R
(\ref{LR}) as well as well-known  safety conditions (\ref{g_prec_ACC}) 
 (see, e.g.,~\cite{Nagel1998}) 
\begin{eqnarray}
\label{RL}
R \rightarrow L: v^{+}(t) \geq v_{\ell}(t)+\delta_{1} \ {\rm and} \ v(t) \geq v_{\ell}(t), \\
L \rightarrow R: v^{+}(t) \geq v_{\ell}(t)+\delta_{2} \ {\rm or} \  v^{+}(t) \geq v(t)+\delta_{2},
\label{LR} \\
g^{+}(t)   \geq\ v(t) \tau_{2}, \quad g^-(t) \geq\ v^{-}(t) \tau_{1},
\label{g_prec_ACC}
\end{eqnarray} 
at which
a vehicle changes to the faster target lane 
with the objective to pass a slower  vehicle in the current lane     if   time headway to  preceding and following vehicles in the target  lane   are not shorter than some given safety time headway $\tau_{1}$
and $\tau_{2}$. In   (\ref{RL})--(\ref{g_prec_ACC}),
superscripts $+$  and $-$ denote, respectively, the preceding and the following vehicles in the target lane;
 $\tau_{1}$, $\tau_{2}$,   $\delta_{1}$, $\delta_{2}$  are positive constants~\footnote{It should be noted that  
in (\ref{RL}), (\ref{LR}) the value $v^{+}$ at $g^{+} > L_{\rm a}$ and the
 value $v_{\ell}$ at $g > L_{\rm a}$ are replaced by $\infty$, where $L_{\rm a}$ is a look-ahead distance; in simulations, we have used
$L_{\rm a}=$ 80 m. }.
  
Open boundary conditions are applied. At the beginning of the two-lane  road $x=0$ vehicles  are generated one after another
in each of the lanes of the road at time instants
$t^{(k)}=k\tau_{\rm in}$, $k=1,2,\ldots$, 
where $\tau_{\rm in}=1/q_{\rm in}$, $q_{\rm in}$ is a given time-independent flow rate   per road lane.
The initial vehicle speed is equal to  
$v_{\rm free}$. After the vehicle has reached the end of the road $x=L$ it is removed.
Before this occurs, the farthest downstream vehicle  maintains its speed and lane.

In the on-ramp model, there is
a merging region of length $L_{\rm m}$ in the right   lane that begins at   location $x=x_{\rm on}$
within which  vehicles can merge from the on-ramp.  
  Vehicles
are generated at the on-ramp one after another at time instants
$t^{(m)}=m\tau_{\rm on}$, $m=1,2,\ldots$, 
where $\tau_{\rm on}=1/q_{\rm on}$, $q_{\rm on}$ is the on-ramp inflow rate.  To reduce
a  local  speed decrease occurring through the vehicle merging at the on-ramp bottleneck, vehicles merge   with the speed of the preceding vehicle $v^{+}$ at a middle location
$x=(x^{+}+x^{-})/2$ between  the preceding and following vehicles in the right lane, when the space gap between the  vehicles 
  exceeds some safety value $g^{\rm (min)}_{\rm target}=\lambda_{\rm b}v^{+}+ d$, i.e., some safety condition
$x^{+}-x^{-}-d>g^{\rm (min)}_{\rm target}$ should be
  satisfied. In accordance with these merging conditions,
 the space gap for a vehicle merging between each pair of consecutive vehicles in the right   lane is checked 
within merging region $L_{\rm m}$, starting from the upstream boundary of the merging region. If there is such a pair of consecutive vehicles, the vehicle merges onto the right   lane;
 if there is no pair of consecutive vehicles, for  which the safety condition   is satisfied at the current time step, the procedure is repeated at the next time step,
and so on.

 Vehicle motion   is found 
from  a system of  equations $a^{\ast}=\min(a, a_{\rm max})$,
$dv^{\ast}/dt=a^{\ast}$,   $v=\max(0, \min(v^{\ast},  v_{\rm free}))$  that results from conditions $0 \leq v \leq v_{\rm free}$, and   $dx/dt=v$ solved with the second-order Runge-Kutta method with time step 
$10^{-2}$ s.  

In the main text of the paper, 
 we use   simple speed-functions $G(v)$ and $g_{\rm safe}(v)$:
\begin{equation}
G(v)=v\tau_{\rm G}, \quad g_{\rm safe}(v)=v\tau_{\rm safe},
\label{G_g_safe_simple}
\end{equation} 
where $\tau_{\rm G}$ is a synchronization time headway.
We choose   dynamic coefficients  $K_{3}$ and $K_{4}$ in (\ref{Helly_st2}), (\ref{Helly_st3}) at which
under conditions (\ref{G_g_safe_simple}) no classical traffic flow instability (no string instability) 
		can occur. Respectively, no wide moving jams (and no cases at which $v\rightarrow 0$) can occur in 
		synchronized flow of the   model (\ref{g_v_g_min1})--(\ref{G_g_safe_simple}).

\section{Overacceleration caused by safety acceleration  \label{Safety_Acc_S}}  

We show that in addition with the
overacceleration mechanisms  in a road lane (\ref{a_OA})~\cite{Kerner2023B} and 
through lane-changing~\cite{Kerner2023C} 
 there are other mechanisms of overacceleration in the road lane caused by safety acceleration $a_{\rm safety}$.
 To understand this,   we should consider the overacceleration mechanisms
 separately from each other. With this aim, in   Sec.~\ref{Safety_Acc_S} we consider a single-lane road and 
	   neglect overacceleration   (\ref{a_OA}) of Ref.~\cite{Kerner2023B}, i.e.,  in (\ref{a_OA})
	we get
	\begin{equation}
	\alpha=0.  
	\label{alpha_0}
	\end{equation}

		\subsection{No overacceleration -- no nucleation nature of F$\rightarrow$S transition \label{1-lane-neglecting}}
		
		First, we consider a case in which under condition (\ref{alpha_0}) safety acceleration behavior does not act as    overacceleration  
	  (Fig.~\ref{LWR_Fig}):  
		There is no 
acceleration behavior  that  could
cause the free flow metastability with respect to traffic breakdown at the
bottleneck. Indeed, in Fig.~\ref{LWR_Fig}, contrary to empirical data,  traffic breakdown  does not exhibit
  the nucleation nature. 

		 \begin{figure} 
\begin{center}
\includegraphics[width = 8 cm]{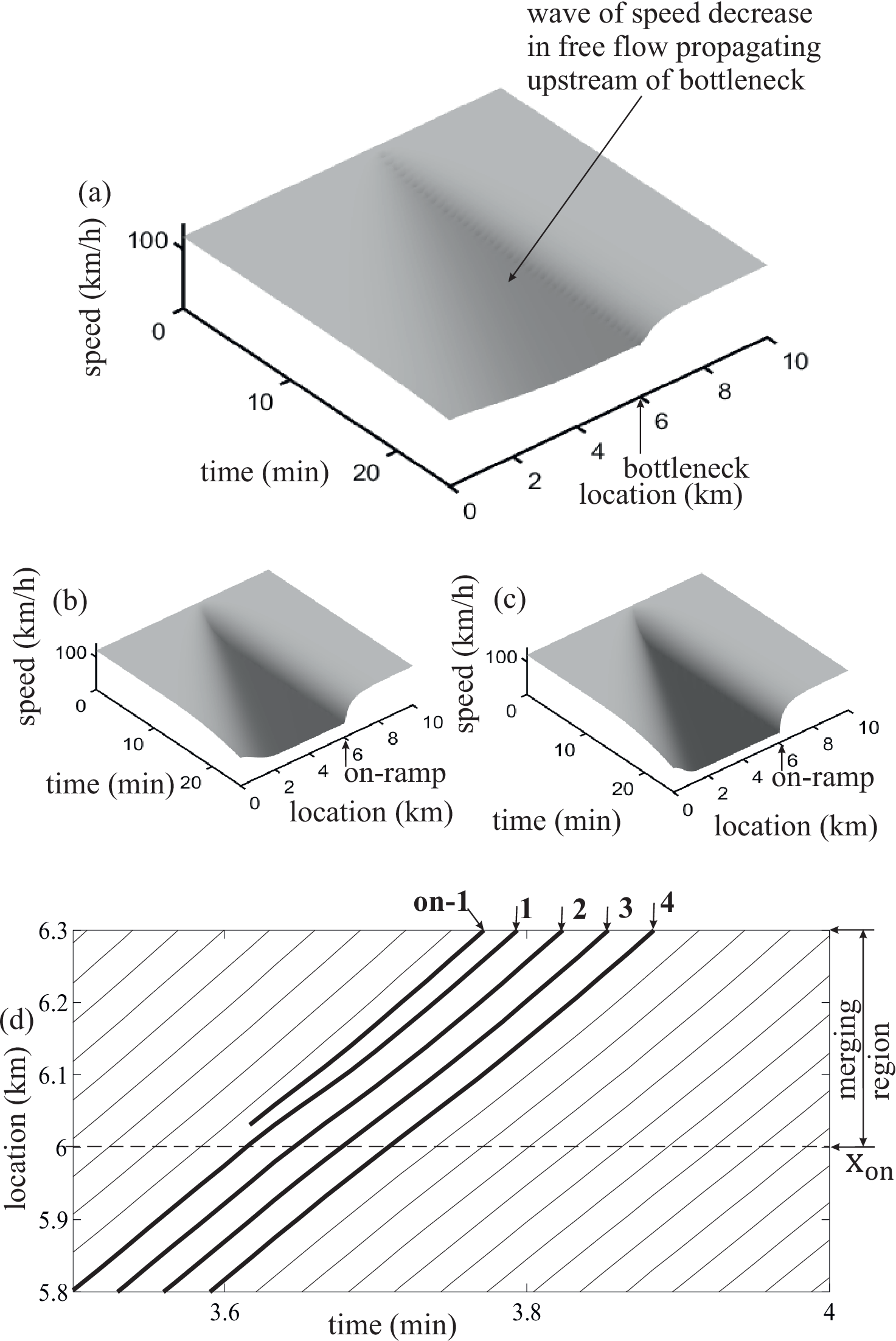}
\end{center}
\caption[]{Explanation to $\lq\lq$no overacceleration -- no nucleation nature of F$\rightarrow$S transition".  
 Simulations of model (\ref{g_v_g_min1})--(\ref{K_Deltav2}), (\ref{G_g_safe_simple}),
		(\ref{alpha_0})
at   $q_{\rm in}=$ 2000 vehicles/h 
and different values of on-ramp inflow rate $q_{\rm on}$ at
 coefficient of overacceleration $\alpha=$ 0 (\ref{alpha_0}), i.e., in (\ref{g_v_g_min1})
and (\ref{a_G-f}) $a_{\rm OA}=$ 0.
(a)--(c) Speed in space and time: (a)  $q_{\rm on}=$ 100 vehicles/h,   (b)  $q_{\rm on}=$ 350 vehicles/h, (c)  $q_{\rm on}=$ 500 vehicles/h.   (d) Simulated vehicle trajectories in some time interval   related to (a).
Other model parameters:  $v_{\rm free}=$ 120 km/h,  $\tau_{\rm safe}=$ 1 s, $\tau_{\rm G}=$ 2 s,
  $v_{\rm syn}=$ 80 km/h,
 $d=$ 7.5 m,  $a_{\rm max}=$ 2.5 $\rm m/s^{2}$,   $K_{1}=   0.3 \ s^{-2}$, $K_{2}= K^{(1)}_{4}= 0.6 \ s^{-1}$, 
$K_{3}= 0.5 \ s^{-2}$, 
 $K^{(2)}_{4}= 1 \ s^{-1}$,   $L=$ 10 km, $x_{\rm on}=$ 6 km, $\lambda_{\rm b}=$ 0.2 s,  $L_{\rm m}=$ 0.3 km.
}
\label{LWR_Fig}
\end{figure}

\begin{figure}
\begin{center}
\includegraphics[width = 8 cm]{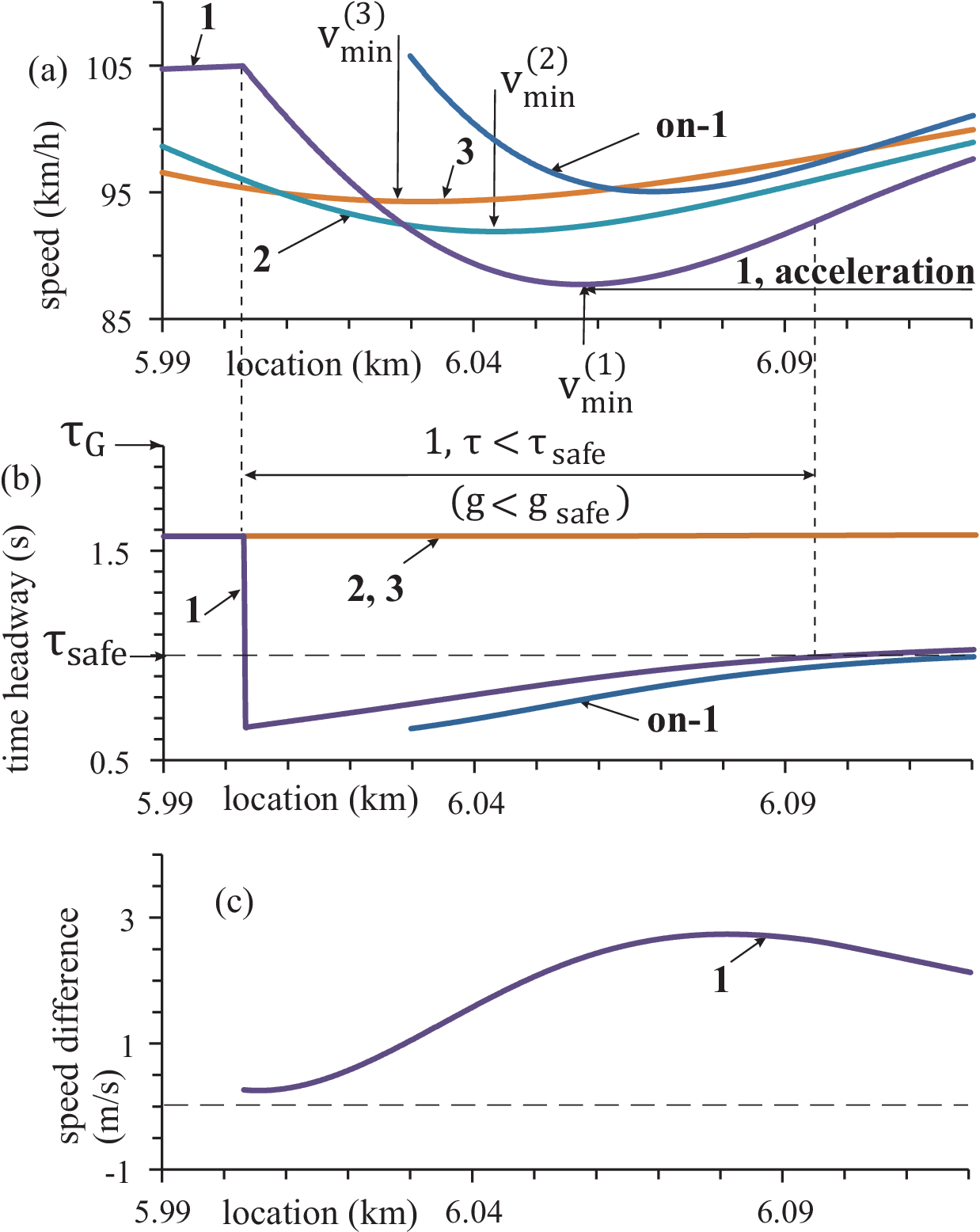}
\end{center}
\caption[]{Continuation of Fig.~\ref{LWR_Fig}(d). (a), (b) 
Location-functions of speed (a) and time-headway (b)  of some vehicles.  (c)
Location-function of speed difference $\Delta v= v^{\rm (on-1)}-v^{(1)}$  between speeds of vehicles on-1 and 1.
Vehicle numbers 
are the same   as those  in Fig.~\ref{LWR_Fig}(d), respectively.
}
\label{No-overacceleration_traj}
\end{figure}

To explain this, we consider Fig.~\ref{No-overacceleration_traj}. After vehicle on-1 has merged from the on-ramp onto the main road
[Fig.~\ref{LWR_Fig}(d)],   time-headway of vehicle 1  following  vehicle on-1 on the main  road
becomes smaller than safe time-headway $\tau_{\rm safe}$
[Fig.~\ref{No-overacceleration_traj}(b)]. Therefore,
according to Eqs.~(\ref{Helly_st2}) and (\ref{Helly_st3}), vehicle  1 should   decelerate strongly resulting in
  deceleration of   following vehicles 2 and 3 [Fig.~\ref{No-overacceleration_traj}(a)].
Deceleration  of  
		vehicles 2 and 3 is realized within the indifferent zone for car-following
		$g_{\rm safe} \leq g \leq G$  [$\tau_{\rm safe} \leq \tau \leq \tau_{\rm G}$ in Fig.~\ref{No-overacceleration_traj}(b)]  in accordance with equation  
		\begin{equation}
a =  K_{\Delta v}\Delta v   \ \textrm{at $g_{\rm safe} \leq g \leq G$} 
\label{g_v_g_min1_SA}  
\end{equation}
resulting from Eqs.~(\ref{g_v_g_min1}) and
		(\ref{alpha_0}).
Eq.~(\ref{g_v_g_min1_SA}) describes vehicle {\it speed adaptation} to the speed of the preceding vehicle: After vehicle 1 begins to decelerate,
 following vehicle 2    decelerates while adapting its speed to the speed of   vehicle 1; the deceleration of vehicle 2
leads to the deceleration of vehicle 3, and so on.
		   Locations of the minimum speed of vehicles 2, 3, $\ldots$
		(e.g., $v^{(2)}_{\rm min}$ and $v^{(3)}_{\rm min}$)
 are {\it  upstream} of the location of the minimum speed of vehicle 1  
			 [Fig.~\ref{No-overacceleration_traj}(a)].
This explains the occurrence of the wave of a speed decrease
propagating upstream of the bottleneck [Fig.~\ref{LWR_Fig}(a)]; at a given $q_{\rm in}$,
		the larger the on-ramp inflow rate $q_{\rm on}$, the lower the speed in this congested traffic [Figs.~\ref{LWR_Fig}(a)
		and~\ref{LWR_Fig}(c)].
		Such spontaneous propagation congestion upstream of the on-ramp bottleneck is
 well-known in standard traffic flow theory~\footnote{Although  
 traffic flow model (\ref{g_v_g_min1})--(\ref{K_Deltav2}), (\ref{G_g_safe_simple}),
		(\ref{alpha_0}) used in simulations presented in Fig.~\ref{LWR_Fig} is qualitatively different
		from the classical Lighthill-Whitham-Richards (LWR)
model of traffic breakdown~\cite{LW,Richards,May,Daganzo,HCM}, we have qualitative the same 
 theoretical conclusion about the occurrence of traffic congestion at the bottleneck:
As the   model (\ref{g_v_g_min1})--(\ref{K_Deltav2}), (\ref{G_g_safe_simple}),
		(\ref{alpha_0}) at model parameters used in Fig.~\ref{LWR_Fig}, the LWR model cannot
explain the empirical nucleation nature of
 traffic breakdown at the bottleneck~\cite{KernerBook1,KernerBook2,KernerBook3,KernerBook4,KernerBook5}.}.
In accordance with overacceleration definition (Sec.~\ref{Int}),
 in the model used in Fig.~\ref{LWR_Fig}, in which no nucleation nature of
 traffic breakdown  is realized,  safety acceleration [labeled by $\lq\lq$1, acceleration" in Fig.~\ref{No-overacceleration_traj}(a)] cannot be considered   overacceleration.

\subsection{ Overacceleration under decrease in time-headway between vehicles in free flow     \label{Close_Safe_S}}  

 We consider 
  the same   model (\ref{g_v_g_min1})--(\ref{Helly_st3}), (\ref{G_g_safe_simple}),
		(\ref{alpha_0})   as used in Fig.~\ref{LWR_Fig}. However, we
		  decrease time-headway between vehicles in free flow   through an increase in the flow rate on the main road
		$q_{\rm in}$. Then, rather than the speed wave propagating {\it upstream} of the bottleneck [Fig.~\ref{LWR_Fig}(a)],
		the   wave of a  speed decrease propagating {\it downstream} of the bottleneck occurs
		[Fig.~\ref{Greater_q-in_induced}(a)]. As a result, free flow remains   upstream of the bottleneck. 
		Moreover, if in simulations shown in Fig.~\ref{Greater_q-in_induced}(a)
		at some time instant $T_{\rm ind}$ we apply   a short-time impulse of 
		addition large enough on-ramp inflow   $\Delta q_{\rm on}$, synchronized flow propagating upstream
		has been induced in the initial free flow at the bottleneck [Fig.~\ref{Greater_q-in_induced}(b)].
		Thus, contrary to Fig.~\ref{LWR_Fig}, in Fig.~\ref{Greater_q-in_induced} free flow becomes
	  metastable  with respect to the F$\rightarrow$S transition.

				 \begin{figure}
\begin{center}
\includegraphics[width = 8 cm]{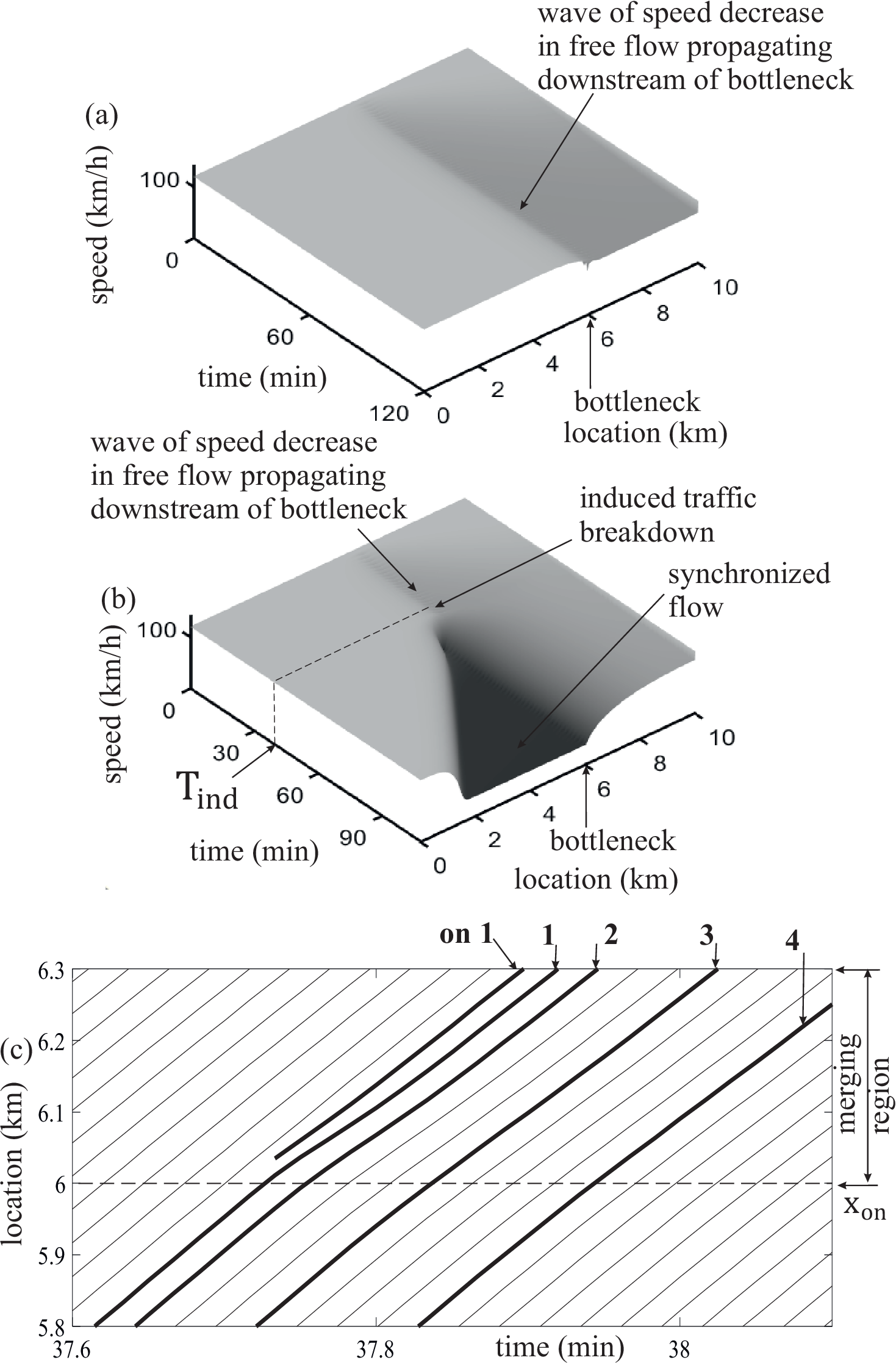}
\end{center}
\caption[]{Overacceleration in single-lane road under decrease in time-headway in free flow upstream of on-ramp bottleneck.  
Simulations of the  model (\ref{g_v_g_min1})--(\ref{K_Deltav2}), (\ref{G_g_safe_simple}),
		(\ref{alpha_0}) under the same model parameters 
   as those in Fig.~\ref{LWR_Fig}(a) with the exceptions of the choice of another
	coefficient   $K^{(2)}_{4}= 0.8 \ s^{-1}$  and a higher flow rate $q_{\rm in}=$
  2250 vehicles. 
 (a), (b) Speed in space and time at $q_{\rm on}=$ 35 vehicles/h. In (b), traffic breakdown   has been induced  
  in   free flow   at $T_{\rm ind}=$ 40 min through application of  addition on-ramp inflow impulse $\Delta q_{\rm on}=$
500 vehicles/h of duration   $\Delta t=$ 1 min.  (c) Some of simulated vehicle trajectories within  local speed decrease in free flow
at  bottleneck at time $t < T_{\rm ind}$. 
}
\label{Greater_q-in_induced}
\end{figure}

To understand the crucial difference between Figs.~\ref{LWR_Fig} and~\ref{Greater_q-in_induced}, first we  note that
 there are common  features in behaviors of vehicles on-1 and 1 following each other
[Figs.~\ref{No-overacceleration_traj} and~\ref{Greater_q-in_traj}]:
After vehicle on-1 has merged from the on-ramp 
[Figs.~\ref{LWR_Fig}(d) and~\ref{Greater_q-in_induced}(c)],   space-gap of vehicle 1 
becomes smaller than safe space-gap $g_{\rm safe}$ (i.e., time-headway is shorter than $\tau_{\rm safe}$)
[Figs.~\ref{No-overacceleration_traj}(b) and~\ref{Greater_q-in_traj}(b)]. Therefore,
according to Eqs.~(\ref{Helly_st2}) and (\ref{Helly_st3}), vehicle  1 should   decelerate strongly
 [Figs.~\ref{No-overacceleration_traj}(a) and~\ref{Greater_q-in_traj}(a)].
Shortly  later, although condition $g<g_{\rm safe}$ is still satisfied,
 this braking of vehicle 1 changes to its acceleration at location labeled by $v^{(1)}_{\rm min}$ in
	Figs.~\ref{No-overacceleration_traj}(a) and~\ref{Greater_q-in_traj}(a).   
	 This safety acceleration of vehicle 1 under  condition $g<g_{\rm safe}$
	occurs because the speed difference $\Delta v= v^{(on-1)}-v^{(1)}$  between   vehicles on-1 and 1
		becomes a large enough  value [Figs.~\ref{No-overacceleration_traj}(c) and~\ref{Greater_q-in_traj}(c)] leading to
   condition  
 \begin{equation}
a_{\rm safety}= K_{3}(g-g_{\rm safe})+ K^{(1)}_{4} \Delta v > 0 \ {\rm at} \ g<g_{\rm safe}.
\label{Helly_st4}
\end{equation}

The   crucial difference  between 
 Figs.~\ref{LWR_Fig} and~\ref{Greater_q-in_induced} is associated with  behaviors
of vehicles 2, 3, and 4 that follow vehicle 1 on the main road
[Figs.~\ref{No-overacceleration_traj}(a) and~\ref{Greater_q-in_traj}(a)].
 In comparison with Fig.~\ref{No-overacceleration_traj}(a),  
		  in Fig.~\ref{Greater_q-in_traj}(a) 
  time-headway between vehicles  in free flow  are considerably shorter.  
As a result,  locations of the speed minimum    of vehicles 2, 3, and 4 ($v^{(2)}_{\rm min}$,
$v^{(3)}_{\rm min}$, and $v^{(4)}_{\rm min}$, respectively) following vehicle
 1    are {\it downstream} of the beginning of the
 on-ramp merging region  $x_{\rm on}$ [Figs.~\ref{Greater_q-in_traj}(a) and~\ref{Greater_q-in_traj}(b)]. Moreover,  the distances between
 the location of the minimum speed of vehicle 1 [$v^{(1)}_{\rm min}$ in Fig.~\ref{Greater_q-in_traj}(a)] and   locations  
			of the minimum speeds of the following vehicles    increase  with the vehicle number 
	[Fig.~\ref{Greater_q-in_traj}(a)]. 	 
	This explains the occurrence of the wave of a   speed decrease propagating  downstream  of the bottleneck
		[Fig.~\ref{Greater_q-in_induced}(a)]. This wave
		  results from   safety acceleration of vehicle 1
		described by Eq.~(\ref{Helly_st4}). Free flow
		is in a metastable state with respect to the F$\rightarrow$S transition   at the bottleneck
		[Fig.~\ref{Greater_q-in_induced}(b)]. Thus, in accordance with the overacceleration definition,
		  safety acceleration   should be considered overacceleration
			[$\lq\lq$1, overacceleration" in Fig.~\ref{Greater_q-in_traj}(a)].

	\begin{figure}
\begin{center}
\includegraphics[width = 8 cm]{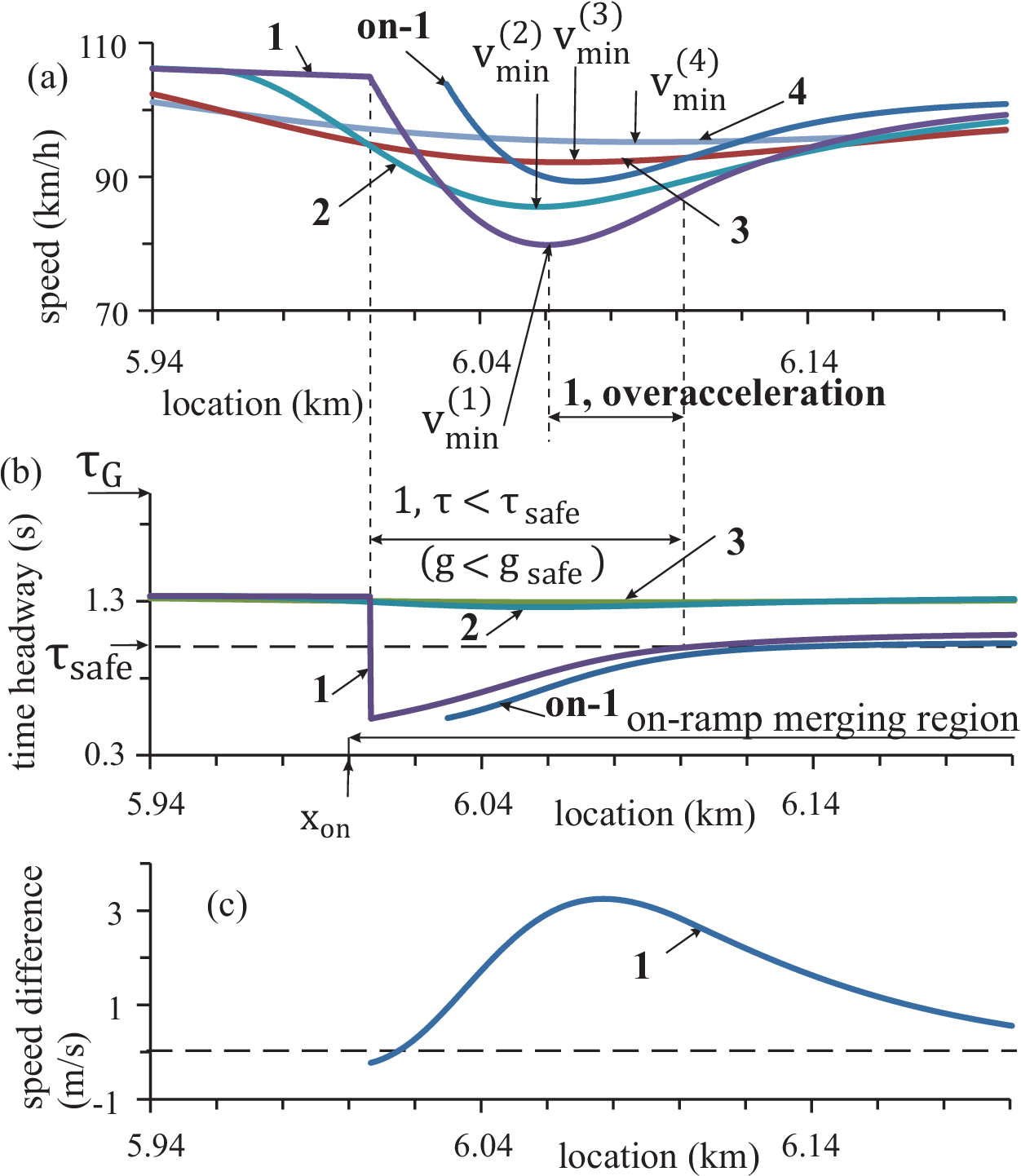}
\end{center}
\caption[]{Continuation of Fig.~\ref{Greater_q-in_induced}(c).   
(a), (b) 
Location-functions of speed (a) and time-headway (b)  of some vehicles.  (c)
Location-function of speed difference $\Delta v= v^{(on-1)}-v^{(1)}$  between speeds of vehicles on-1 and 1.
 Vehicle numbers
are labeled by the same numbers as those  in Fig.~\ref{Greater_q-in_induced}(c).     
}
\label{Greater_q-in_traj}
\end{figure}

		\subsection{Overacceleration under increase in time-headway between vehicles in free flow above
synchronization time-headway   
     \label{Greater_Syn_S}}
  
   Under condition (\ref{alpha_0})
	and at the same flow rates $q_{\rm in}$
and $q_{\rm on}$ as those in Fig.~\ref{LWR_Fig}(b), another overacceleration mechanism caused by safety acceleration
	has been found  (Fig.~\ref{Greater_G_induced}). This overacceleration mechanism is realized
	under
condition
\begin{equation}
g_{\rm free}>G, \ {\rm i.e.,} \ \tau_{\rm free}> \tau_{\rm G},
\label{g-free_G}
 \end{equation}
where $g=g_{\rm free}=\frac{v_{\rm free}}{q_{\rm in}}-d$ and $\tau=\tau_{\rm free}=g_{\rm free}/v_{\rm free}$ are, respectively, 
	the average space-gap $g$ and time-headway $\tau$ between vehicles
in free flow moving at the maximum speed $v_{\rm free}$.
Indeed, under condition (\ref{g-free_G}) rather than congestion propagation upstream 
of the bottleneck [Fig.~\ref{LWR_Fig}(b)], free flow remains at the bottleneck.
In this free flow,
  only a    small speed decrease localized at the bottleneck is realized
[Fig.~\ref{Greater_G_induced}(a)]. This free flow is in a metastable state with respect to
an F$\rightarrow$S transition at the bottleneck [Fig.~\ref{Greater_G_induced}(b)].
To understand the physics of this overacceleration effect, first we should note that
 there are common  features in behaviors of vehicles on-1 and 1  in
Figs.~\ref{No-overacceleration_traj} and~\ref{Greater_G_traj}. These common features are exactly the same as those 
explained above in Sec.~\ref{Close_Safe_S} when we have compared Figs.~\ref{No-overacceleration_traj} and~\ref{Greater_q-in_traj}.
In particular, under  condition $g<g_{\rm safe}$   safety acceleration of vehicle 1 
[Figs.~\ref{Greater_G_traj}(a) and~\ref{Greater_G_traj}(b)]
occurs in accordance with   
Eq.~(\ref{Helly_st4}).

				 \begin{figure}
\begin{center}
\includegraphics[width = 8 cm]{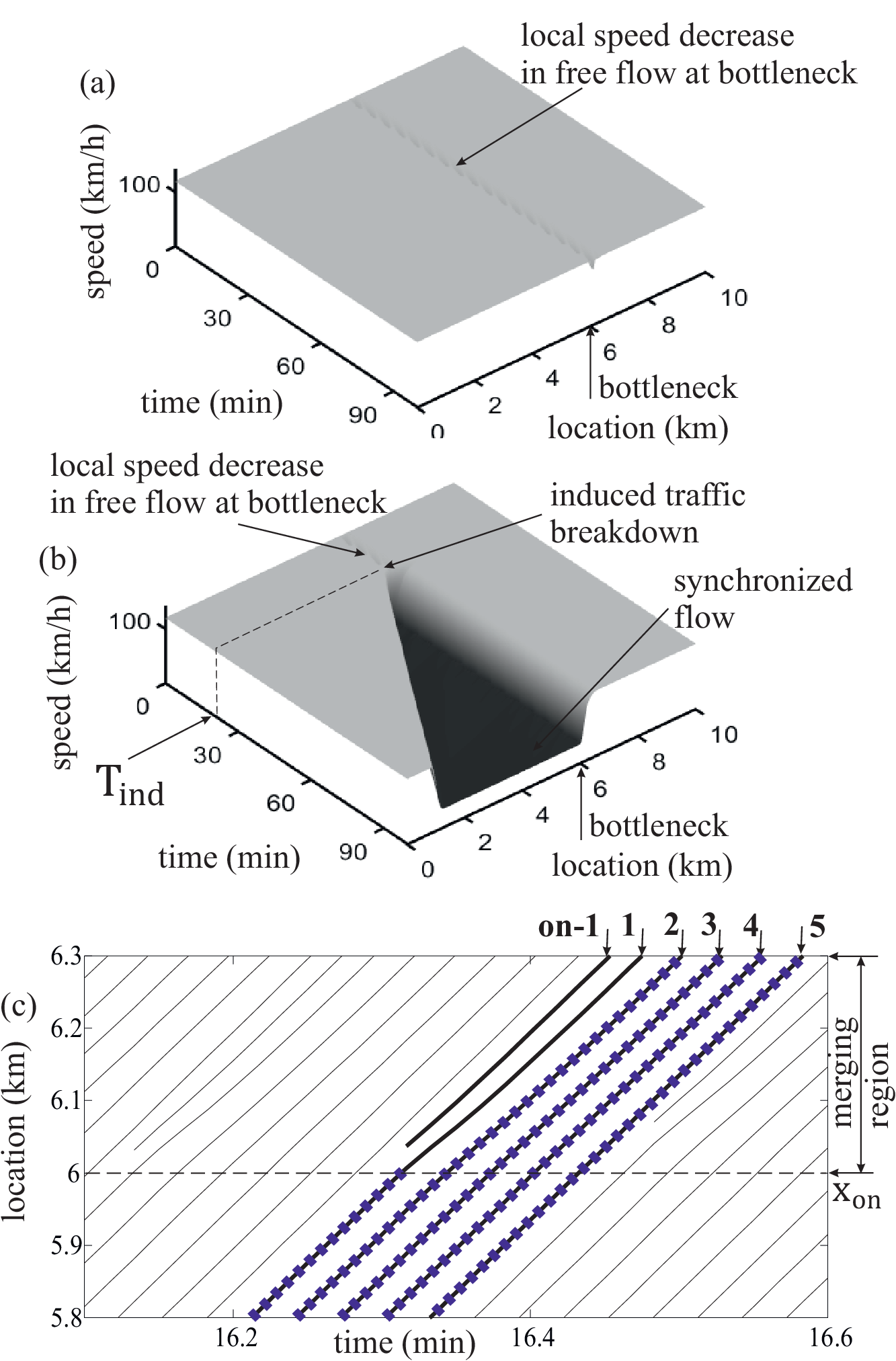}
\end{center}
\caption[]{Induced traffic breakdown on single-lane road with on-ramp bottleneck
 under condition   (\ref{g-free_G}).
 Simulations of model (\ref{g_v_g_min1})--(\ref{K_Deltav2}), (\ref{G_g_safe_simple}),
		(\ref{alpha_0}) at  $\tau_{\rm G}=$ 1.4 s,   $q_{\rm in}=$ 2000 vehicles/h, and $q_{\rm on}=$ 350 vehicles/h. 
Other model parameters are the same as those in Fig.~\ref{LWR_Fig}(a).
 (a), (b) Speed in space and time: Local speed decrease at bottleneck in free flow (a) and synchronized flow pattern (SP) (b)
induced in  free flow in (a) at $T_{\rm ind}=$ 20 min through application of  addition on-ramp inflow impulse $\Delta q_{\rm on}=$
600 vehicles/h of duration   $\Delta t=$ 1 min.  (c) Some of simulated vehicle trajectories within  local speed decrease in free flow
at  bottleneck at time $t < T_{\rm ind}$; parts of trajectories within which condition $g>G$ is satisfied
  are marked through colored blue rectangles.  
}
\label{Greater_G_induced}
\end{figure}

	\begin{figure}   
\begin{center}
\includegraphics[width = 8 cm]{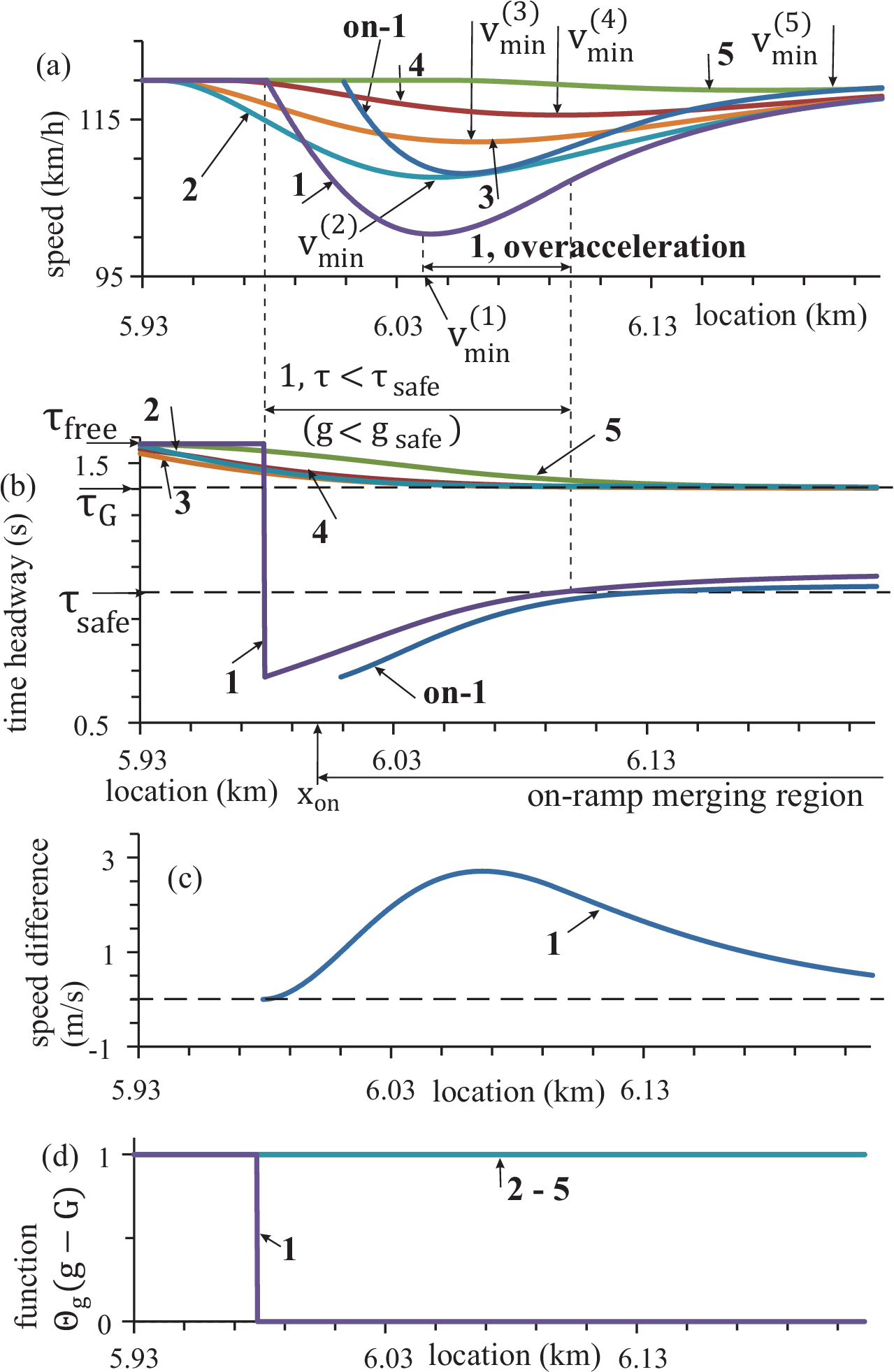}
\end{center}
\caption[]{Continuation of Fig.~\ref{Greater_G_induced}(c).   
(a), (b) 
Location-functions of speed (a) and time-headway (b)  of some vehicles.  (c)
Location-function of speed difference $\Delta v= v^{\rm (on-1)}-v^{(1)}$  between speeds of vehicles on-1 and 1.
(d) Dependencies of function $\Theta_{\rm g}(g-G)$ on road location for vehicles 1--5, where
$\Theta_{\rm g} (z) =0$ at $z<0$ and $\Theta (z) =1$ at $z\geq 0$.
 Vehicle numbers
are labeled by the same numbers as those  in Fig.~\ref{Greater_G_induced}(c).     
}
\label{Greater_G_traj}
\end{figure}

 There is the following   crucial difference  between 
 the case under consideration (Fig.~\ref{Greater_G_induced})
and the cases studied   in Secs.~\ref{1-lane-neglecting} and~\ref{Close_Safe_S}
[Figs.~\ref{LWR_Fig} and~\ref{Greater_q-in_induced}]:    
In   Figs.~\ref{LWR_Fig} and~\ref{Greater_q-in_induced},
  deceleration and acceleration of the
		vehicles 2 and 3 following vehicle 1  are realized within the indifferent zone for car-following 
		$g_{\rm safe} \leq g \leq G$ [region $\tau_{\rm safe} \leq \tau \leq \tau_{\rm G}$ in
		Figs.~\ref{No-overacceleration_traj}(b) and~\ref{Greater_q-in_traj}(b)], i.e., it corresponds to solutions of
		Eq.~(\ref{g_v_g_min1_SA}).  
	Contrarily, under condition (\ref{g-free_G}) for which Fig.~\ref{Greater_G_induced} is valid,
		vehicles 2--5 following vehicle 1 [Fig.~\ref{Greater_G_traj}(b)]
		decelerate and accelerate in accordance with Eqs.~(\ref{a_G-f}) and
		(\ref{alpha_0}) corresponding to solutions of equation
		\begin{equation}
a_{\rm G}=   K_{1}(g-v\tau_{\rm G})+ K_{2}\Delta v \ {\rm at} \ g>v\tau_{\rm G}. 
\label{g_v_g_min1_SA-2}  
\end{equation}
When vehicle 1 decelerates, then both Eqs.~(\ref{g_v_g_min1_SA}) and  (\ref{g_v_g_min1_SA-2}) 
 describe   speed adaptation of following vehicles   to the speed of   vehicle 1.
However, as long as condition $g>v\tau_{\rm G}$ is satisfied,
due to positive term $K_{1}(g-v\tau_{\rm G})$ in Eq.~(\ref{g_v_g_min1_SA-2}), at the same
negative speed difference
  $\Delta v$ in Eqs.~(\ref{g_v_g_min1_SA}) and~(\ref{g_v_g_min1_SA-2})
  the deceleration and  respective speed decrease of vehicles 2--5 in Fig.~\ref{Greater_G_traj}(a)  are considerable lower 
	than those of vehicles 2 and 3 in Fig.~\ref{No-overacceleration_traj}(a).
	
	As mentioned in Sec.~\ref{1-lane-neglecting}, a  
large speed decrease of vehicles 2 and 3 in Fig.~\ref{No-overacceleration_traj}(a)
occurs already upstream of the bottleneck leading to the wave of a  speed decrease
propagating {\it upstream} of the bottleneck [Fig.~\ref{LWR_Fig}(a)]. Contrarily, 
due to a weaker deceleration (\ref{g_v_g_min1_SA-2}), vehicles 2--5 in Fig.~\ref{Greater_G_traj}(a) reach
the bottleneck   while moving still in free flow, i.e.,
free flow remains at the bottleneck. Deceleration of vehicles 2--5 causes only
a weak wave of a  speed decrease propagating {\it downstream} of the bottleneck. Indeed, 
the locations of the speed minimum    of vehicles 2--5 [$v^{(2)}_{\rm min}$,
$v^{(3)}_{\rm min}$, $v^{(4)}_{\rm min}$, and $v^{(5)}_{\rm min}$ in Fig.~\ref{Greater_G_traj}(a), respectively] following vehicle
 1    are  downstream of the location of the speed minimum    of vehicle 1
 [$v^{(1)}_{\rm min}$ in Fig.~\ref{Greater_G_traj}(a)]; however, contrary to Fig.~\ref{Greater_q-in_induced}(a), due to the weak
deceleration of vehicle 2--5
	the wave   disappears almost fully at a short distance 
		 downstream of the beginning of the on-ramp   [Fig.~\ref{Greater_G_traj}(a)].
		For this reason, a
		speed decrease caused by vehicle merging from the on-ramp can be considered
		as a  local  speed decrease   at the bottleneck [$\lq\lq$local speed decrease" in Fig.~\ref{Greater_G_induced}(a)].

In    Fig.~\ref{Greater_G_induced}, contrary to   Fig.~\ref{LWR_Fig},  
   safety acceleration   
does cause the metastability of free flow with respect of F$\rightarrow$S transition
at the bottleneck. Thus,  due to condition (\ref{g-free_G}),    safety acceleration  (\ref{Helly_st2}),
(\ref{Helly_st3})     becomes overacceleration [$\lq\lq$1, overacceleration" in Fig.~\ref{Greater_G_traj}(a)].

 \section{Cooperation of different mechanisms of overacceleration    \label{Cooperation_Sec}}

				 \begin{figure}
\begin{center}
\includegraphics[width = 8 cm]{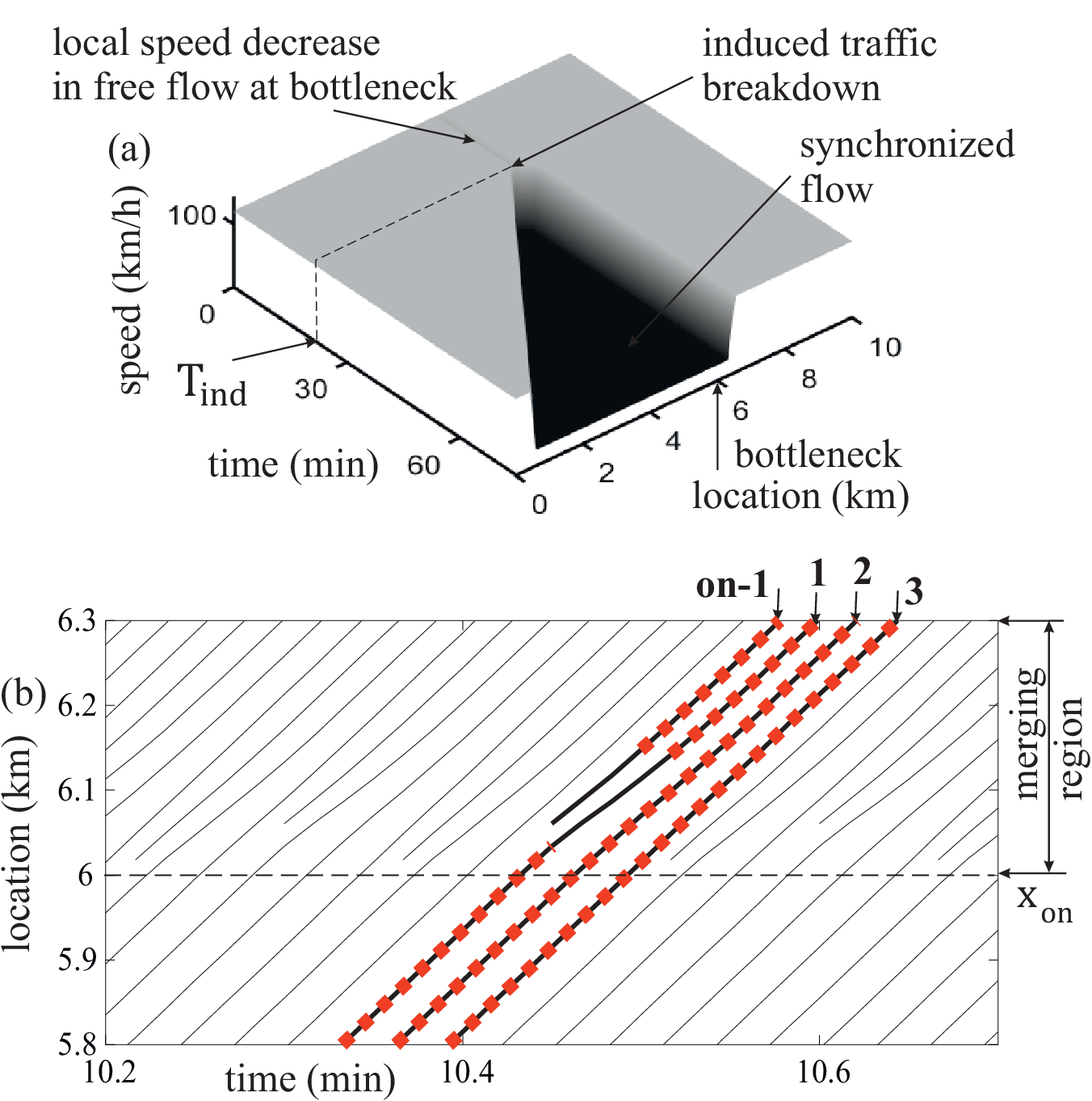}
\end{center}
\caption[]{Simulations of cooperation of different mechanisms of overacceleration on single-lane road.
Induced traffic breakdown on single-lane road with on-ramp bottleneck.
Simulations of model (\ref{g_v_g_min1})--(\ref{K_Deltav2}), (\ref{G_g_safe_simple})
with model parameters:   $\alpha_{0}=$ 2 $\rm m/s^{2}$,
$\alpha_{1}=$ 0.1 $\rm m/s^{2}$, $k=$ 1, $\tau_{\rm G}=$ 1.4 s, $q_{\rm in}=$ 2000 vehicles/h, $q_{\rm on}=$ 800 vehicles/h; other model parameters are the same as those in
 Fig.~\ref{LWR_Fig}. 
 (a)  Speed in space and time: Local speed decrease at bottleneck in free flow  and synchronized flow pattern (SP) 
induced in the free flow   at $T_{\rm ind}=$ 20 min through application of  addition on-ramp inflow impulse $\Delta q_{\rm on}=$
900 vehicles/h of duration   $\Delta t=$ 2 min.  (b) Some of simulated vehicle trajectories within  local speed decrease in free flow
at  bottleneck at time $t < T_{\rm ind}$
  before traffic breakdown has been induced; parts of trajectories within which condition $a_{\rm OA}>0$ is satisfied
  are marked through colored red squares.  
}
\label{Greater_G_induced_alpha}
\end{figure}

  We    consider 
  the  model  (\ref{g_v_g_min1})--(\ref{K_Deltav2}),  (\ref{G_g_safe_simple}) on single-lane road when condition
(\ref{alpha_0}) used in Sec.~\ref{Safety_Acc_S} is not applied. As in~\cite{Kerner2023B}, we have found that
 the application of
overacceleration mechanism    given by   (\ref{a_OA}), (\ref{a_OA_gap})  in the model
(\ref{g_v_g_min1})--(\ref{K_Deltav2}),  (\ref{G_g_safe_simple}) leads to the  free flow metastability with respect
to an F$\rightarrow$S transition (traffic breakdown) at the bottleneck.  
Here we have found that there can be   a
  cooperation of  the
overacceleration mechanism     (\ref{g_v_g_min1}), (\ref{a_OA}) of~\cite{Kerner2023B}
with the   overacceleration mechanism due to safety acceleration revealed in
Sec.~\ref{Greater_Syn_S} [Figs.~\ref{Greater_G_induced_alpha} and~\ref{Greater_G_alpha_traj}].

				 \begin{figure}  
\begin{center}
\includegraphics[width = 8 cm]{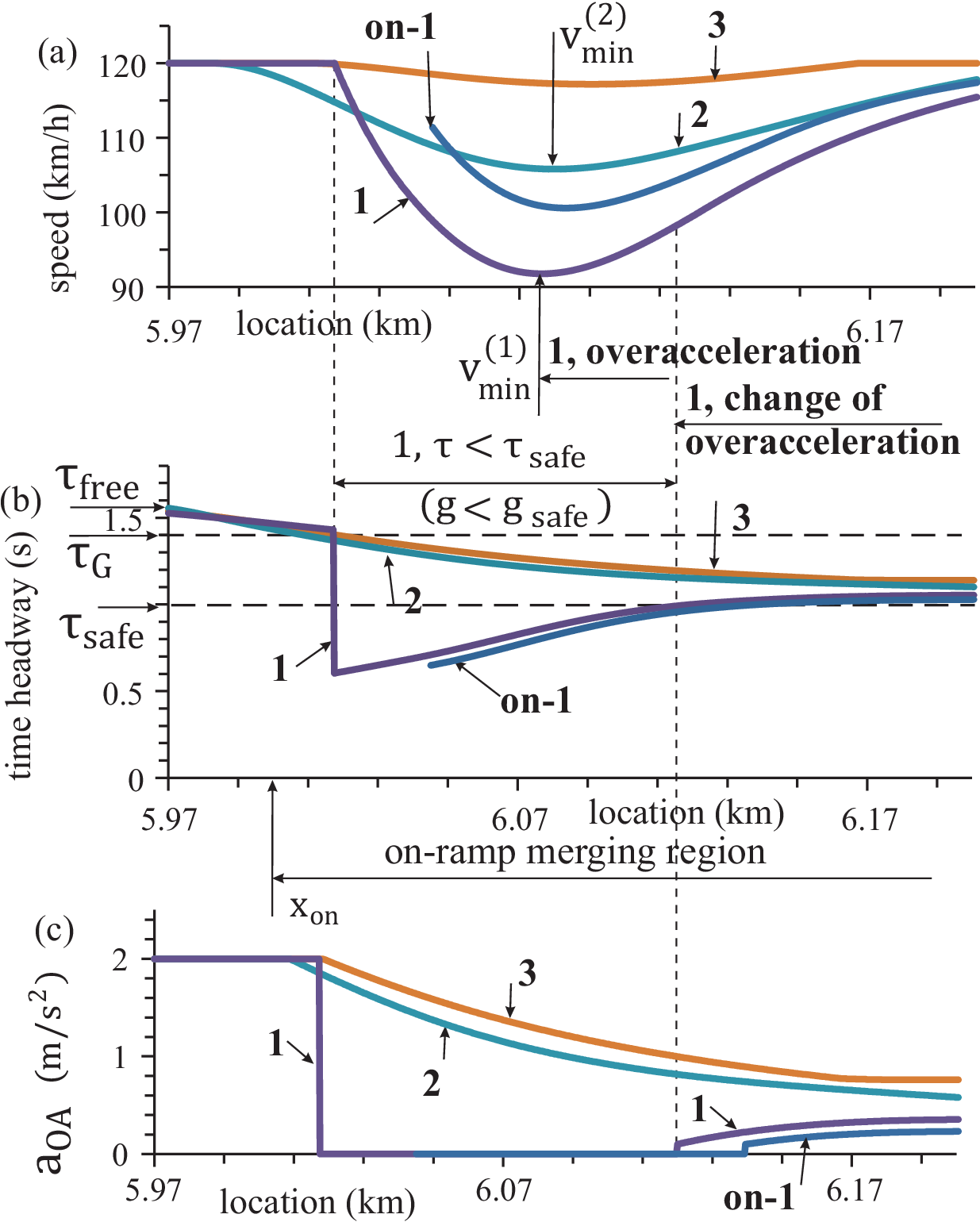}
\end{center}
\caption[]{Continuation of  Fig.~\ref{Greater_G_induced_alpha}(b).
(a)--(c) Location-dependencies of
microscopic characteristics for some of the vehicles whose numbers are the same as those in  Fig.~\ref{Greater_G_induced_alpha}(b), respectively: (a) vehicle speeds; (b) vehicle time-headway;  (c) overacceleration  $a_{\rm OA}$  (\ref{a_OA}),
 (\ref{a_OA_gap}). 
 }
\label{Greater_G_alpha_traj}
\end{figure}

   First,   overacceleration   due to safety acceleration
of vehicle 1  
[$\lq\lq$1, overacceleration" in Fig.~\ref{Greater_G_alpha_traj}(a)] following on-ramp vehicle on-1 
occurs. This overacceleration mechanism is  the same as   explained
in Sec.~\ref{Greater_Syn_S}. Later,
as can be seen from Figs.~\ref{Greater_G_alpha_traj}(b) and~\ref{Greater_G_alpha_traj}(c), 
overacceleration cooperation is realized  due to 
addition overacceleration mechanism $a_{\rm OA}$ (\ref{g_v_g_min1}), (\ref{a_OA}) 
[labeled by $\lq\lq$1, change of overacceleration" in Fig.~\ref{Greater_G_alpha_traj}(a)].
For vehicles   2 and 3 following vehicle 1, overacceleration is caused by overacceleration mechanism $a_{\rm OA}$
[Figs.~\ref{Greater_G_alpha_traj}(b) and~\ref{Greater_G_alpha_traj}(c)]
 only. In other words, overacceleration cooperation can occur over time
[time-sequence  of the effects of   different overacceleration mechanisms on vehicle 1
and the effect of overacceleration   $a_{\rm OA}$ on vehicles 2 and 3 shown
 in Fig.~\ref{Greater_G_alpha_traj}].
   
	Contrary to Fig.~\ref{Greater_G_traj}(a), due to  overacceleration
  cooperation,  the speed of
  vehicle 3 that follows vehicle 2 reaches its maximum value $v_{\rm free}$ already within the local speed decrease at the bottleneck
[Fig.~\ref{Greater_G_alpha_traj}(a)].  Thus, overacceleration   cooperation   increases  the tendency to free flow.
Indeed, in Sec.~\ref{Suppression_S}  we will show that overacceleration cooperation 
increases the maintenance of free flow   upstream at the bottleneck.
 
\section{Effect of overacceleration in road lane on overacceleration through lane-changing \label{2-lane_Coop_Sec2}}

\subsection{Prevention of overacceleration through lane-changing due to speed adaptation within the indifferent zone
\label{Prevention-2lane-Sec}}

\begin{figure}
\begin{center}
\includegraphics[width = 8 cm]{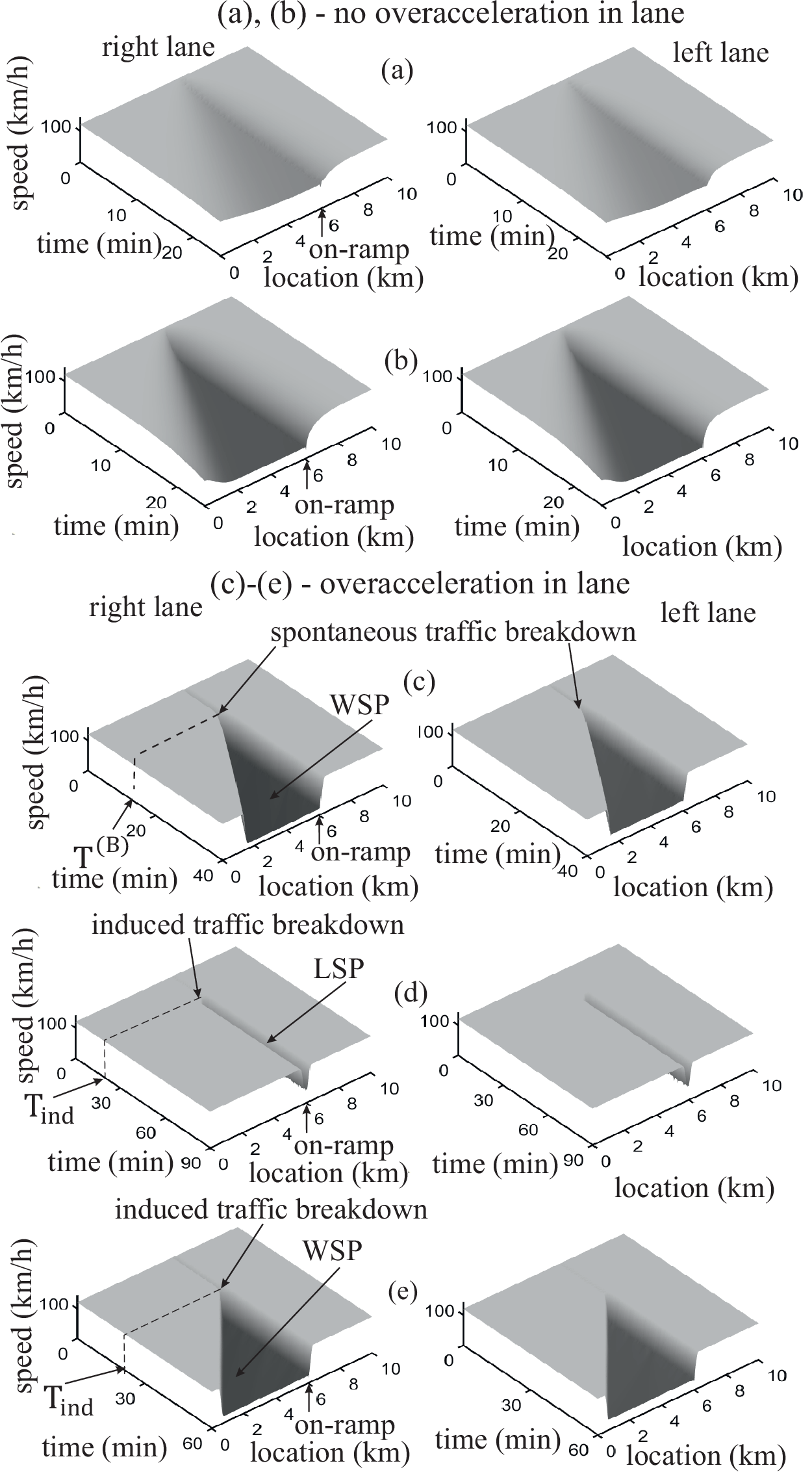}
\end{center}
\caption[]{Simulations  of the effect of overacceleration in road lane on overacceleration through lane-changing
on two-lane road with on-ramp bottleneck 
at $q_{\rm in}=$ 2000 (vehicles/h)/lane, $\tau_{\rm G}=$ 2 s and different on-ramp inflow rates.
 Speed in space and time in the right lane (left column)
and in the left lane (right column). (a), (b) No overacceleration in road lane -- no overacceleration through
lane-changing; simulations of model   (\ref{g_v_g_min1})--(\ref{alpha_0}), i.e., at  
     $a_{\rm OA}=$ 0    
		in (\ref{g_v_g_min1}), (\ref{a_G-f});  
     (a)  $q_{\rm on}=$ 300 vehicles/h,   (b)  $q_{\rm on}=$ 900 vehicles/h.
(c)--(e) Traffic breakdown  in  model
(\ref{g_v_g_min1})--(\ref{G_g_safe_simple});  parameters of overacceleration in road lane:
 $\alpha_{0}=$ 2 $\rm m/s^{2}$,
$\alpha_{1}=$ 0.1 $\rm m/s^{2}$, $k=$ 1.
(c) Spontaneous traffic breakdown with WSP formation at $q_{\rm on}=$ 1144 vehicles/h  
that exceeds slightly the maximum flow rate  $q_{\rm on}=q_{\rm on, \ max}=$ 1142 vehicles/h in free flow.
(d) LSP induced at bottleneck
 at the minimum flow rate $q_{\rm on}=q_{\rm on, \ min}=$ 517 vehicles/h   at which SP can still be induced in free flow.
(e) Induced traffic breakdown at $q_{\rm on}=$ 980 vehicles/h.
  In (d) and (e), 
F$\rightarrow$S transition  has been induced  at $T_{\rm ind}=$ 20 min  through 
 addition  on-ramp inflow-rate impulse  
$\Delta q_{\rm on}=$ 1400 vehicles/h of duration   $\Delta t=$ 2 min.  
Parameters of lane-changing: $\tau_{1}=$ 0.5 s, $\tau_{2}=$ 0.3 s, $\delta_{1}=$ 1 m/s,  $\delta_{2}=$ 5 m/s.
 Other  parameters are the same as those in
 Fig.~\ref{LWR_Fig}. WSP is widening SP, LSP is localized SP, SP is synchronized flow pattern.
}
\label{2-lane-alpha-1-patterns}
\end{figure}
 
First, we   consider traffic breakdown on two-lane road under assumption  $\alpha=$ 0 (\ref{alpha_0}).
  Then, contrary to Helly's model for motion of automated-driving vehicles used in~\cite{Kerner2023C},
in  model (\ref{g_v_g_min1})--(\ref{alpha_0}) we find that   no overacceleration through lane-changing is realized
 [Figs.~\ref{2-lane-alpha-1-patterns}(a) and~\ref{2-lane-alpha-1-patterns}(b)]. This result is explained as follows.
As shown in Sec.~\ref{1-lane-neglecting} for single-lane road (Fig.~\ref{LWR_Fig}),   vehicle 1 following on-ramp vehicle on-1 
 should   decelerate strongly resulting in
 deceleration of   following vehicles  [vehicles 2 and 3 in Fig.~\ref{No-overacceleration_traj}(a)]
due to speed adaptation (\ref{g_v_g_min1_SA}). The same effect we observe on two-lane road:
 In the right lane, a wave of a   speed decrease
propagating upstream  occurs [Figs.~\ref{2-lane-alpha-1-patterns}(a) and~\ref{2-lane-alpha-1-patterns}(b)].

The  speed adaptation effect (\ref{g_v_g_min1_SA}) is also realized  
 for vehicles in the left lane that follow a slower moving vehicle that has just
changed from the right lane to the  left lane. For this reason,
in the left lane we also observe   a wave of the   speed decrease
propagating upstream  [Figs.~\ref{2-lane-alpha-1-patterns}(a) and~\ref{2-lane-alpha-1-patterns}(b)].   
Under used  parameters in   model (\ref{g_v_g_min1})--(\ref{alpha_0}), i.e., when no overacceleration exists
in   road lane, there is also no overacceleration 
  through lane-changing and, respectively,    traffic breakdown does not exhibit 
the nucleation character. This result can be explained as follows: Speed adaptation within the indifferent zone for car-following
		   (\ref{g_v_g_min1_SA}) is stronger than vehicle acceleration to free flow downstream  
			and, therefore, the   acceleration cannot maintain free flow upstream of the bottleneck.

 \subsection{Microscopic features of overacceleration   on two-lane road \label{2-lane_Coop_Sec}}

If   assumption    $\alpha=$ 0 (\ref{alpha_0}) is not
 applied, then
 contrary to Figs.~\ref{2-lane-alpha-1-patterns}(a) and~\ref{2-lane-alpha-1-patterns}(b), 
due to the existence of  overacceleration in road lane
$a_{\rm OA}$ (\ref{a_OA})
   traffic on two-lane road becomes 
 metastable   with respect to
an F$\rightarrow$S transition [Figs.~\ref{2-lane-alpha-1-patterns}(c)--\ref{2-lane-alpha-1-patterns}(e)].
There is the following common feature of vehicle overacceleration found in~\cite{Kerner2023C}
for automated-driving vehicles moving based on Helly's model  
 and in  model (\ref{g_v_g_min1})--(\ref{G_g_safe_simple}):
When free flow
[Figs.~\ref{2-lane-alpha-1-patterns_traj}(a) and~\ref{2-lane-alpha-1-patterns_traj}(b)] 
transforms into synchronized flow [Figs.~\ref{2-lane-alpha-1-patterns_traj}(c) and~\ref{2-lane-alpha-1-patterns_traj}(d)], there is a  discontinuity  in the rate of R$\rightarrow$L lane-changing
denoted by $R_{\rm RL}$ (Fig.~\ref{2-lane-alpha-1-OA-rate}). This causes the discontinuity in
the acceleration rate       $R_{\rm OA}$ in accordance with equation $R_{\rm OA}=R_{\rm RL}$
(Fig.~\ref{2-lane-alpha-1-OA-rate}).
 There are the following differences in  
features of overacceleration in Helly's model for
 automated-driving vehicles~\cite{Kerner2023C} and     model  (\ref{g_v_g_min1})--(\ref{G_g_safe_simple})  
[Figs.~\ref{2-lane-alpha-1-patterns_traj2}--\ref{2-lane-alpha-1-patterns_traj-on-2}]:

\begin{figure}
\begin{center}
\includegraphics[width = 8 cm]{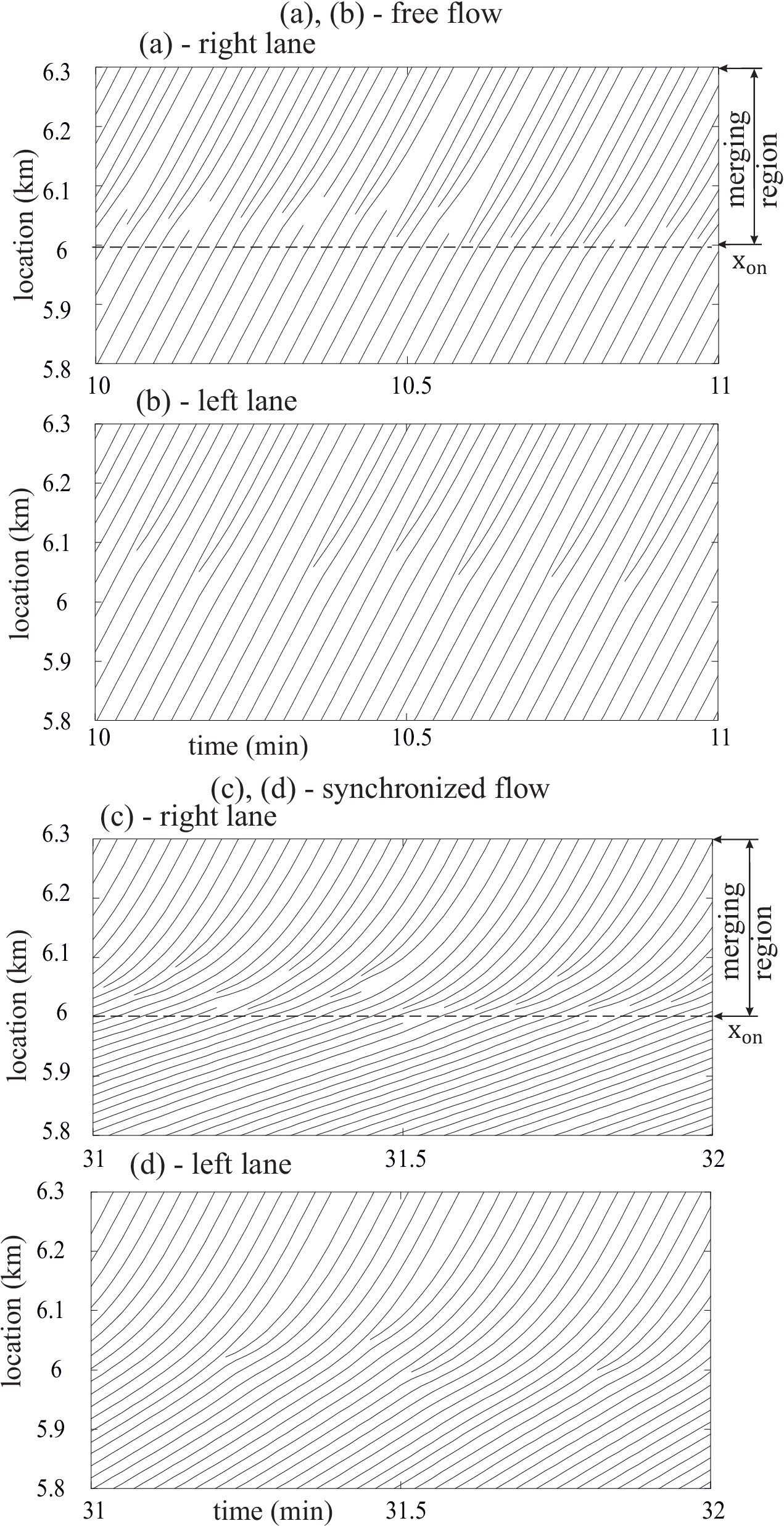}
\end{center}
\caption[]{Continuation of Fig.~\ref{2-lane-alpha-1-patterns}(e). (a), (b)
Some of simulated vehicle trajectories within  local speed decrease in free flow
at  bottleneck at time $t < T_{\rm ind}=$ 20 min  in the right (a) and left (b) lanes.
	(c), (d)
Some of simulated vehicle trajectories within    synchronized  flow
at  bottleneck at time $t > T_{\rm ind} + \Delta t$.
}
\label{2-lane-alpha-1-patterns_traj}
\end{figure}

(i)  When due to  lane-changing of vehicle 2  (Fig.~\ref{2-lane-alpha-1-patterns_traj2})
space gap $g$ between vehicle 1 and 3 becomes larger than   synchronization gap $G$ [i.e., $\tau>\tau_{\rm G}$
in Fig.~\ref{2-lane-alpha-1-patterns_traj-on-1}(b)], acceleration of   vehicle 3
given by Eq.~(\ref{a_G-f}) increases considerably due to overacceleration in road lane $a_{\rm OA}$ (\ref{a_OA}),  (\ref{a_OA_gap}) 
[Fig.~\ref{2-lane-alpha-1-patterns_traj-on-1}(c)]. As a result,   in   model 
	  (\ref{g_v_g_min1})--(\ref{G_g_safe_simple})
		vehicle 3 accelerates strongly while reaching the maximum speed $v_{\rm free}$ quickly
		[Fig.~\ref{2-lane-alpha-1-patterns_traj-on-1}(a)]~\footnote{This acceleration of vehicle 3 is realized {\it only} after vehicle 2 has changed to the left lane.
		The rate of lane-changing exhibits the discontinuous character (Fig.~\ref{2-lane-alpha-1-OA-rate}).
		For this reason, we can also call the acceleration of vehicle 3 as overacceleration [labeled as $\lq\lq$3, overacceleration" in
		Fig.~\ref{2-lane-alpha-1-patterns_traj-on-1}(a)]. This conclusion is also related to traffic flow consisting of
			automated-driving vehicles  studied in~\cite{Kerner2023C}.
		This means
		that label $\lq\lq$3, acceleration" in Fig.~8(a) of~\cite{Kerner2023C} should be replaced by $\lq\lq$3, overacceleration".}.

\begin{figure}
\begin{center}
\includegraphics[width = 8 cm]{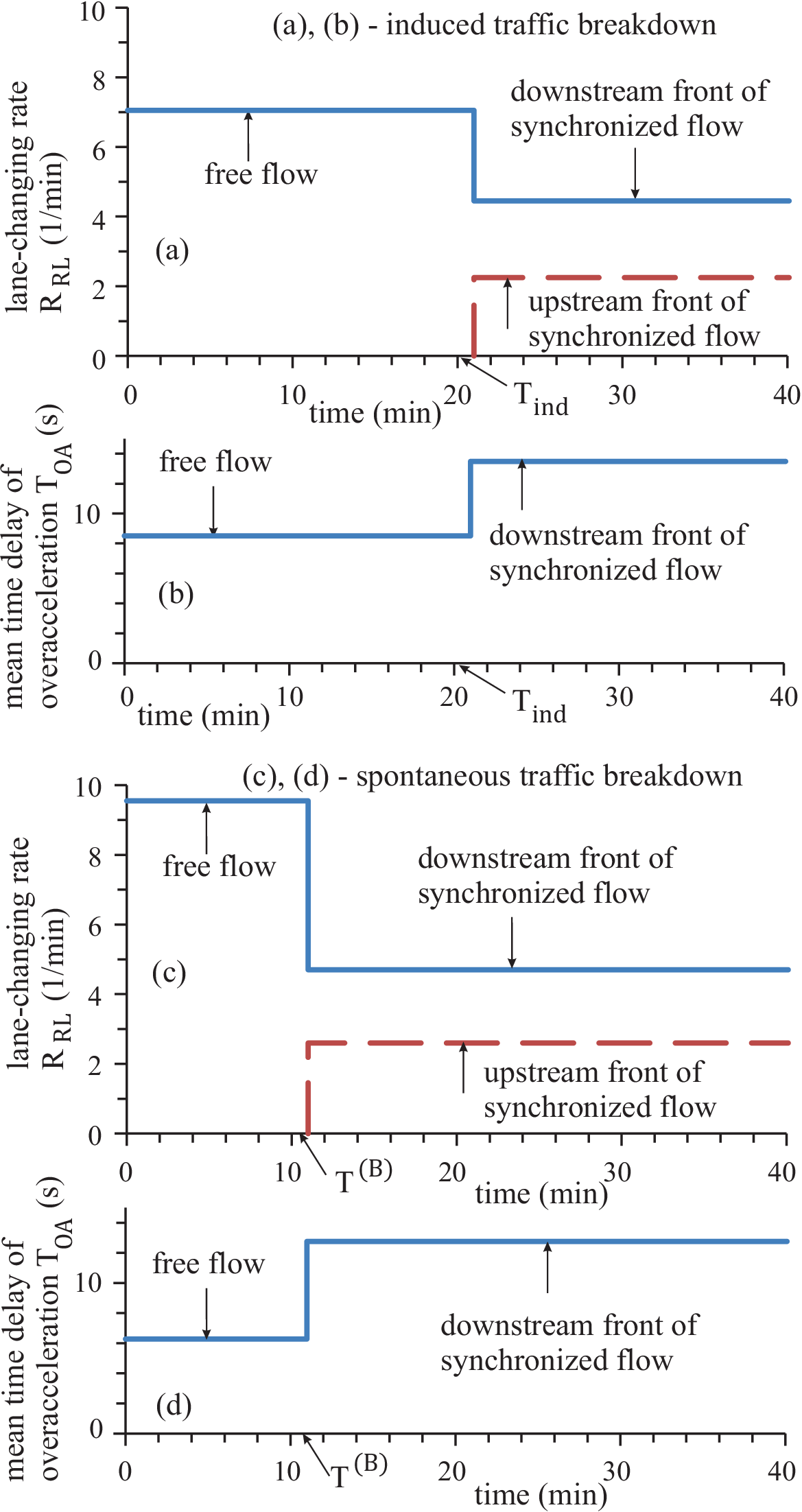}
\end{center}
\caption[]{Continuation of Figs.~\ref{2-lane-alpha-1-patterns}(c) and~\ref{2-lane-alpha-1-patterns}(e).
   Time-dependencies of the lane-changing rate $R_{\rm RL}$ (a), (c)  
 and the mean time delay in overacceleration $T_{\rm OA}=1/R_{\rm OA}$ (b), (d).
Solid blue colored curves are related to cases in which $R_{\rm OA}=R_{\rm RL}$, where   $R_{\rm OA}$ is
the overacceleration rate.
In (a), (c), dashed red colored curves are time-dependencies of 
 the lane-changing rate $R^{\rm (up)}_{\rm RL}$ at the upstream front of synchronized flow. (a), (b) For induced traffic breakdown shown in Fig.~\ref{2-lane-alpha-1-patterns}(e).
(c), (d) For spontaneous traffic breakdown shown in Fig.~\ref{2-lane-alpha-1-patterns}(c).
 Values $R_{\rm OA}$, $R^{\rm (up)}_{\rm RL}$, and $T_{\rm OA}$ have been averaged in free
flow (during time intervals $0 \leq t < T_{\rm ind}$, $T_{\rm ind}=$ 20 min
and $0 \leq t < T^{\rm (B)}$, $T^{\rm (B)}\approx$ 10.8 min for induced and spontaneous breakdowns, respectively)
and in synchronized flow
(during time intervals $T_{\rm ind}+\Delta t< t \leq 40$  min
and $T^{\rm (B)} < t \leq 40$   min for induced and spontaneous breakdowns, respectively).
}
\label{2-lane-alpha-1-OA-rate}
\end{figure}

\begin{figure} 
\begin{center}
\includegraphics[width = 8 cm]{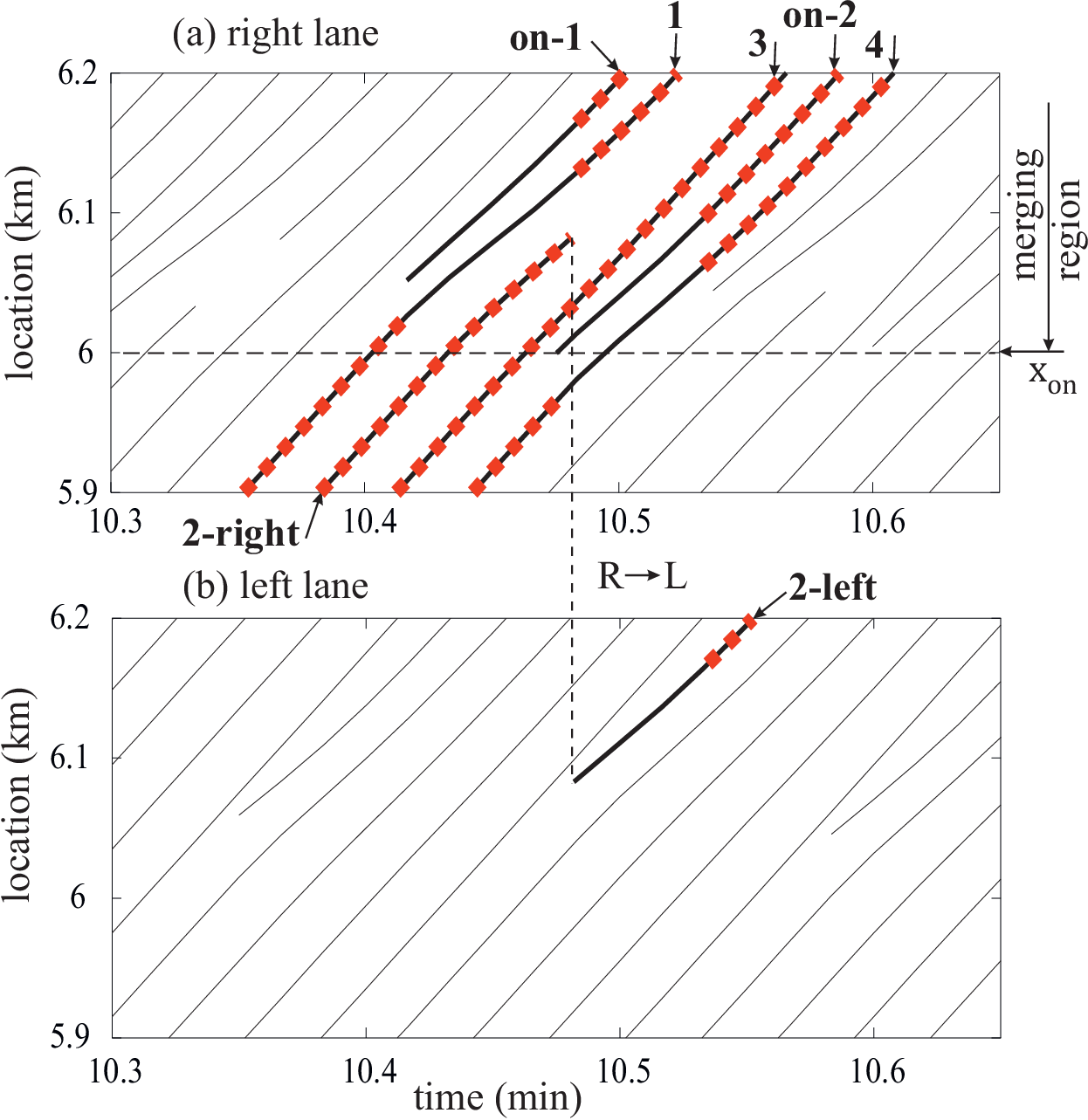}
\end{center}
\caption[]{Continuation of Fig.~\ref{2-lane-alpha-1-patterns}(e).
(a), (b) Simulated vehicle
trajectories within local speed decrease in free flow at bottleneck in
the right lane (a) and left lane (b) at time $t < T_{\rm ind}=$ 20 min. 
	R$\rightarrow$L lane-changing
of vehicle 2 is marked by dashed vertical lines R$\rightarrow$L.
Part of trajectories  within which condition $a_{\rm OA}>0$ is satisfied
  are marked through colored red squares. 
}
\label{2-lane-alpha-1-patterns_traj2}
\end{figure}

\begin{figure}
\begin{center}
\includegraphics[width = 8 cm]{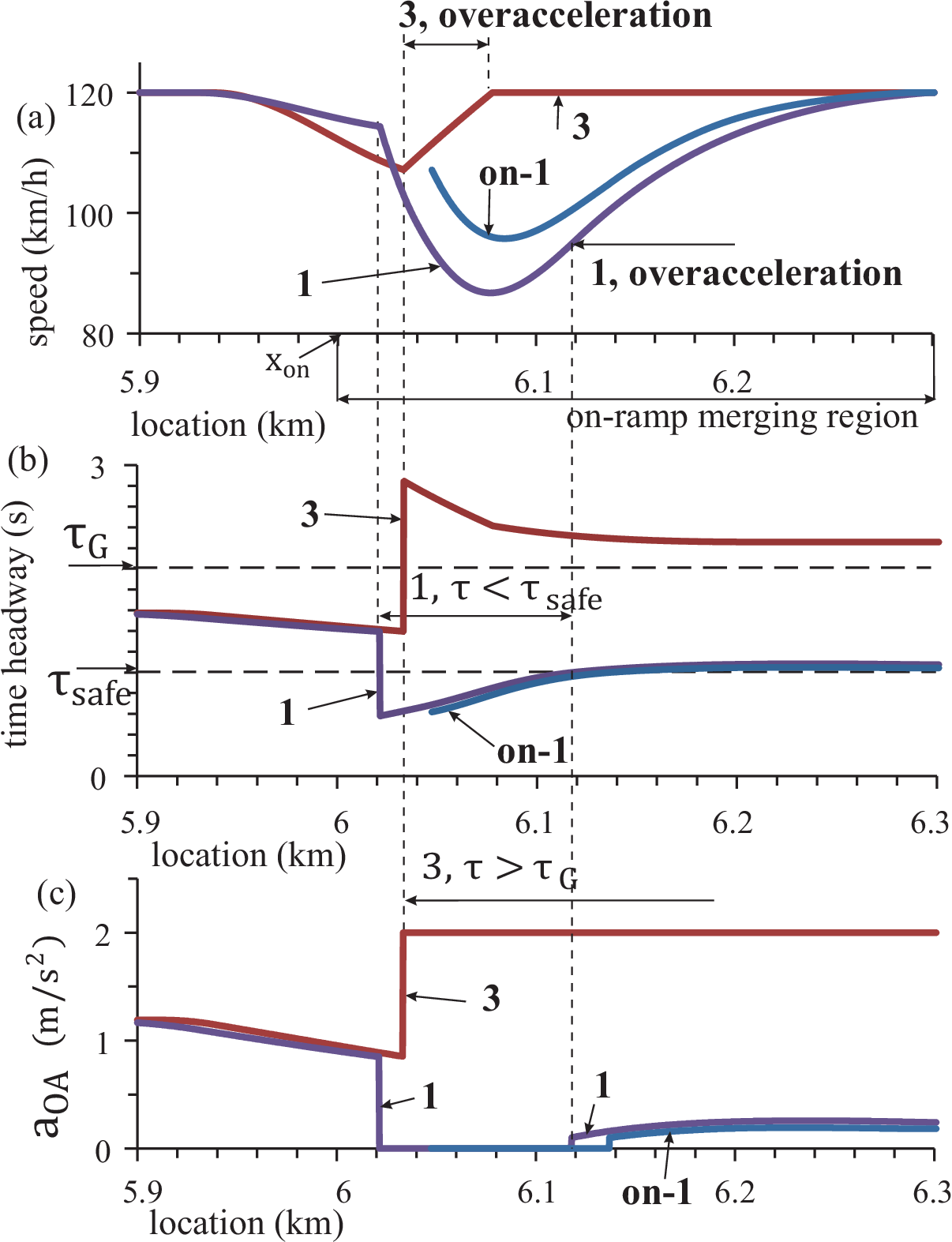}
\end{center}
\caption[]{Continuation of Fig.~\ref{2-lane-alpha-1-patterns_traj2}.
(a)--(c) Location-dependencies of
microscopic characteristics for some of the vehicles whose numbers are the same as those 
in  Fig.~\ref{2-lane-alpha-1-patterns_traj2}, respectively: (a) vehicle speeds; (b) vehicle time-headway;
(c)   overacceleration   $a_{\rm OA}$  (\ref{a_OA}),
 (\ref{a_OA_gap}). 
}
\label{2-lane-alpha-1-patterns_traj-on-1}
\end{figure}

\begin{figure}
\begin{center}
\includegraphics[width = 8 cm]{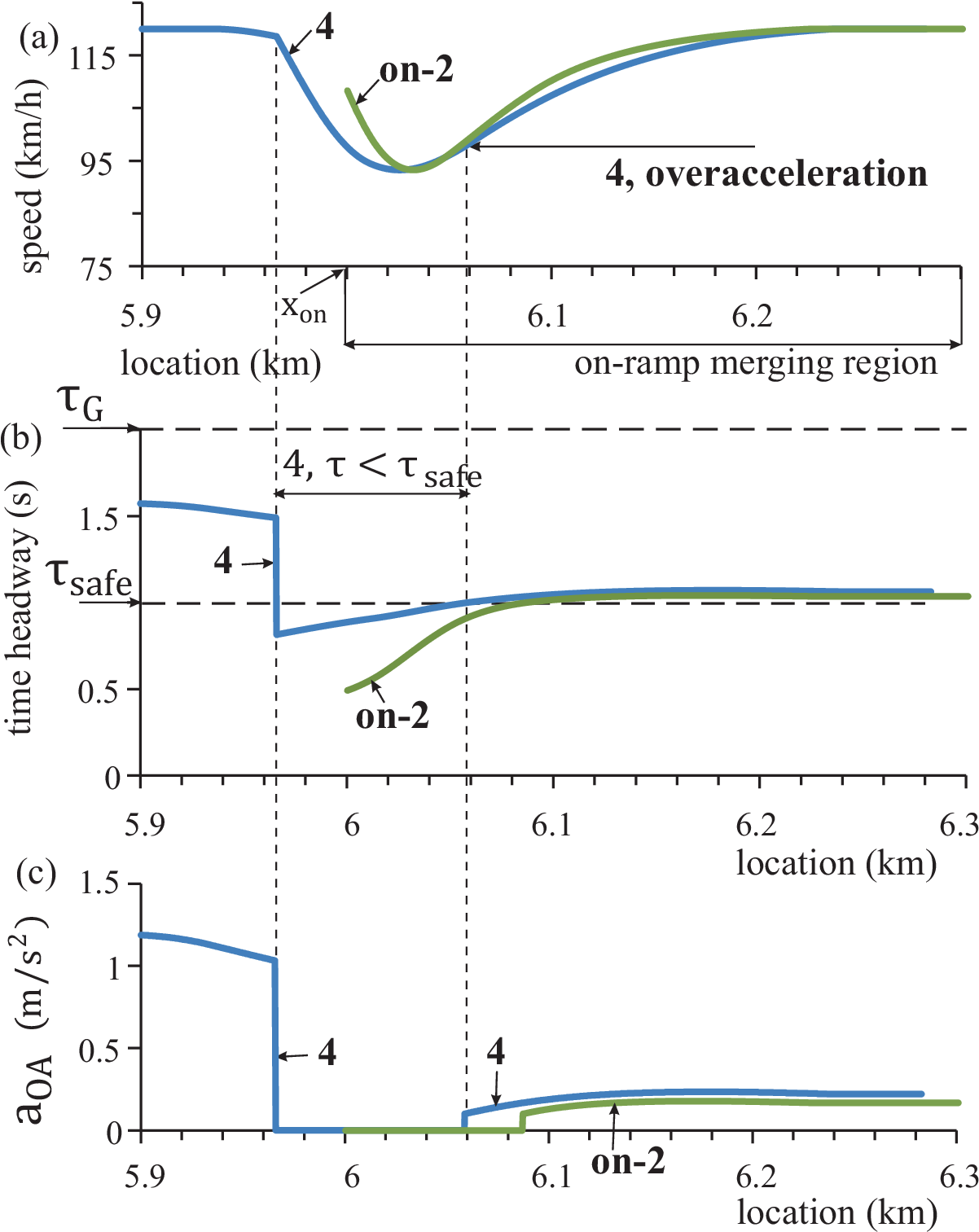}
\end{center}
\caption[]{Continuation of Fig.~\ref{2-lane-alpha-1-patterns_traj2}.
(a)--(c) Location-dependencies of
microscopic characteristics for some of the vehicles whose numbers are the same as those in
  Fig.~\ref{2-lane-alpha-1-patterns_traj2}, respectively: (a) vehicle speeds; (b) vehicle time-headway; 
		(c) overacceleration   $a_{\rm OA}$  (\ref{a_OA}),
 (\ref{a_OA_gap}).   
}
\label{2-lane-alpha-1-patterns_traj-on-2}
\end{figure}

		(ii) There can be overacceleration mechanism(s) in road lane
		that maintains (in addition to lane-changing) free flow at the bottleneck:
		After time-headway of vehicle 4 in Fig.~\ref{2-lane-alpha-1-patterns_traj2}  following the on-ramp vehicle on-2
		satisfies condition $\tau\geq \tau_{\rm safe}$, vehicle acceleration increases through the term
		$a_{\rm OA}$ [Figs.~\ref{2-lane-alpha-1-patterns_traj-on-2}(b) and~\ref{2-lane-alpha-1-patterns_traj-on-2}(c)].
Overacceleration of vehicle 4 
		  leads to recovering of free flow already within on-ramp
		merging region   [$\lq\lq$4, overacceleration" in Fig.~\ref{2-lane-alpha-1-patterns_traj-on-2}(a)].
		  
			\subsection{Effect of safety acceleration on overacceleration on two-lane road}

\begin{figure}
\begin{center}
\includegraphics[width = 8 cm]{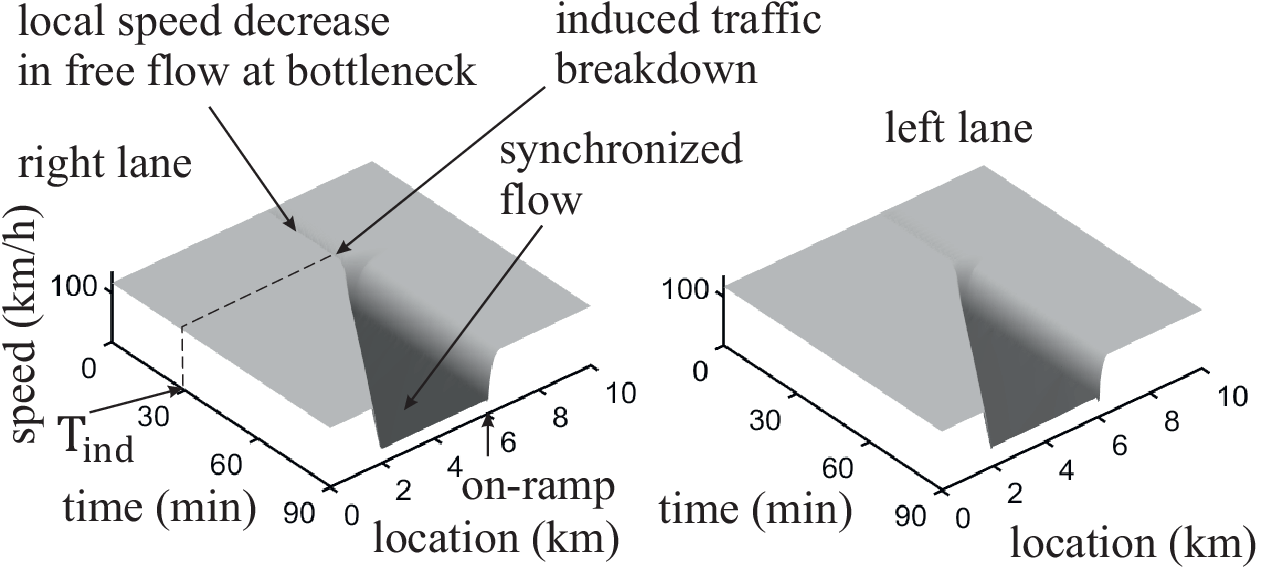}
\end{center}
\caption[]{Effect of overacceleration through safety acceleration
 in road lane at $\tau_{\rm G}=$ 1.4 s on overacceleration through lane-changing on two-lane road with bottleneck;
simulations of model   (\ref{g_v_g_min1})--(\ref{alpha_0}), i.e., at  
     $a_{\rm OA}=$ 0    
		in (\ref{g_v_g_min1}), (\ref{a_G-f}).   $q_{\rm on}=$ 700 vehicles/h.
F$\rightarrow$S transition  has been induced   at $T_{\rm ind}=$ 30 min  through  
 addition  on-ramp inflow-rate impulse  
$\Delta q_{\rm on}=$ 900 vehicles/h of duration   $\Delta t=$ 2 min.
 Other model parameters are the same as those in Figs.~\ref{2-lane-alpha-1-patterns}(a) and~\ref{2-lane-alpha-1-patterns}(b).  
}
\label{Effect-safety-2-lane}
\end{figure}

			The situation considered in Figs.~\ref{2-lane-alpha-1-patterns}(a) and~\ref{2-lane-alpha-1-patterns}(b) changes basically if we decrease the synchronization
			time-headway from $\tau_{\rm G}=$ 2 s
			 to shorter value $\tau_{\rm G}=$ 1.4 s
			that satisfies condition (\ref{g-free_G}).
			As we know from Sec.~\ref{Greater_Syn_S}, in this case   there is overacceleration in road lane through
			safety acceleration. Then, traffic breakdown does  exhibit  
the nucleation character (Fig.~\ref{Effect-safety-2-lane}).
			We have found (not shown) that, as in the case shown in Fig.~\ref{2-lane-alpha-1-OA-rate},
a discontinuous character of the overacceleration rate  
 $R_{\rm OA}$ is   realized, i.e., there is also   overacceleration 
  through lane-changing.

\section{Critical nucleus for spontaneous traffic breakdown    \label{Physics_v_cr}}

 As for Helly's model used for automated-driving vehicles in~\cite{Kerner2023B},
in   model of Sec.~\ref{Model-Sec} at a given $q_{\rm in}$ when the on-ramp inflow rate $q_{\rm on}$ increases, then
 at some $q_{\rm on}> q_{\rm on, \ max}$ the minimum speed within a local speed decrease at the bottleneck decreases over time
   and 
after a time-delay  $T^{\rm (B)}$ spontaneous traffic breakdown
 (spontaneous F$\rightarrow$S transition) occurs  [Figs.~\ref{2-lane-alpha-1-patterns}(c)
and~\ref{Greater_G_spont_alpha}(a)].  Here we reveal microscopic features of the critical nucleus for spontaneous traffic breakdown.

\subsection{Microscopic   characteristics of  critical nucleus \label{Veh_dyn_Sec}}

\begin{figure}
\begin{center}
\includegraphics[width = 8 cm]{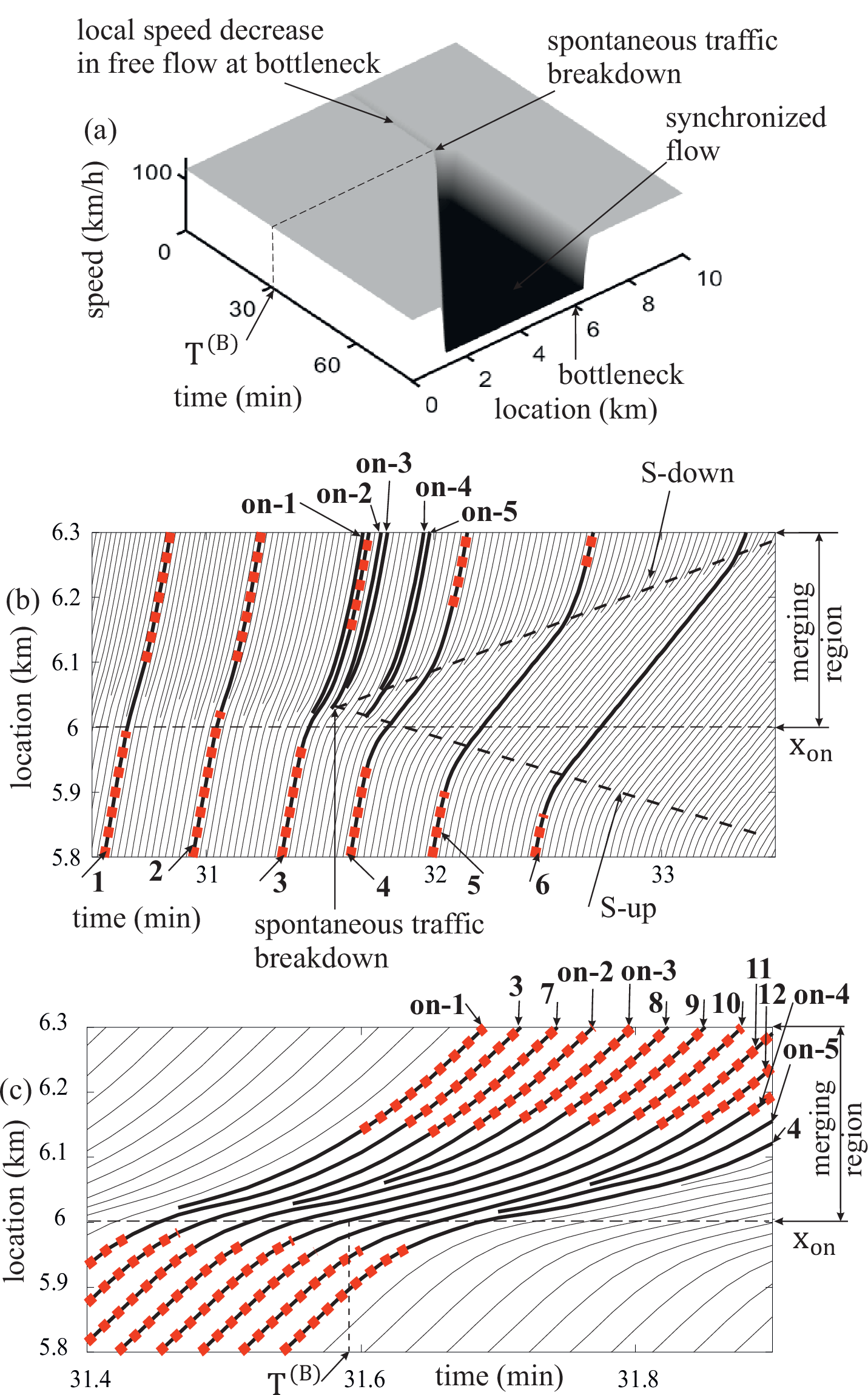} 
\end{center}
\caption[]{Spontaneous traffic breakdown on single-lane road with on-ramp bottleneck. 
Simulations of the same model (\ref{g_v_g_min1})--(\ref{K_Deltav2}), (\ref{G_g_safe_simple})
	  at  $\tau_{\rm G}=$ 1.4 s,   $q_{\rm in}=$ 2000 vehicles/h as that used in Fig.~\ref{Greater_G_induced_alpha},
		however, at
  $q_{\rm on}=$ 809 vehicles/h that is only slightly higher than $q_{\rm on}=q_{\rm on, \ max}=$ 807 vehicles/h. 
Other model parameters are the same as those in Fig.~\ref{Greater_G_induced_alpha}.
 (a), (b) Speed in space and time: Local speed decrease at bottleneck in free flow (a) and synchronized flow pattern (SP) (b)
that occurs spontaneously in   free flow after time-delay   $T^{\rm (B)}\approx$ 31.6 min.  (c) Some of simulated vehicle trajectories within  local speed decrease in free flow
at  bottleneck in a vicinity of time $t =T^{\rm (B)}$; part of trajectories within which condition $a_{\rm OA}>0$ is satisfied
  are marked through colored red squares. 
}
\label{Greater_G_spont_alpha}
\end{figure}

On  
 single-lane road [Figs.~\ref{Greater_G_spont_alpha} and~\ref{Greater_G_spont_alpha_traj}] 
and  on two-lane road [Figs.~\ref{2-lane-alpha-1-spont_traj} and~\ref{2-lane-alpha-1-spont_traj2}],
during interval
$0 < t < T^{\rm (B)}$   a minimum
speed   $v_{\rm min}$  within the local speed decrease at the bottleneck  is a  
decreasing time-function [minimum speeds of vehicles 1--3 in  
Fig.~\ref{Greater_G_spont_alpha_traj}(a) for single-lane road and minimum speeds of vehicles 1--3 and 7--9 in  
Fig.~\ref{2-lane-alpha-1-spont_traj2} for two-lane road].
 When $v_{\rm min}$ becomes less than   $v_{\rm syn}$ (\ref{a_OA}),  
a state within the local speed decrease
 can be considered   synchronized flow. Thus, free flow   persists
upstream of the bottleneck, while synchronized flow is only within the  local speed decrease
at the bottleneck.

  We have found that time-delay of spontaneous traffic breakdown  $T^{\rm (B)}$ is determined by  
 a saturation  in   decrease of $v_{\rm min}$: At $t> T^{\rm (B)}$ without further noticeable decrease in $v_{\rm min}$, 
the upstream front  of synchronized flow
   moves upstream of the bottleneck, i.e.,
  free flow
upstream of the bottleneck transforms over time   into synchronized
flow propagating upstream [vehicles 4--6 in  
Fig.~\ref{Greater_G_spont_alpha_traj}  and   vehicles 4--6 and 10--12 in  
Fig.~\ref{2-lane-alpha-1-spont_traj2}]. The beginning of the upstream propagation
of synchronized flow can be seen from a comparison of microscopic speeds along trajectories 4 and 8
in Fig.~\ref{Greater_G_spont_alpha_traj}(b)
for single-lane road as well as a comparison of microscopic speeds along trajectories 16 and 20 for two-lane road in Fig.~\ref{2-lane-alpha-1-spont_traj2}(c).

	\begin{figure}
\begin{center}
\includegraphics[width = 8 cm]{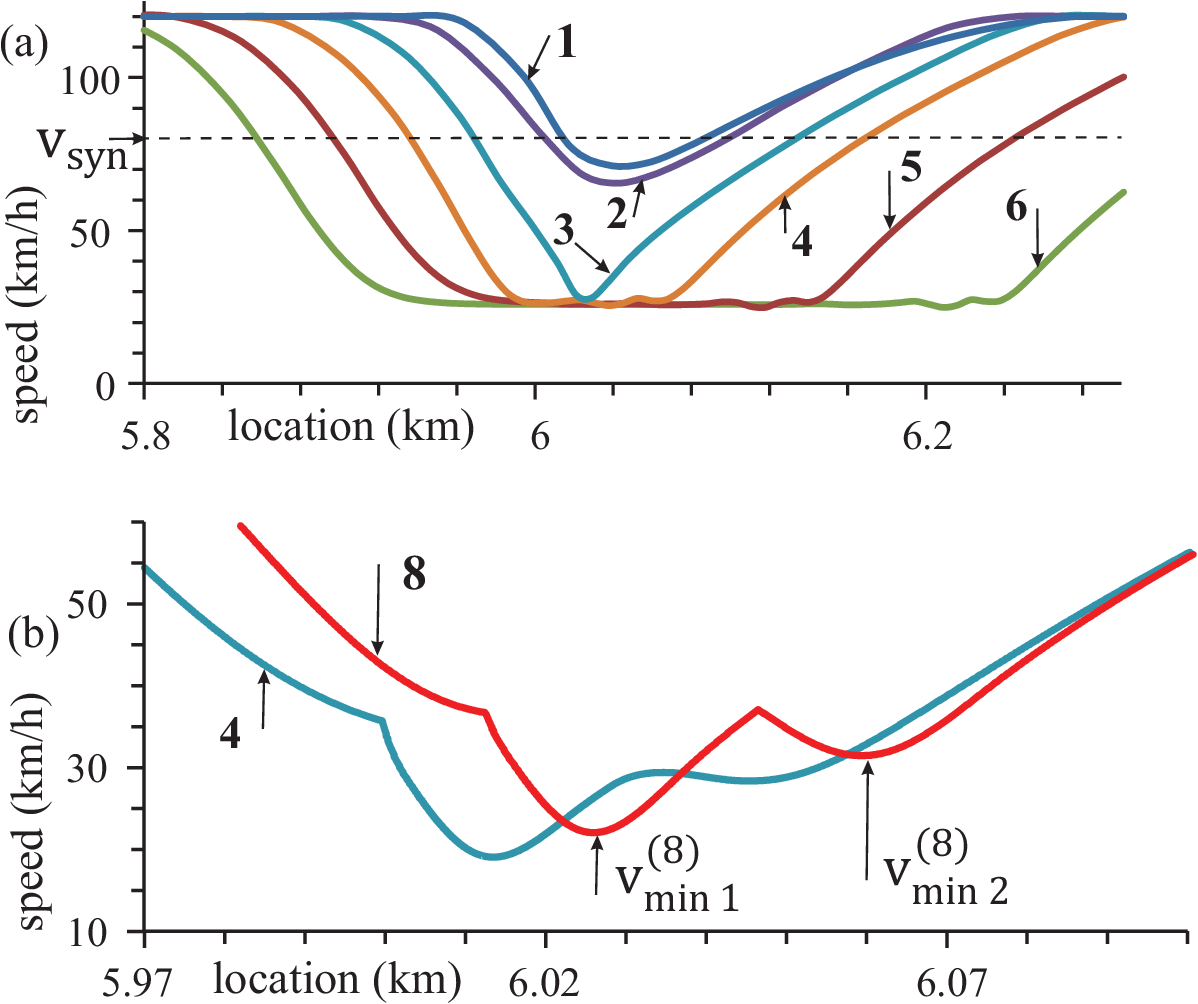} 
\end{center}
\caption[]{Location-dependence of vehicle speeds
those trajectories are shown by the same vehicle numbers in Figs.~\ref{Greater_G_spont_alpha}(b) and~\ref{Greater_G_spont_alpha}(c):
(a) Averaged speed along trajectories (1-sec averaged data)  shown in 
Fig.~\ref{Greater_G_spont_alpha}(b);
dashed line
$v=v_{\rm syn}$ separates road regions on vehicle trajectories within which overacceleration
$a_{\rm OA}$ is either more than zero
(on and above the line) or it is equal to zero (below the line).  (b) Microscopic (not averaged  data) speeds along trajectories 4 and 8    shown in 
Figs.~\ref{Greater_G_spont_alpha}(b) and~\ref{Greater_G_spont_alpha}(c). 
}
\label{Greater_G_spont_alpha_traj}
\end{figure}

\begin{figure}
\begin{center}
\includegraphics[width = 7.9 cm]{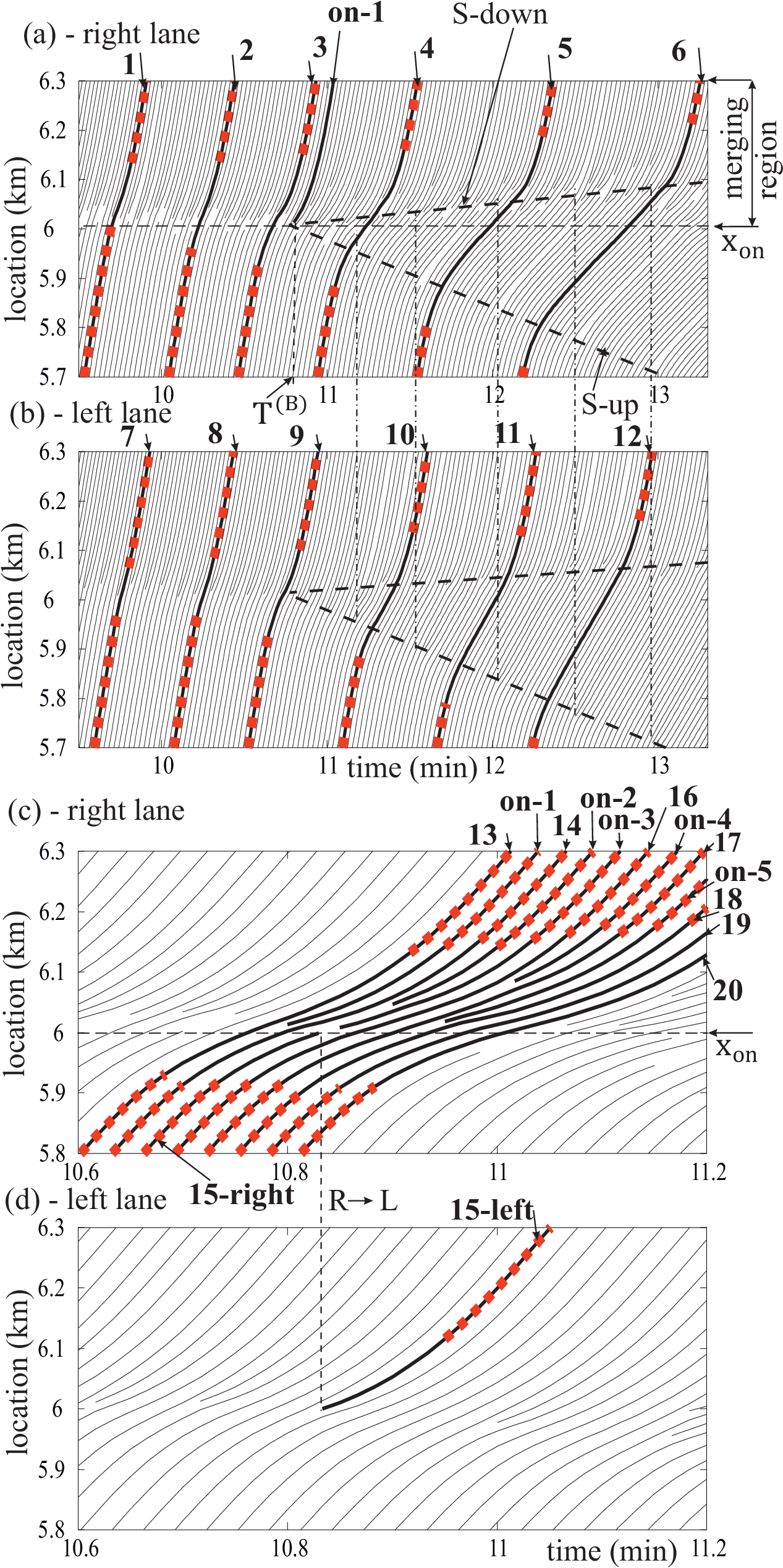}
\end{center}
\caption[]{Continuation of Fig.~\ref{2-lane-alpha-1-patterns}(c). 
(a), (b) Simulated vehicle trajectories within  local speed decrease in free flow on two-lane road 
with  bottleneck showing  spontaneous traffic breakdown
occurring after time-delay  at $T^{\rm (B)}\approx$ 10.8    min. 
(c), (d) Some of simulated vehicle trajectories of (a), (b) shown in a shorter time-scale; 
parts of trajectories upstream of the bottleneck within which condition $a_{\rm OA}>0$ is satisfied
  are marked through colored red squares.
}
\label{2-lane-alpha-1-spont_traj}
\end{figure}

	\begin{figure}
\begin{center}
\includegraphics[width = 8 cm]{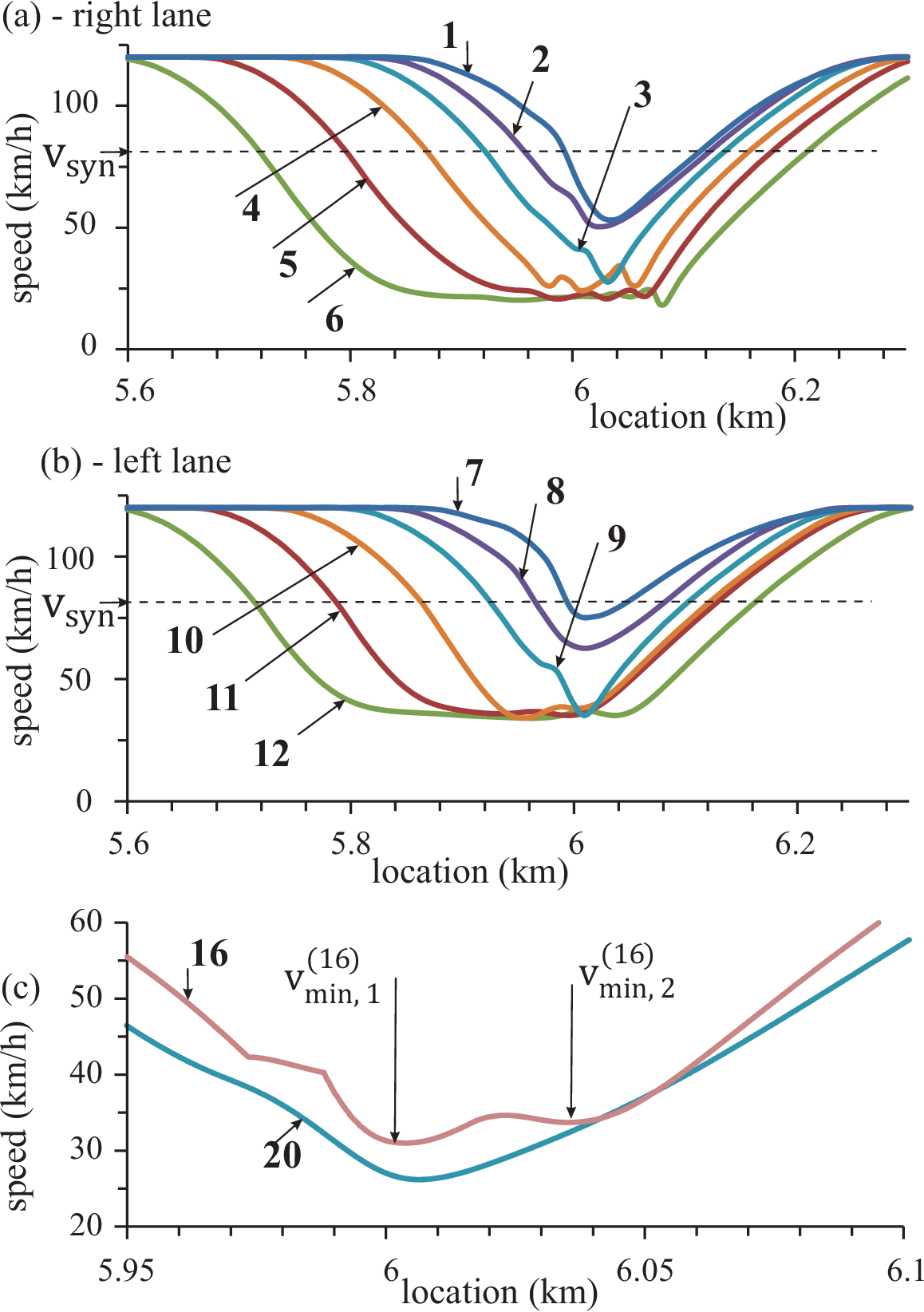} 
\end{center}
\caption[]{Continuation of Figs.~\ref{2-lane-alpha-1-spont_traj}(a)--\ref{2-lane-alpha-1-spont_traj}(c). (a), (b)
Location-dependence of averaged vehicle speeds (1-sec averaged data); dashed lines
$v=v_{\rm syn}$ separate road regions on vehicle trajectories within which overacceleration
$a_{\rm OA}$ is either more than zero
(on and above the lines) or it is equal to zero (below the lines). 
(c)  Location-dependence of  speeds (not averaged data) of vehicles 16 and 20. 
Trajectories are shown with the same vehicle numbers as those in 
Figs.~\ref{2-lane-alpha-1-spont_traj}(a)--~\ref{2-lane-alpha-1-spont_traj}(c),
respectively.     
}
\label{2-lane-alpha-1-spont_traj2}
\end{figure}

 Thus, the critical nucleus  at the bottleneck is 
 a local speed decrease at the bottleneck
 caused by the last on-ramp vehicle or the last sequence on-ramp vehicles merging from the on-ramp at  
$t < T^{\rm (B)}$ at which the speed decrease is still localized at the bottleneck.  This   means that
the next on-ramp vehicle or sequence of on-ramp vehicles that have merged from the on-ramp
at $t =T^{\rm (B)}$   cause another local speed decrease that cannot be localized
at the bottleneck any more: At $t >T^{\rm (B)}$ 
	the irreversible upstream propagation of synchronized flow occurs.
	 
	There can be one, two, or more minimums
	of  the synchronized flow 
   speed within  the nucleus.
In Fig.~\ref{Greater_G_spont_alpha}(c), two following each other on-ramp vehicles have merged onto the main road
(vehicles on-2 and on-3). This results in two speed minimums for  vehicle 8
($v^{\rm (8)}_{\rm min, \ 1}$ and $v^{\rm (8)}_{\rm min, \ 2}$ in Fig.~\ref{Greater_G_spont_alpha_traj}(b)) that follows
vehicles on-2 and on-3. A   similar result  is realized on
two-lane road for vehicle 16 following
two on-ramp vehicles (vehicles on-2 and on-3 in
Fig.~\ref{2-lane-alpha-1-spont_traj}(c)). This leads to 
two speed minimums for vehicle 16 ($v^{\rm (16)}_{\rm min, \ 1}$ and $v^{\rm (16)}_{\rm min, \ 2}$ in 
Fig.~\ref{2-lane-alpha-1-spont_traj2}(c)).
The lowest speed minimum within the critical nucleus can be considered a microscopic {\it critical speed}
for traffic breakdown (F$\rightarrow$S transition).

\subsection{Effect of 
 overacceleration in road lane $a_{\rm OA}$ on critical nucleus development \label{OA_Effect_Spont_Sec}} 

When time $t$ is not  very close to  $T^{\rm (B)}$,    the minimum speed
$v_{\rm min}$ within the local speed decrease at the bottleneck
is   not considerably less than $v_{\rm syn}$ [e.g., $v^{(1)}_{\rm min}\approx$ 72.2 km/h
for vehicles 1 on single-lane road
  in Fig.~\ref{Greater_G_spont_alpha_traj}(a)
and $v^{(1)}_{\rm min}\approx$ 53.3 km/h for vehicles 1 on two-lane road  in Fig.~\ref{2-lane-alpha-1-spont_traj2}(a)].    
As a result, upstream of the bottleneck overacceleration   $a_{\rm OA}>0$ with the exception of
 a narrow road region at the bottleneck   within which
  $v<v_{\rm syn}$ [Figs.~\ref{Greater_G_spont_alpha_traj}(a),
	\ref{2-lane-alpha-1-spont_traj2}(a) and~\ref{2-lane-alpha-1-spont_traj2}(b)] and, 
	therefore, $a_{\rm OA}=0$.
 
When    $t\rightarrow T^{\rm (B)}$, then
the effect of overacceleration $a_{\rm OA}$ on the maintenance of free flow at the bottleneck becomes weaker. Indeed,
due to a decrease in $v_{\rm min}$, the size of a road region in the bottleneck vicinity
 within which the speed $v<v_{\rm syn}$, i.e., overacceleration $a_{\rm OA}=$ 0,
 increases [this can be seen    on trajectories 1--3  for single-lane road
in Fig.~\ref{Greater_G_spont_alpha_traj}(a) and on trajectories 1--3 and 7--9 for two-lane road
in Figs.~\ref{2-lane-alpha-1-spont_traj2}(a),(b)].  
Thus, at $t\rightarrow T^{\rm (B)}$ the effect of 
 overacceleration   $a_{\rm OA}$ on the critical nucleus  development   is caused by
   the    increase in  the road region size in the bottleneck vicinity within which
  $a_{\rm OA}=$ 0.

 \section{A proof of  effect of overacceleration
 on vehicular traffic \label{Suppression_S}}

\begin{figure} 
\begin{center}
\includegraphics[width = 8 cm]{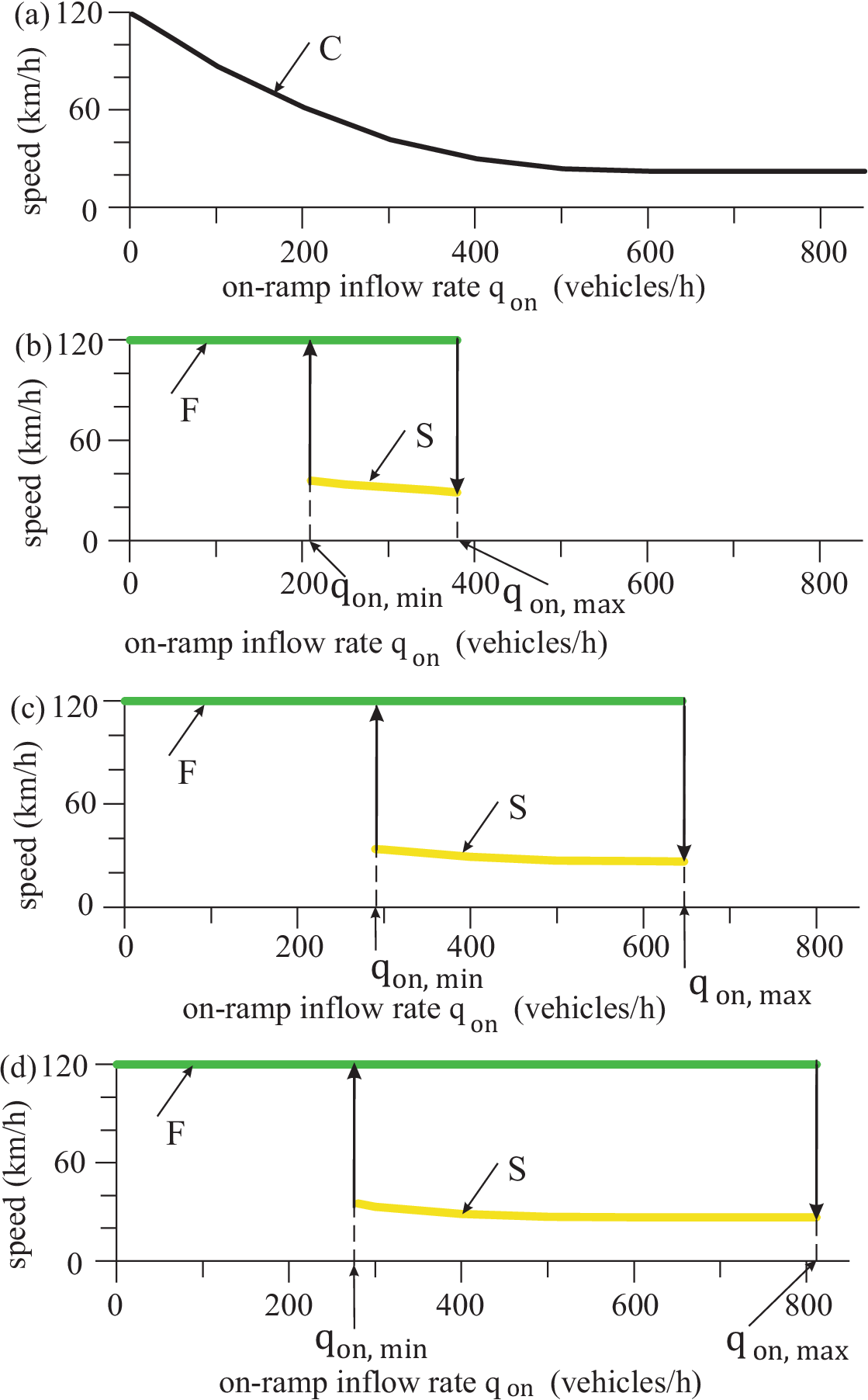}
\end{center}
\caption[]{Proof of effect of overacceleration on vehicular traffic on single-lane road
with the use of speed--on-ramp-inflow characteristics calculated at   $q_{\rm in}=$ 2000 vehicles/h:
 (a) No overacceleration, model and  parameters of Fig.~\ref{LWR_Fig}.
(b) Overacceleration mechanism through safety acceleration only,  model and parameters of Fig.~\ref{Greater_G_induced}.
(c) Overacceleration  due to $a_{\rm OA}$ (\ref{a_OA}),  (\ref{a_OA_gap}) in model
 (\ref{g_v_g_min1})--(\ref{K_Deltav2}), (\ref{G_g_safe_simple})
with   parameters:   $\alpha_{0}=$ 2 $\rm m/s^{2}$,
$\alpha_{1}=$ 0.1 $\rm m/s^{2}$, $k=$ 1; 
 other model parameters are the same as those in Fig.~\ref{LWR_Fig}. 
(d) Cooperation of overacceleration mechanisms through safety acceleration  at $g_{\rm free}>G$ (\ref{g-free_G})
and due to overacceleration mechanism
 $a_{\rm OA}$ (\ref{a_OA}),  (\ref{a_OA_gap}); model and parameters of Fig.~\ref{Greater_G_induced_alpha}.
Contrary to (d), in (c) there is only overacceleration mechanism
    $a_{\rm OA}$; this is because due to $\tau_{\rm G}=$ 2 s
used in (c)   condition (\ref{g-free_G}) is not satisfied and, therefore, 
safety acceleration cannot be considered 
 overacceleration  at the given $q_{\rm in}$.    
C -- congested traffic, F -- free flow, S -- synchronized flow.
$q_{\rm on}=q_{\rm on, \ min}$ is a minimum on-ramp flow rate
related to  a minimum highway capacity
$C_{\rm min}=q_{\rm in}+q_{\rm on, \ min}$,
whereas   $q_{\rm on}=q_{\rm on, \ max}$
is related to  a maximum highway capacity
$C_{\rm max}=q_{\rm in}+q_{\rm on, \ max}$. 
Within the range
$q_{\rm on, \ min}\leq q_{\rm on}< q_{\rm on, \ max}$
either free flow or an SP can exist at the bottleneck; 
 at $q_{\rm on}<q_{\rm on, \ min}$ no congested patterns can be induced in free flow, whereas
at $q_{\rm on}>q_{\rm on, \ max}$ the SP occurs spontaneously at the bottleneck after a time delay $T^{\rm (B)}$.
Calculated parameters: $(q_{\rm on, \ min}, \ q_{\rm on, \ max})=$ (217, 372) vehicles/h (b),
(293, 648) vehicles/h (c), (280, 807) vehicles/h (d).  
}
\label{2Z_Fig}
\end{figure}

\begin{figure} 
\begin{center}
\includegraphics[width = 8 cm]{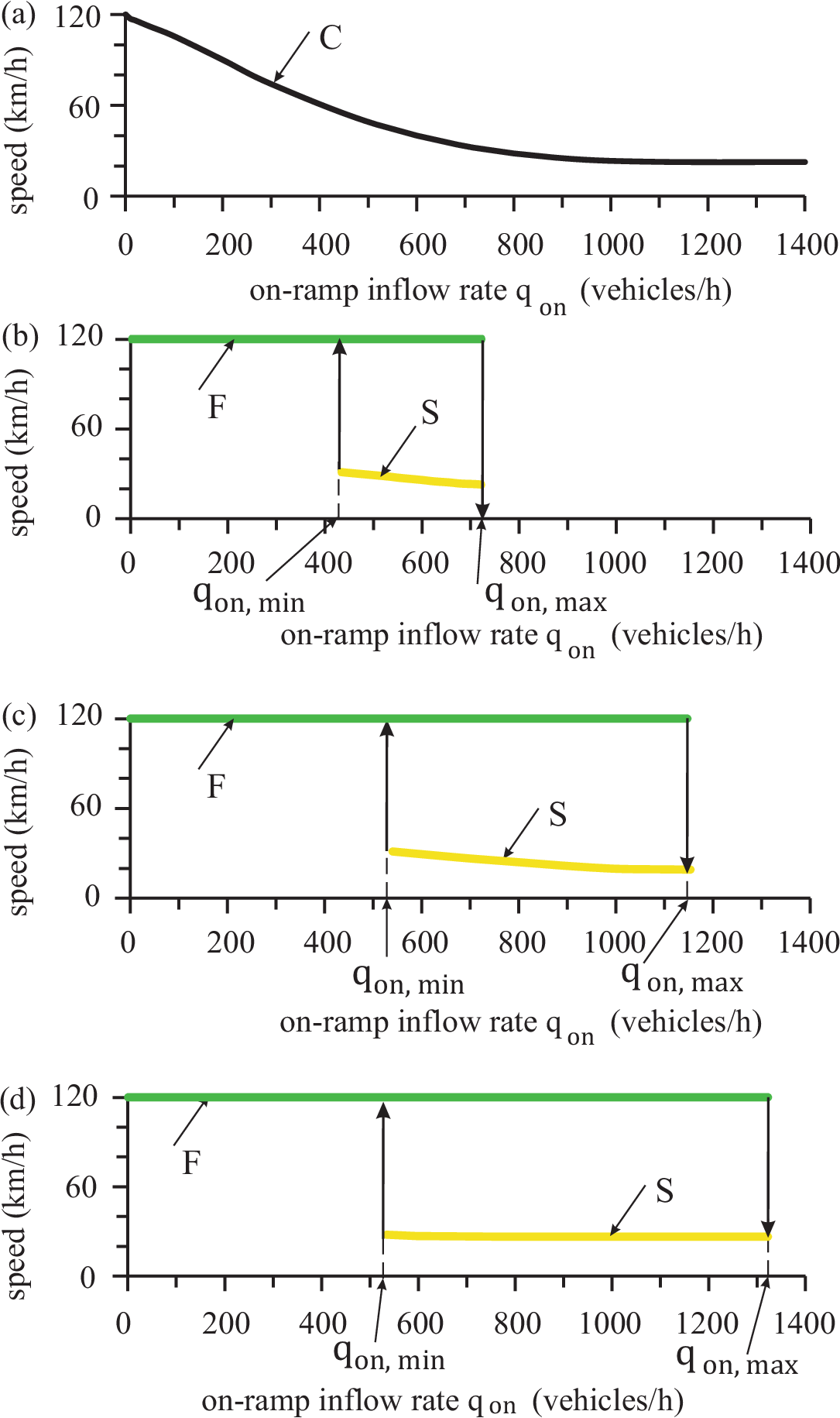}   
\end{center}
\caption[]{Proof of   effect of overacceleration on vehicular traffic on two-lane road
with the use of speed--on-ramp-inflow characteristics calculated at  
 $q_{\rm in}=$ 2000 (vehicles/h)/lane: (a) No overacceleration, model and  parameters of 
 Figs.~\ref{2-lane-alpha-1-patterns}(a) and~\ref{2-lane-alpha-1-patterns}(b).
(b) Model and parameters of Fig.~\ref{Effect-safety-2-lane}.
(c) Model and parameters of  Figs.~\ref{2-lane-alpha-1-patterns}(c)--\ref{2-lane-alpha-1-patterns}(e).
(d) Model   (\ref{g_v_g_min1})--(\ref{G_g_safe_simple})
at $\tau_{\rm G}=$
1.4 s; other parameters are the same as those in Figs.~\ref{2-lane-alpha-1-patterns}(c)--\ref{2-lane-alpha-1-patterns}(e).    
C -- congested traffic, F -- free flow, S -- synchronized flow. Minimum and maximum highway capacities, respectively, are:  
 $C_{\rm min} =2q_{\rm in} + q_{\rm on, \ min}$ and  $C_{\rm max} =2q_{\rm in} + q_{\rm on, \ max}$.
	Calculated parameters: $(q_{\rm on, \ min}, \ q_{\rm on, \ max})=$
	(430, 721) vehicles/h (b), (517, 1142) vehicles/h (c),
(525, 1324) vehicles/h (d).
}
\label{2Z_2-lane}
\end{figure}

\begin{itemize}
\item [--] The main effect of overacceleration
 on vehicular traffic at a bottleneck 
is the maintenance of   free flow  upstream of the bottleneck, i.e.,
the prevention of traffic breakdown (F$\rightarrow$S transition).
\end{itemize}
To provide a proof of this statement, we   show than   for the same chosen flow rate in free flow $q_{\rm in}$
upstream of the bottleneck, the stronger the effect of overacceleration
 on vehicular traffic, the larger the maximum on-ramp inflow rate $q_{\rm on}=q_{\rm on, \ max}$ 
at which free flow  upstream of the bottleneck is still maintained.
Through the revealing of different overacceleration mechanisms as well as their
 cooperation   (Secs.~\ref{Cooperation_Sec} and~\ref{2-lane_Coop_Sec}), such a proof    is possible to do as
explained below with the use of
  Fig.~\ref{2Z_Fig} for single-lane road and   Fig.~\ref{2Z_2-lane} for two-lane road.

In Figs.~\ref{2Z_Fig}(a) and~\ref{2Z_2-lane}(a), there is no overacceleration (Secs.~\ref{1-lane-neglecting} 
and~\ref{Prevention-2lane-Sec}).
Then, vehicles on the main road moving initially at free flow speed $v_{\rm free}$ upstream of the bottleneck
must decelerate due to  vehicles  merging from the on-ramp. 
Through this speed adaptation effect (\ref{g_v_g_min1_SA}), at the given  
$q_{\rm in}$   the
larger the on-ramp inflow-rate $q_{\rm on}$, the lower the
  average vehicle speed upstream of the bottleneck. This means that   speed adaptation   is the cause
for the emergence of congested traffic at the bottleneck.
	
Overacceleration is the opposite effect to speed adaptation. Indeed,	contrary to Figs.~\ref{2Z_Fig}(a) and~\ref{2Z_2-lane}(a), in Figs.~\ref{2Z_Fig}(b)--\ref{2Z_Fig}(d) and~\ref{2Z_2-lane}(b)--\ref{2Z_2-lane}(d)  when   overacceleration is realized  at the bottleneck, free flow can be self-maintained upstream of the bottleneck up to some maximum on-ramp inflow rate
 $q_{\rm on}= q_{\rm on, \ max}$, i.e.,
up to the  maximum highway capacity
$C_{\rm max}=q_{\rm in}+q_{\rm on, \ max}$.
 Thus, as long as $q_{\rm on} < q_{\rm on, \ max}$ (i.e., highway capacity is less than $C_{\rm max}$)
 overacceleration occurring in Figs.~\ref{2Z_Fig}(b)--\ref{2Z_Fig}(d) and~\ref{2Z_2-lane}(b)--\ref{2Z_2-lane}(d) prevents   traffic congestion at the bottleneck caused by speed adaptation in Figs.~\ref{2Z_Fig}(a) and~\ref{2Z_2-lane}(a).

On single-lane road, when overacceleration is caused by safety acceleration  at $g_{\rm free}>G$ (\ref{g-free_G})
 only (Sec.~\ref{Greater_Syn_S}), then
the maximum on-ramp inflow-rate $q_{\rm on, \ max}$ and, therefore, the maximum capacity
$C_{\rm max}$ [Fig.~\ref{2Z_Fig}(b)] are, respectively, less than  these values are 
 when overacceleration is caused by  the term
$a_{\rm OA}$  (\ref{a_OA})  
 [Fig.~\ref{2Z_Fig}(c)]. This means that at model parameters under consideration
 the effect of overacceleration $a_{\rm OA}$  (\ref{a_OA}) 
 on the maintenance of free flow at the bottleneck is stronger than the overacceleration effect   due to safety acceleration.
 The stronger the overacceleration effect on free flow is, the larger the on-ramp inflow rate
	$q_{\rm on, \ max}$ at which free flow is still self-maintained at the bottleneck.
	This statement is   confirmed by the result that
the greatest values 
 $q_{\rm on, \ max}$ and $C_{\rm  max}$ are realized when overacceleration  is further increased
through cooperation of overacceleration $a_{\rm OA}$  (\ref{a_OA}) and overacceleration
caused by safety acceleration at $g_{\rm free}>G$ (\ref{g-free_G}) [Fig.~\ref{2Z_Fig}(d)].

		As on single-lane road,   
	the stronger the overacceleration effect on free flow  on two-lane road is, the larger the on-ramp inflow rate
	$q_{\rm on, \ max}$ at which free flow is still self-maintained at the bottleneck. Indeed, in Fig.~\ref{2Z_2-lane}(b)   overacceleration in road lane is determined
	by safety acceleration at $g_{\rm free}>G$ (\ref{g-free_G})  only;  
	as explained above [Fig.~\ref{2Z_Fig}(b)], this overacceleration mechanism is  
		weaker than that determined   by overacceleration $a_{\rm OA}$  (\ref{a_OA}). 
	Therefore,   we have found that
  cooperation of  overacceleration  due to safety acceleration at $g_{\rm free}>G$ (\ref{g-free_G})
	and overacceleration through lane-changing
	can self-maintain free flow at the bottleneck only up to a relative small on-ramp inflow rate
	$q_{\rm on, \ max}$ [Fig.~\ref{2Z_2-lane}(b)].  In Fig.~\ref{2Z_2-lane}(c),   overacceleration in road lane is determined
	by $a_{\rm OA}$  (\ref{a_OA}) that, as above-mentioned,   is stronger than overacceleration due 
	to safety acceleration at $g_{\rm free}>G$ (\ref{g-free_G}).
	As a result, we have found that cooperation of two overacceleration mechanisms caused by $a_{\rm OA}$ and through lane-changing
	 maintains free flow at the bottleneck up to a considerably larger value $q_{\rm on, \ max}$ [Fig.~\ref{2Z_2-lane}(c)]
	than that in 
	Fig.~\ref{2Z_2-lane}(b).
	
	Finally, in  Fig.~\ref{2Z_2-lane}(d) related to model   (\ref{g_v_g_min1})--(\ref{G_g_safe_simple})
at $\tau_{\rm G}=$  1.4 s there is
cooperation of   three mechanisms of overacceleration (due to $a_{\rm OA}$  (\ref{a_OA}), through safety acceleration at $g_{\rm free}>G$ (\ref{g-free_G}), and through lane-changing). Then, we have found the greatest
		maximum on-ramp inflow-rate
		$q_{\rm on, \ max}$ (and, therefore, the greatest   maximum highway capacity $C_{\rm max}$ at  the given $q_{\rm in}$). 
		This shows that 
the	cooperation of   three mechanisms of overacceleration	exhibits the  strongest effect on the self-maintaining of free flow at the bottleneck.

\section{Microscopic effect of overacceleration cooperation on S$\rightarrow$F instability       \label{Competition_SF_Sec}}

\subsection{Overacceleration cooperation and S$\rightarrow$F instability on  
 two-lane road without bottlenecks \label{Effect_SF_hom_Sec}}

In~\cite{Kerner2023C}, we have shown that overacceleration $a_{\rm OA}$  (\ref{a_OA})
on  single-lane road without bottlenecks can cause an S$\rightarrow$F instability that exhibits
the nucleation nature. It can be expected that on two-lane road overacceleration $a_{\rm OA}$  (\ref{a_OA})
can also cause the S$\rightarrow$F instability (Fig.~\ref{2-lane-S-F-patterns}). Indeed, a local disturbance in initial homogeneous synchronized flow
caused by an addition acceleration  of one of the vehicles during $\delta t =$ 5 s
   initiates the S$\rightarrow$F instability   
[Figs.~\ref{2-lane-S-F-patterns}(a) and~\ref{2-lane-S-F-patterns}(b)]. Contrarily, if the same vehicle  accelerates during $\delta t =$ 4.5 s
(i.e., only 0.5 s 
shorter), then no S$\rightarrow$F instability occurs [Fig.~\ref{2-lane-S-F-patterns}(c)].
   Here we show that  
cooperation of overacceleration in road lane $a_{\rm OA}$  (\ref{a_OA}) and overacceleration caused by lane-changing
  (Sec.~\ref{2-lane_Coop_Sec})  leads to 
  spatiotemporal lane-changing effects that qualitatively change the dynamics of  the S$\rightarrow$F instability
	in comparison with that found in~\cite{Kerner2023C}.

 	\begin{figure} 
\begin{center}
\includegraphics[width = 8 cm]{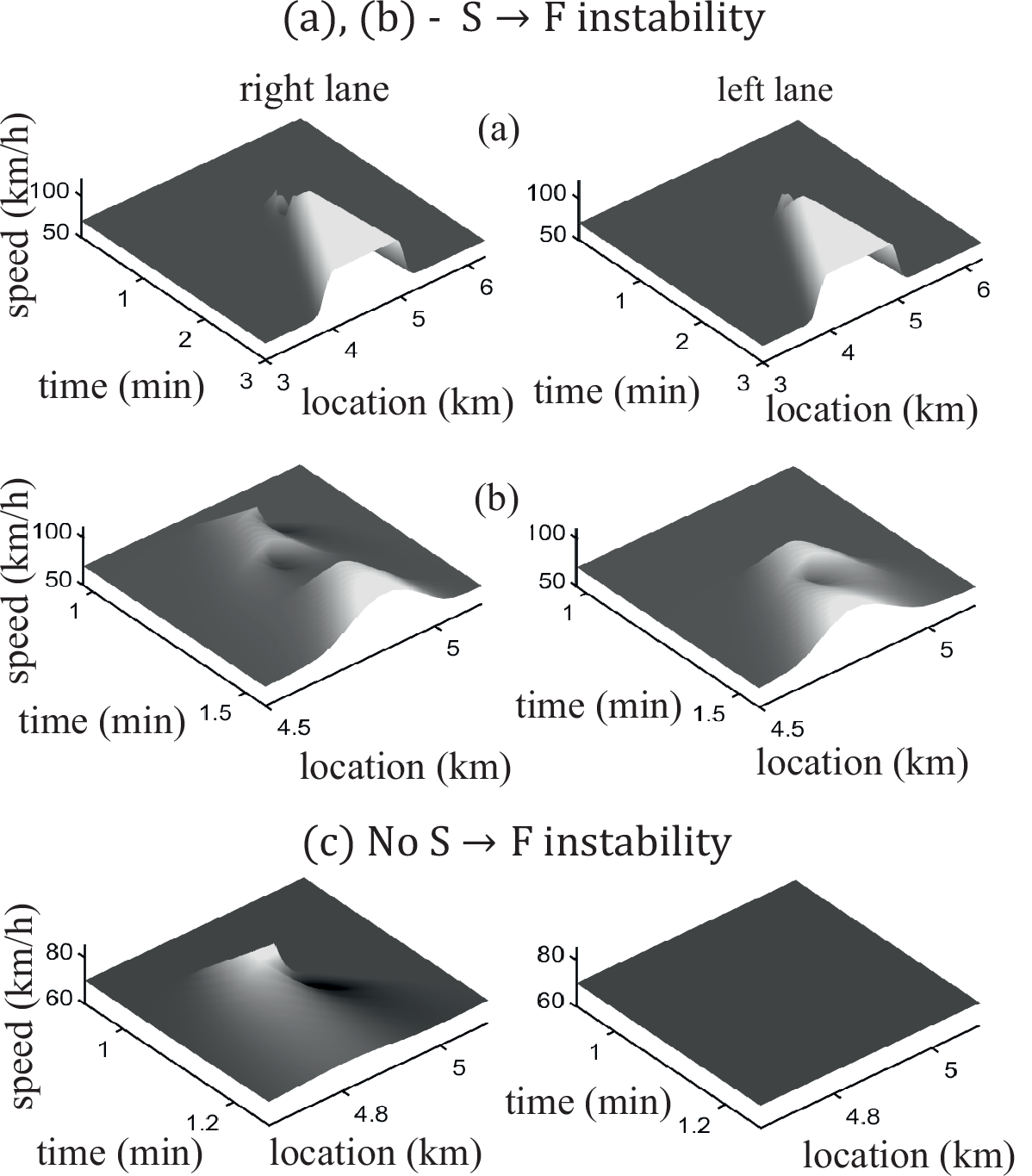}
\end{center}
\caption[]{Nucleation nature of S$\rightarrow$F instability on   two-lane road 
caused by cooperation of overacceleration in road lane and due to lane-changing.
Simulations  
with model (\ref{g_v_g_min1})--(\ref{G_g_safe_simple}) at $\tau_{\rm G}=$ 1.4 s made  
on two-lane  road of length 8 km    without bottlenecks with initial steady synchronized flow state  
 at speed $v=$ 70 km/h and space gap between vehicles $g=$ 33 m. Vehicle speed
 is space and time in the right lane (left subplots) and in   left lane (right subplots):
(a)  S$\rightarrow$F instability is   caused by
  initial local speed increase of a vehicle [vehicle 1 in Fig.~\ref{S-F-instability-hom_traj}(a)]
	  moving in the right lane   simulated through 
	short-time acceleration of the vehicle   with $a=$ 1
 $\rm m/s^{2}$ during    $\delta t=$ 5 s. (b) The beginning of the
S$\rightarrow$F instability in (a) shown in a larger   scale. (c) S$\rightarrow$F instability does not occur
if the same vehicle as that in (a), (b) accelerates with the same acceleration, however,
during $\delta t=$ 4.5 s, i.e., 0.5 s shorter. 
Other model parameters are the same as those in Figs.~\ref{2-lane-alpha-1-patterns}(c)--\ref{2-lane-alpha-1-patterns}(e).
}
\label{2-lane-S-F-patterns}
\end{figure}

\begin{figure}
\begin{center}
\includegraphics[width = 8 cm]{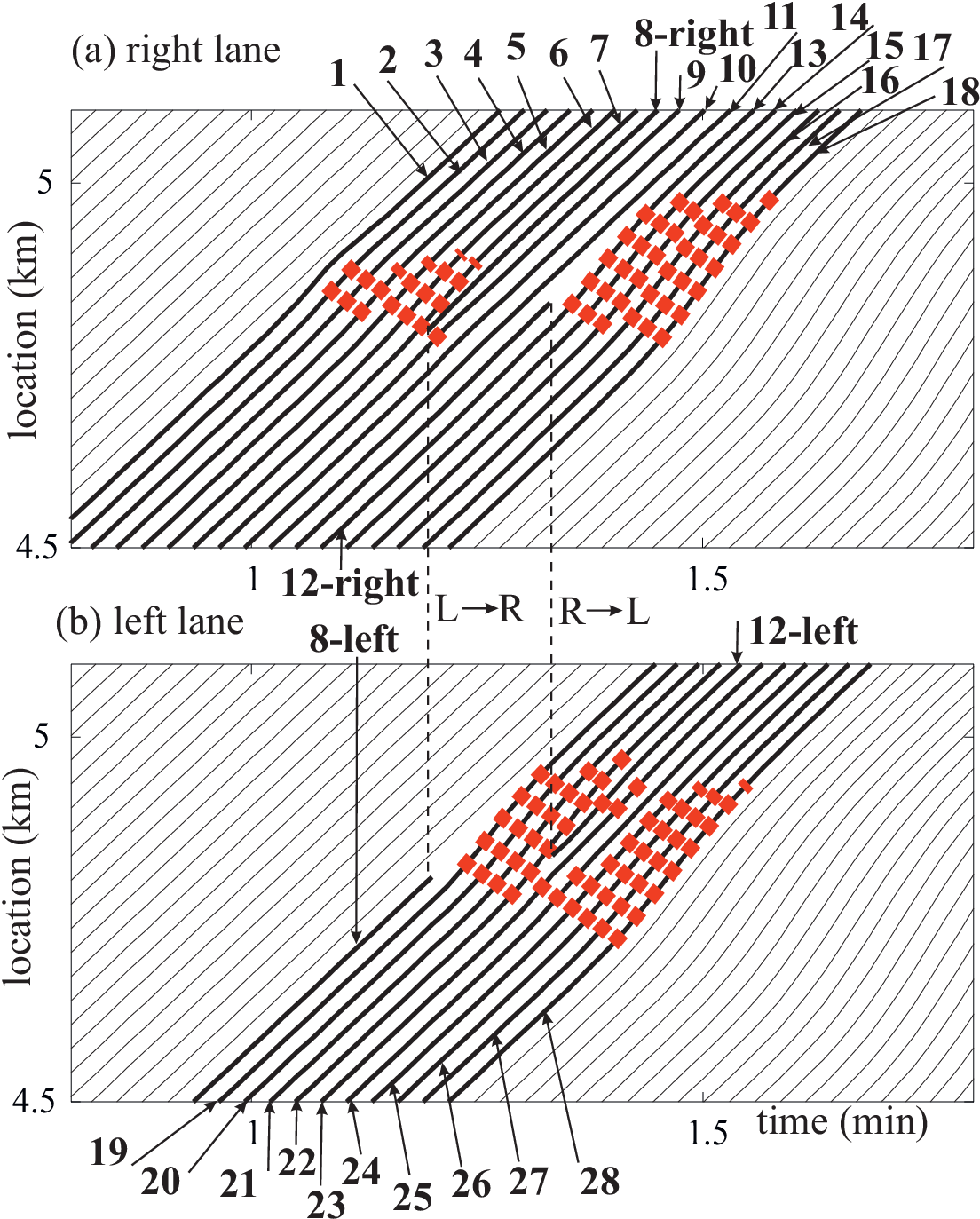}
\end{center}
\caption[]{Continuation of Figs.~\ref{2-lane-S-F-patterns}(a) and~\ref{2-lane-S-F-patterns}(b).
(a), (b) Simulated vehicle
trajectories within local speed increase in synchronized flow in
the right lane (a) and left lane (b). L$\rightarrow$R lane-changing
of vehicle 8  is marked by dashed vertical line L$\rightarrow$R.
	R$\rightarrow$L lane-changing
of vehicle  12 is marked by dashed vertical line R$\rightarrow$L.
Part of trajectories  within which condition $a_{\rm OA}>0$ is satisfied
  are marked through colored red squares. 
}
\label{S-F-instability-hom_traj}
\end{figure}

\begin{figure}
\begin{center}
\includegraphics[width = 8 cm]{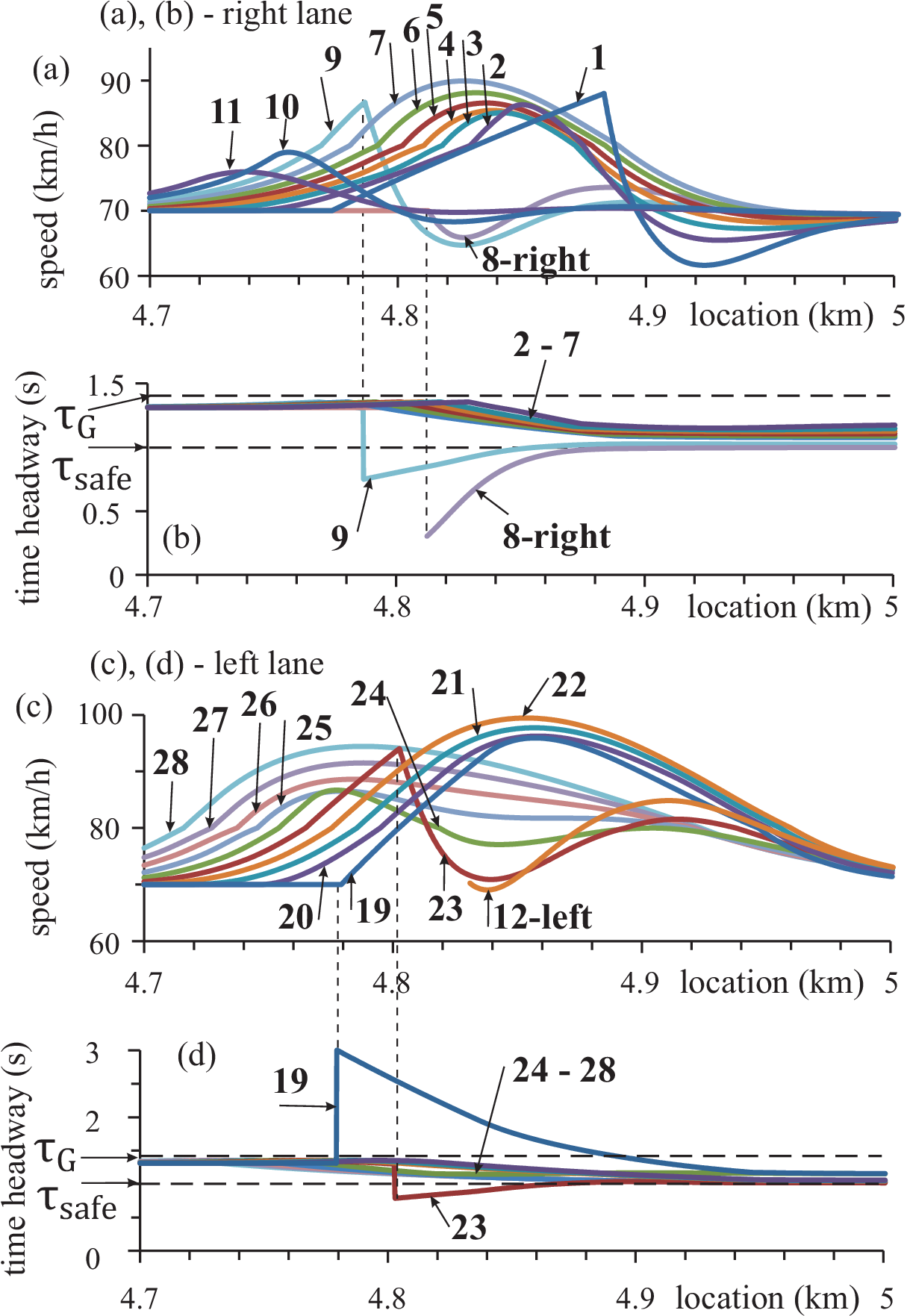}
\end{center}
\caption[]{Continuation of Fig.~\ref{S-F-instability-hom_traj}.
 Microscopic vehicle speeds (a), (c) and time headway (b), (d) for right lane (a), (b) and left lane (c), (d)
for some of the vehicles whose trajectories are labeled by the same numbers as those  
in Fig.~\ref{S-F-instability-hom_traj}, respectively.  
}
\label{S-F-instability-hom_traj_1}
\end{figure}

Acceleration of vehicle 1 [Figs.~\ref{S-F-instability-hom_traj}(a) and~\ref{S-F-instability-hom_traj_1}(a)]
causes a continuous increase in the speed of following vehicles 2--7 in the right lane
[Fig.~\ref{S-F-instability-hom_traj_1}(a)]. As in~\cite{Kerner2023C},   this S$\rightarrow$F instability causing the
 avalanche-like local speed increase in the initial synchronized flow is due to overacceleration 
$a_{\rm OA}$  (\ref{a_OA}) in the right lane: Indeed, regions on  trajectories,  in which condition $a_{\rm OA}>0$
is satisfied, become broader along the successive sequence of vehicles 2--7
[Fig.~\ref{S-F-instability-hom_traj}(a)]. 

\subsubsection{Interruption of S$\rightarrow$F instability \label{Interruption_S-F_Sec}}

However, contrary to  single-lane road studied in~\cite{Kerner2023C}, vehicles in the left lane that  move  initially  in synchronized flow at $v=$ 70 km/h
search for the opportunity to change to   the higher speed region  caused by the S$\rightarrow$F instability in the right lane.
After vehicle 8 in the left lane [vehicle 8-left in Fig.~\ref{S-F-instability-hom_traj}(b)]
 has changed to the right lane [vehicle 8-right in Figs.~\ref{S-F-instability-hom_traj}(a) and~\ref{S-F-instability-hom_traj_1}(a)],
the situation changes basically: Vehicles 9--11 in the right lane that follow vehicle 8-right 
begin to decelerate [Fig.~\ref{S-F-instability-hom_traj_1}(a)]. 
Because the speeds of vehicles 10 and 11 become less than  $v_{\rm syn}=$ 80 km/h in  (\ref{a_OA}), along   trajectories of vehicle 10 and 11 overacceleration
 $a_{\rm OA}= $ 0 [Fig.~\ref{S-F-instability-hom_traj}(a)]. This means that the S$\rightarrow$F instability in the right lane
is abrupt interrupted through L$\rightarrow$R lane-changing. The interruption of the development of
the S$\rightarrow$F instability in the right lane  occurs because just after  lane-changing the
 speed of vehicle 8-right is considerably lower than the speed of vehicle 9 whereas
 time headway of   vehicle   9 to  preceding vehicle 8-right
 is considerably shorter than
the safe time-headway $\tau_{\rm safe}$; therefore, vehicle 9 must decelerate strongly
 [Figs.~\ref{S-F-instability-hom_traj_1}(a) and~\ref{S-F-instability-hom_traj_1}(b)].

\subsubsection{S$\rightarrow$F instability in left lane causing repetition of S$\rightarrow$F instability 
in right lane \label{Redevelopment_S-F_Sec}}
 
There are three   dynamic effects following L$\rightarrow$R lane-changing: (i) The interruption of the S$\rightarrow$F instability in the right lane  (Sec.~\ref{Interruption_S-F_Sec}), (ii) the beginning of the S$\rightarrow$F instability in the left lane,
and (iii) the repetition of the S$\rightarrow$F instability in the right lane.

The S$\rightarrow$F instability in the left lane is caused by overacceleration through
L$\rightarrow$R lane-changing of vehicle 8:
Time headway of vehicle 19 in the left lane
  increases considerably   [Figs.~\ref{S-F-instability-hom_traj}(b), 
	 \ref{S-F-instability-hom_traj_1}(c) and~\ref{S-F-instability-hom_traj_1}(d)]. This results in overacceleration of vehicle 19  
	[$a_{\rm OA}>$ 0 on trajectory 19 in Fig.~\ref{S-F-instability-hom_traj}(b)]. This overacceleration causes 
the S$\rightarrow$F instability in the left lane   [vehicles 20--22   in 
Fig.~\ref{S-F-instability-hom_traj}(b)].

\begin{figure}
\begin{center}
\includegraphics[width = 8 cm]{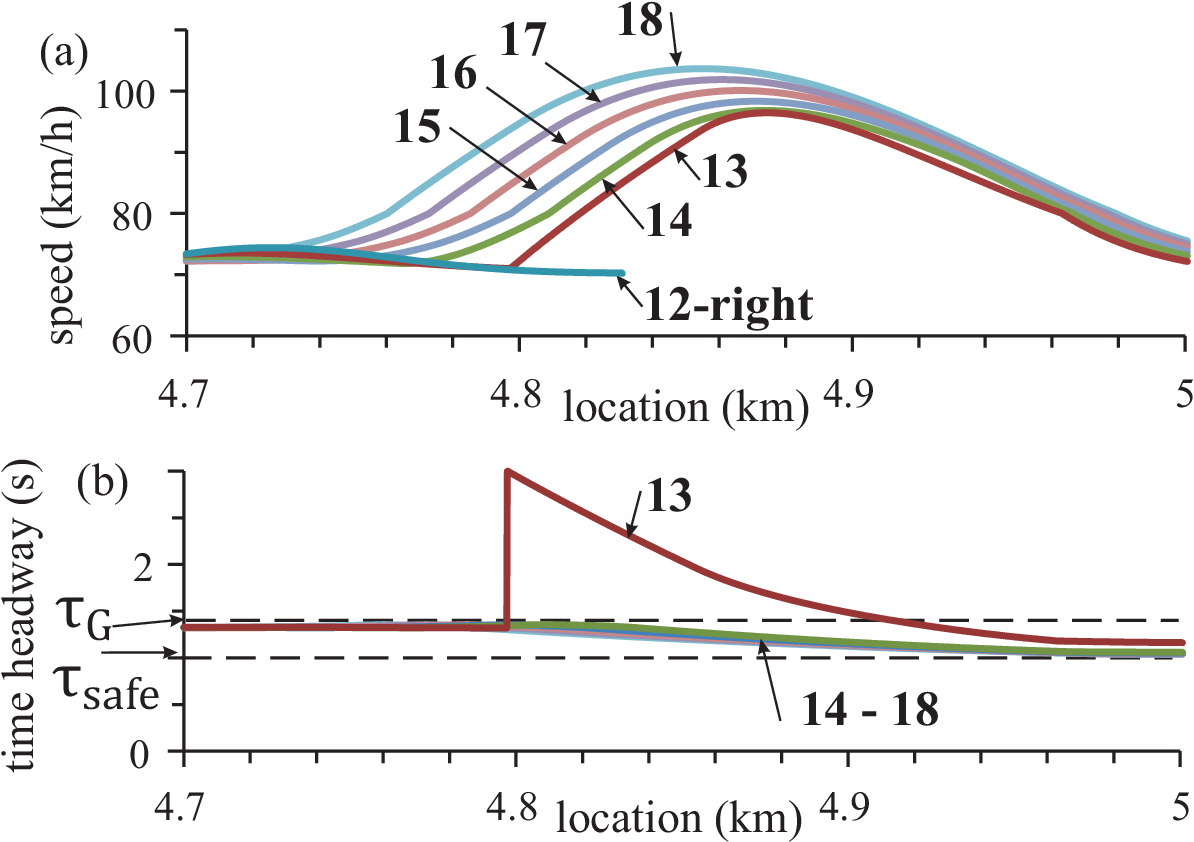}
\end{center}
\caption[]{Continuation of Fig.~\ref{S-F-instability-hom_traj}. Repetition of S$\rightarrow$F instability in right lane.
 Microscopic vehicle speeds (a) and time headway (b) for right lane 
for some of the vehicles whose trajectories are labeled by the same numbers as those  
in Fig.~\ref{S-F-instability-hom_traj}, respectively.  
}
\label{S-F-instability-hom_traj_2}
\end{figure}

Now vehicles following vehicle 8-right
 in the right lane, which due to the interruption of  the S$\rightarrow$F instability in the right lane should
 move    in synchronized flow,  
search for the opportunity to change to   a higher speed region  caused by the S$\rightarrow$F instability in the left lane:
After vehicle 12 in the right lane [vehicle 12-right in Fig.~\ref{S-F-instability-hom_traj}(a)]
 has changed to the left lane [vehicle 12-left in
 Figs.~\ref{S-F-instability-hom_traj}(b) and~\ref{S-F-instability-hom_traj_1}(c))],
 two   effects occur: (i) Because vehicle 12-right changes the lane, time headway of vehicle 13 in the right lane
  increases considerably   [Fig.~\ref{S-F-instability-hom_traj_2}(b)]. (ii) Vehicle 23 
	that follows vehicle 12-left in the left lane
 decelerates strongly [Fig.~\ref{S-F-instability-hom_traj_1}(c)].

The first effect of R$\rightarrow$L lane-changing causes  repetition of
 the S$\rightarrow$F instability in the right lane (Fig.~\ref{S-F-instability-hom_traj_2}).  
Indeed, time headway of vehicle 13 increases; this leads to  overacceleration  
  of vehicle 13     [$a_{\rm OA}>$ 0 on trajectory 13 in Fig.~\ref{S-F-instability-hom_traj}(a)]. This causes 
the S$\rightarrow$F instability in the right lane   [vehicles 14--18   in 
Fig.~\ref{S-F-instability-hom_traj_2}(a)].

The second effect of R$\rightarrow$L lane-changing, which leads to deceleration of vehicle 23 following vehicle 12-left, does not cause
interruption of
the S$\rightarrow$F instability in the left lane. This is
  in contrast to L$\rightarrow$R lane-changing of vehicle 8 causing the interruption of the
S$\rightarrow$F instability in the right lane (Sec.~\ref{Interruption_S-F_Sec}). 
To explain it, we note that along
trajectory of vehicle 24 following vehicle 23 in the left lane 
overacceleration $a_{\rm OA}$ does not disappear  [trajectory 24 in Fig.~\ref{S-F-instability-hom_traj}(b)].
	For this reason, even due to speed decrease of vehicle 23, the S$\rightarrow$F 
	instability in the left lane continues to develop [Fig.~\ref{S-F-instability-hom_traj}(b)]. 
	Contrarily, after L$\rightarrow$R lane-changing of vehicle 8 has occurred, along trajectory of vehicle 10
	overacceleration $a_{\rm OA}=$ 0 
[vehicle 10  in Fig.~\ref{S-F-instability-hom_traj}(a)]; this causes the interruption 
  of the S$\rightarrow$F instability in the right lane.
	 Finally, due to
		the S$\rightarrow$F instability in the left and right lanes, local regions of free flow are built in both lanes
		[Fig.~\ref{2-lane-S-F-patterns}(a)].

\subsection{Microscopic characteristics of overacceleration cooperation under  S$\rightarrow$F instability 
at bottleneck \label{Effect_SF_onramp_Sec}}

As found  from 
 stochastic three-phase traffic flow models~\cite{Kerner2015_SF}, simulations of    deterministic model (\ref{g_v_g_min1})--(\ref{G_g_safe_simple}) show also that
there can   be a spontaneous S$\rightarrow$F transition at the bottleneck
(Fig.~\ref{S-F-onramp}), which is governed by 
 an  S$\rightarrow$F instability [Figs.~\ref{S-F-onramp_traj}--\ref{S-F-onramp_traj_2}].
The uninterrupted development of the S$\rightarrow$F instability leads to an S$\rightarrow$F transition at the bottleneck, i.e.,
free flow is recovered at the bottleneck. There are two possibilities: (i) After a short time interval,
an F$\rightarrow$S transition occurs spontaneously at the bottleneck, i.e., the recovered free flow  
  persists only the short time interval 
[Fig.~\ref{S-F-onramp}(a)]. (ii) The recovered free flow  persists during a long enough time interval at the bottleneck
[Fig.~\ref{S-F-onramp}(b)].

 \begin{figure}
\begin{center}
\includegraphics[width = 8 cm]{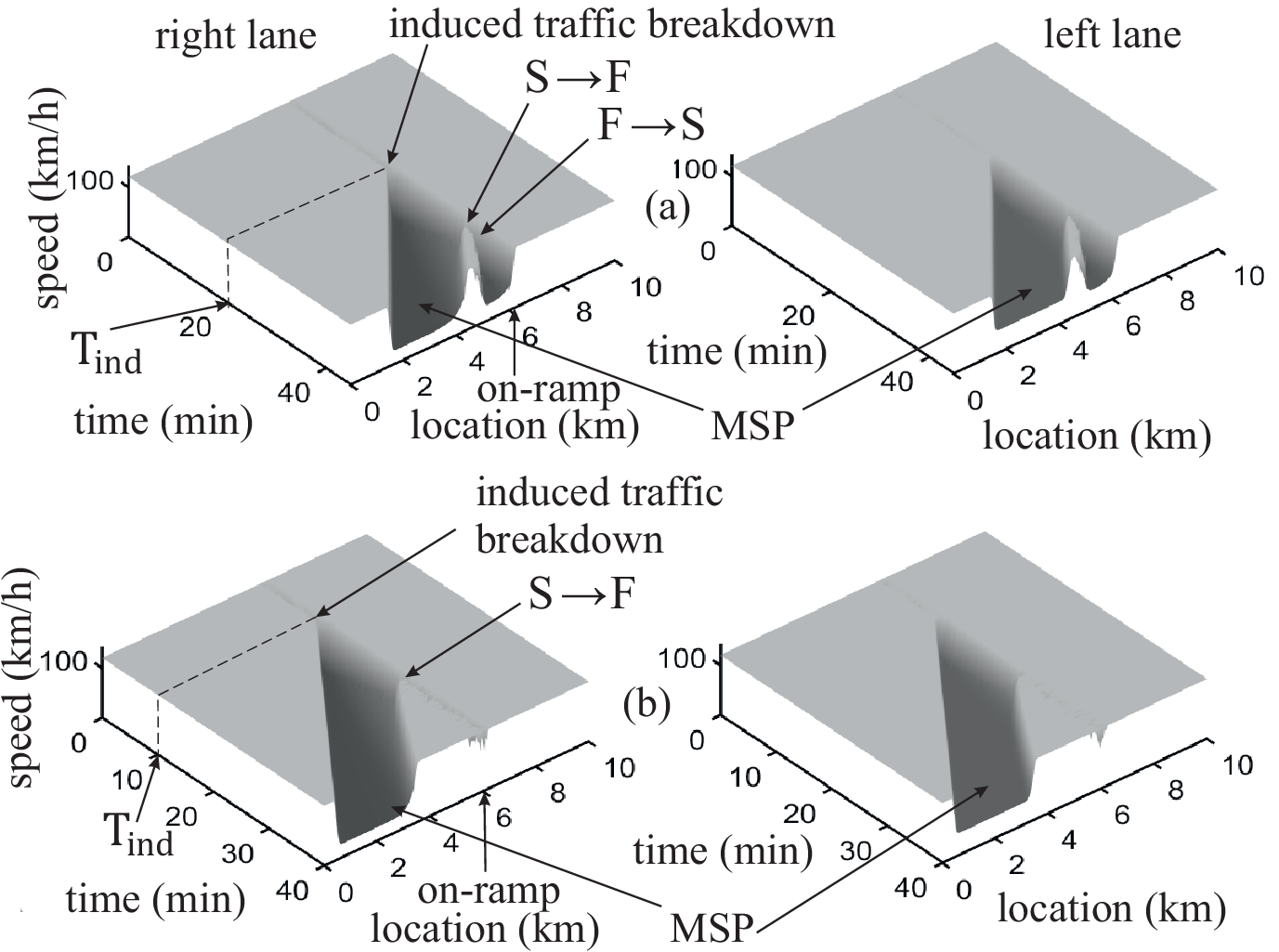}
\end{center}
\caption[]{Cooperation of overacceleration in road lane and due to lane-changing
leading to S$\rightarrow$F instability on   two-lane road with on-ramp bottleneck at $x_{\rm on}=$ 6 km.
Simulations  
with model (\ref{g_v_g_min1})--(\ref{G_g_safe_simple}) at $\tau_{\rm G}=$ 1.4 s made  
on two-lane  road  at $q_{\rm in}=$ 2400 (vehicles/h)/lane, $q_{\rm on}=$ 495 vehicles/h (a), and
$q_{\rm on}=$ 500 vehicles/h (b). Vehicle speed
 is space and time in the right lane (left subplots) and in the left lane (right subplots).
 Traffic breakdown   has been induced  
  in   free flow   at $T_{\rm ind}=$ 20 min (a) and $T_{\rm ind}=$ 10 min (b) through application of  
	addition on-ramp inflow impulse $\Delta q_{\rm on}=$
1000 vehicles/h of duration   $\Delta t=$ 1 min. Spontaneous S$\rightarrow$F   and F$\rightarrow$S transitions
at the bottleneck are labeled by
arrows S$\rightarrow$F and F$\rightarrow$S, respectively. MSP   emerges at the bottleneck
due to the development of the S$\rightarrow$F instability.
Other model parameters are the same as those in Figs.~\ref{2-lane-alpha-1-patterns}(c)--\ref{2-lane-alpha-1-patterns}(e).
}
\label{S-F-onramp}
\end{figure}

\begin{figure}
\begin{center}
\includegraphics[width = 8 cm]{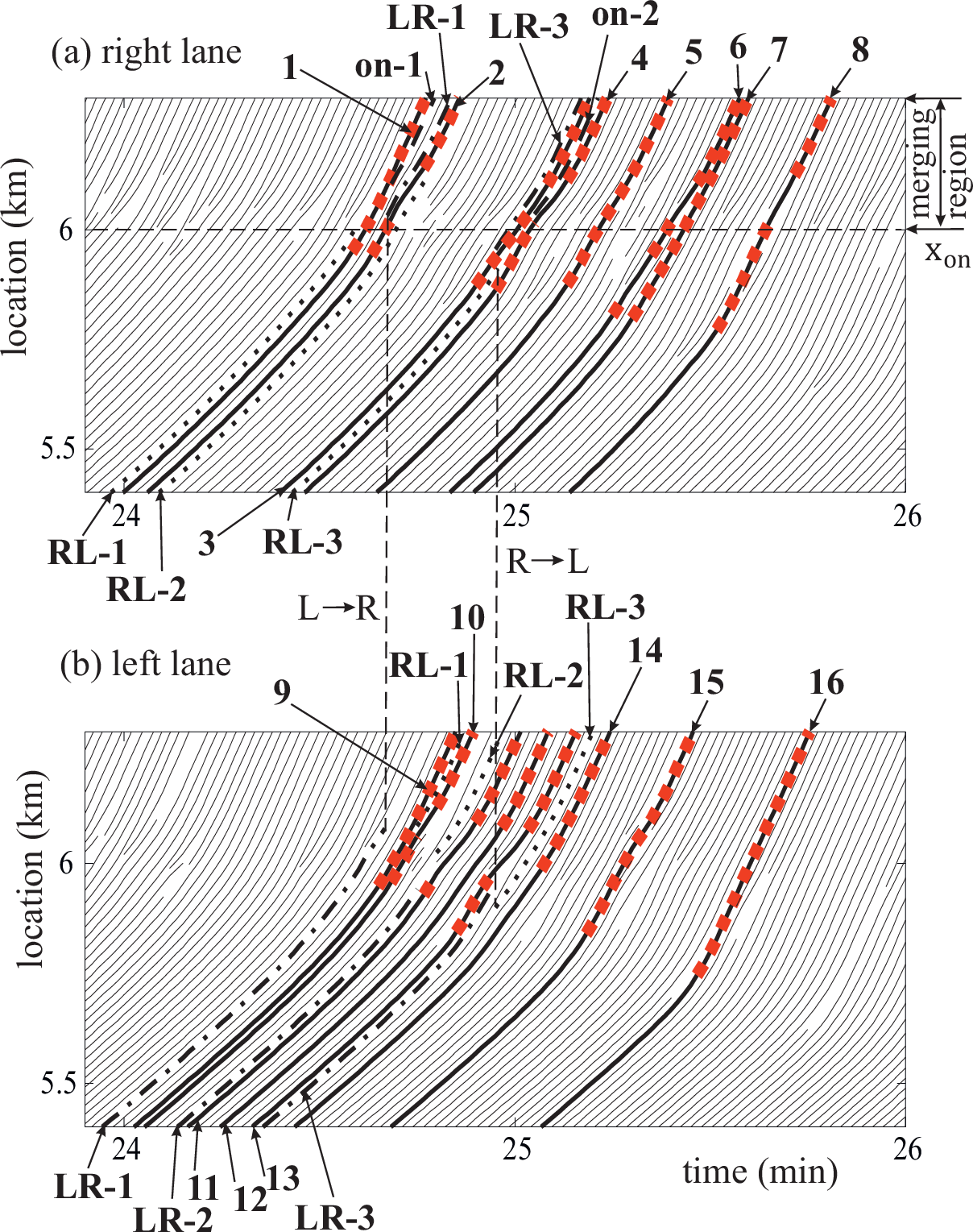}
\end{center}
\caption[]{Continuation of Fig.~\ref{S-F-onramp}(b).
(a), (b) Simulated vehicle
trajectories  in synchronized flow in
the right lane (a) and left lane (b). One of the L$\rightarrow$R lane-changing
effects (for vehicle LR-1)
   is marked by dashed vertical line L$\rightarrow$R. One of the
	R$\rightarrow$L lane-changing effects
(for vehicle RL-3)  is marked by dashed vertical line R$\rightarrow$L.
Part of trajectories  within which condition $a_{\rm OA}>0$ is satisfied
  are marked through colored red squares. Some of the vehicle trajectories used for explanations in text
	have been highlighted by bold as follows: solid trajectories -- vehicles that do not change lane,
	dotted  trajectories -- R$\rightarrow$L lane-changing,
	dashed-dotted trajectories -- L$\rightarrow$R lane-changing, dashed trajectories -- on-ramp vehicles.
	}
\label{S-F-onramp_traj}
\end{figure}

\begin{figure}
\begin{center}
\includegraphics[width = 8 cm]{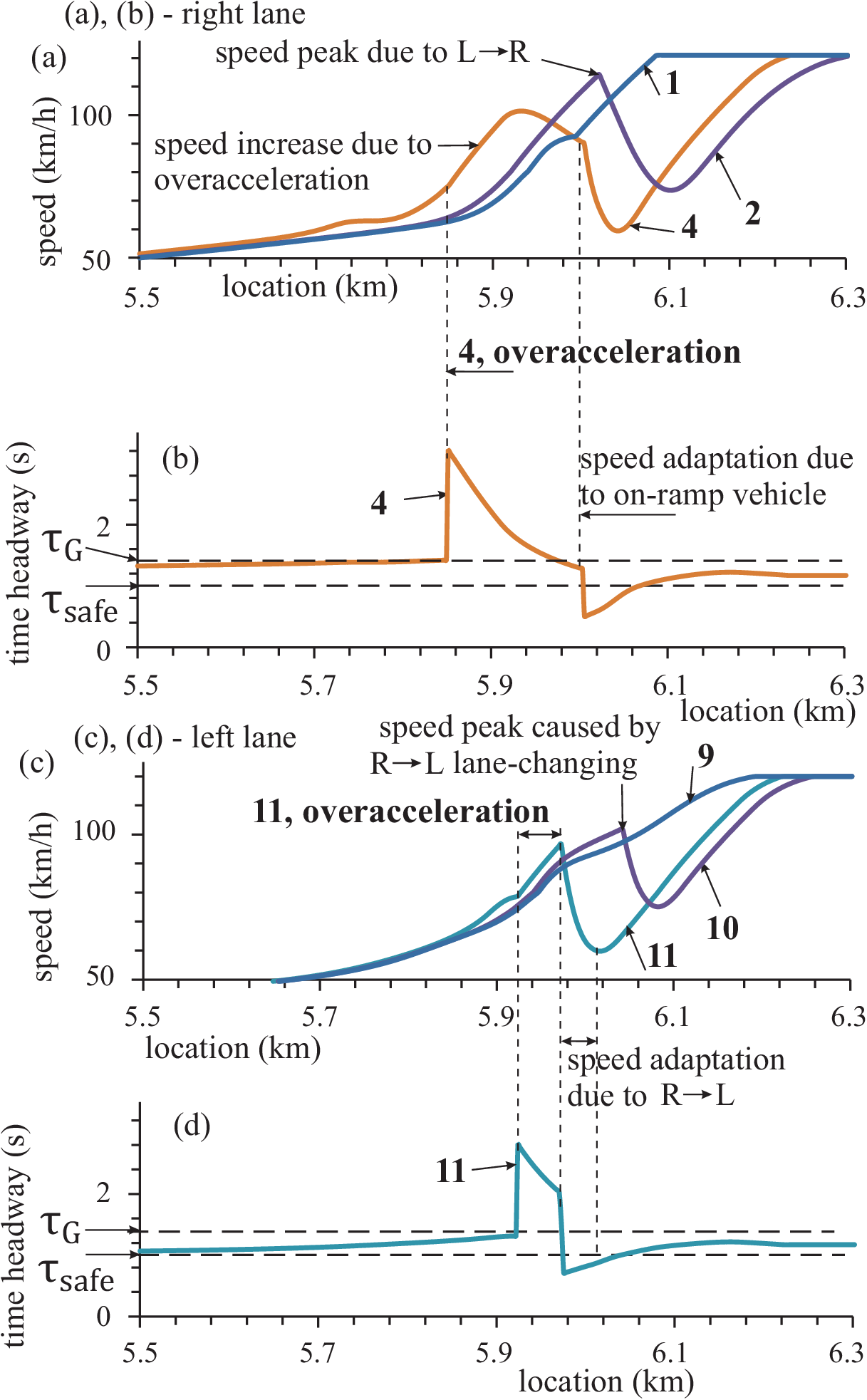}
\end{center}
\caption[]{Continuation of Fig.~\ref{S-F-onramp_traj}. Speed peak in the right lane initiating
  S$\rightarrow$F instability in the right lane and the beginning of S$\rightarrow$F instability in the left lane
	through overacceleration caused by L$\rightarrow$R lane-changing.
 Microscopic vehicle speeds (a), (c) and time headway (b), (d) for right lane (a), (b) and left lane (c), (d)
for some of the vehicles whose trajectories are labeled by the same numbers as those  
in Fig.~\ref{S-F-onramp_traj}, respectively.  
}
\label{S-F-onramp_traj_1}
\end{figure}

As in  stochastic models~\cite{Kerner2015_SF},
  the S$\rightarrow$F instability at the bottleneck is  initiated spontaneously by a
	$\lq\lq$speed peak" (Fig.~\ref{S-F-onramp_traj_1}(a)).
Vehicle 1 is
moving in the right lane and has reached the downstream boundary
of  synchronized flow. Therefore, vehicle 1  accelerates from the
speed inside synchronized flow to   free flow.
Vehicle 2, following vehicle 1, also starts to accelerate. However,  vehicle LR-1 that has changed from the left lane 
(Fig.~\ref{S-F-onramp_traj}(a))    forces vehicle 2 to decelerate, creating
the speed peak (labeled in Fig.~\ref{S-F-onramp_traj_1}(a) by $\lq\lq$speed peak due to L$\rightarrow$R").
	
In this paper, we have revealed that overacceleration cooperation   (Secs.~\ref{Cooperation_Sec}
and~\ref{2-lane_Coop_Sec}) effects
 crucially on microscopic  traffic dynamics   of
the S$\rightarrow$F instability.
We limit a further consideration of the case shown in Fig.~\ref{S-F-onramp}(b).

\subsubsection{Effect of  lane-changing and merging of on-ramp vehicles in right lane \label{Sequences_LC_ON_S}}

 In the right lane, the
S$\rightarrow$F instability is caused by overacceleration cooperation through
 R$\rightarrow$L lane-changing and overacceleration $a_{\rm OA}$ in road lane. Contrarily,  speed adaptation that occurs 
through the merging of on-ramp vehicles together
with L$\rightarrow$R lane-changing tries   to prevent the S$\rightarrow$F instability development
  [Figs.~\ref{S-F-onramp_traj}--\ref{S-F-onramp_traj_2}].

\begin{figure}
\begin{center}
\includegraphics[width = 8 cm]{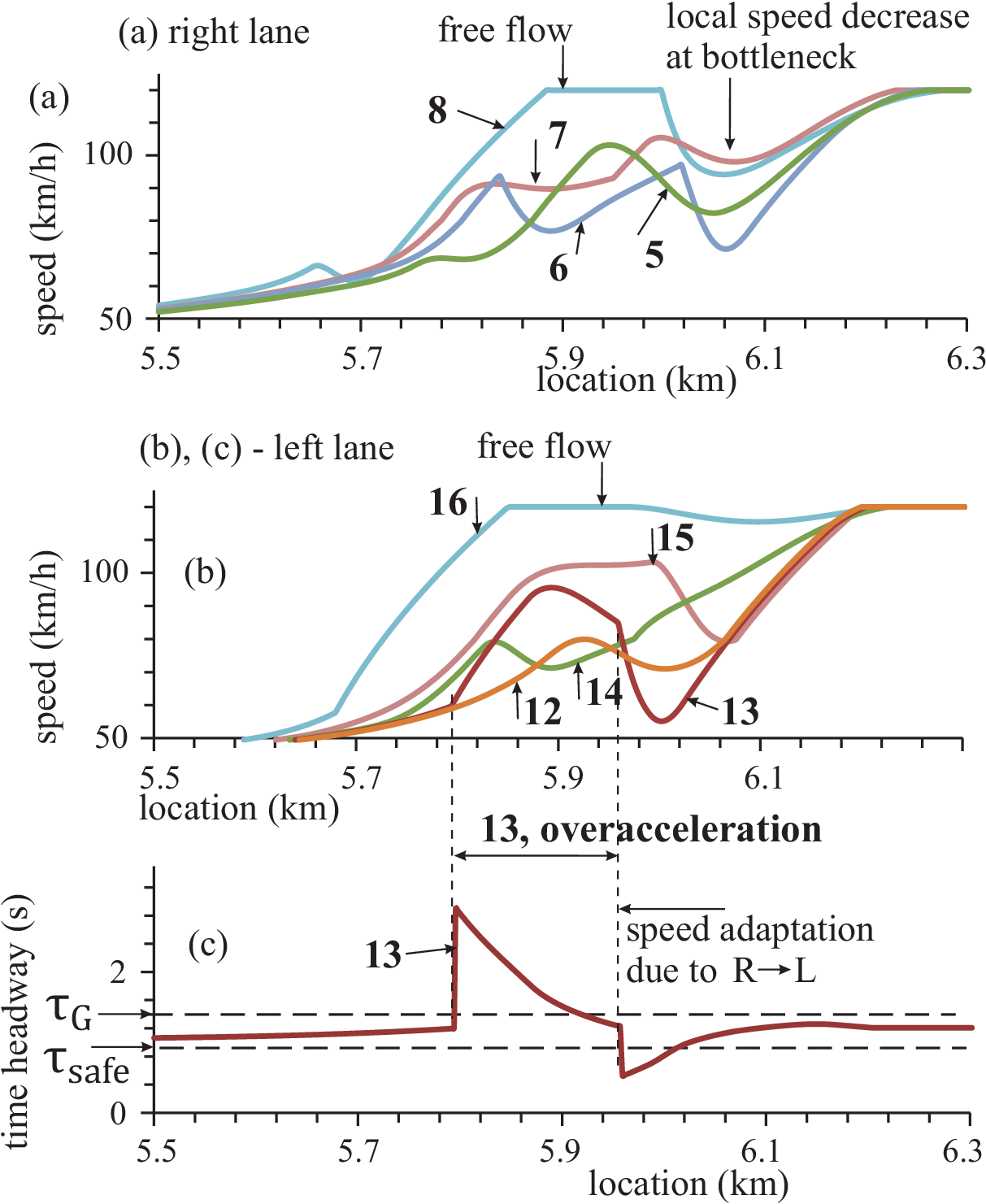}
\end{center}
\caption[]{Continuation of Fig.~\ref{S-F-onramp_traj}.  Development of S$\rightarrow$F instability
in right lane (a)  and left lane (b), (c):
  Microscopic vehicle speeds (a), (b) and time headway (c) 
for some of the vehicles whose trajectories are labeled by the same numbers as those  
in Fig.~\ref{S-F-onramp_traj}, respectively.  
}
\label{S-F-onramp_traj_2}
\end{figure}

Indeed, after L$\rightarrow$R lane-changing [vehicle LR-3 in Fig.~\ref{S-F-onramp_traj}(a)] 
   vehicles 3 and RL-3 following vehicle LR-3 in the right lane
	should decelerate   adapting their speeds to the speed of vehicle LR-3. This causes
  R$\rightarrow$L lane-changing of   vehicle RL-3
[Fig.~\ref{S-F-onramp_traj}(a)]. In its turn,   this R$\rightarrow$L lane-changing
 results in time headway increase for
  vehicle 4 in the right lane [Fig.~\ref{S-F-onramp_traj_1}(b)] and, therefore,
in overacceleration  of   vehicle 4 
[$\lq\lq$4, overacceleration"  in Figs.~\ref{S-F-onramp_traj_1}(a) and~\ref{S-F-onramp_traj_1}(b)]. 
Overacceleration of   vehicle 4
as well as    the subsequent acceleration of some following vehicles 
 causes a local speed increase upstream of the on-ramp merging region
[$\lq\lq$speed increase due to overacceleration" in Fig.~\ref{S-F-onramp_traj_1}(a)].
 However, later due to   on-ramp vehicle merging [vehicle  on-2 in Fig.~\ref{S-F-onramp_traj}(a)],
 vehicle 4 must decelerate, i.e.,   speed adaptation effect is realized 
[$\lq\lq$speed adaptation due to   on-ramp vehicle" in Figs.~\ref{S-F-onramp_traj_1}(a) and~\ref{S-F-onramp_traj_1}(b)].
	For this reason, 
 the   local speed increase upstream of the on-ramp merging region
changes to a local speed decrease within the on-ramp merging region.
Rather than the interruption  of
 the S$\rightarrow$F instability, this competition of overacceleration and speed adaptation
leads only to a decrease in the growth of the S$\rightarrow$F instability~\footnote{However,
 in some other cases we have found (not shown) that
sequences of lane-changing and merging of on-ramp in the right lane can lead to a short time interruption of the development of
 the S$\rightarrow$F instability; this interruption effect is qualitatively
 similar to   S$\rightarrow$F instability interruption explained in Sec.~\ref{Interruption_S-F_Sec}
[Fig.~\ref{S-F-instability-hom_traj_1}(a)]. }.

The decrease in the growth of a sequence of the 
local speed increase and decrease   explained   for vehicle 4  occurs several times upstream of the bottleneck
[vehicles 5--7 in Fig.~\ref{S-F-onramp_traj_2}(a)]. Finally, 
the  S$\rightarrow$F instability is finished by the  S$\rightarrow$F transition at the bottleneck
[vehicle 8 in Fig.~\ref{S-F-onramp_traj_2}(a)] and the formation of  a local speed decrease
in free flow upstream of the bottleneck   caused mostly by on-ramp vehicles
[$\lq\lq$local speed decrease at bottleneck"  in Fig.~\ref{S-F-onramp_traj_2}(a)]. 

 \subsubsection{Emergence of speed peak   in left lane  \label{Speed_peak_left lane_S}}

Vehicle 9     accelerates within the downstream boundary
of the synchronized flow   to  free flow downstream [Fig.~\ref{S-F-onramp_traj_1}(c)].
Vehicle 10 that follows vehicle 9  also starts to accelerate within the downstream boundary
of the synchronized flow. However,  
a slower moving vehicle RL-1 that has changed from the right lane  
[vehicle RL-1  
in Fig.~\ref{S-F-onramp_traj}(b)]    forces vehicle 10 to decelerate, creating
the speed peak on vehicle trajectory 10
 [$\lq\lq$speed peak caused by R$\rightarrow$L lane-changing" in Fig.~\ref{S-F-onramp_traj_1}(c)].

 \subsubsection{Effect of lane-changing on S$\rightarrow$F instability in left lane  \label{Dynamics_left lane_S}}

In the left lane, the
S$\rightarrow$F instability is caused by overacceleration cooperation through
 L$\rightarrow$R lane-changing and overacceleration $a_{\rm OA}$ in road lane. Contrarily,  speed adaptation 
through R$\rightarrow$L lane-changing tries   to interrupt the S$\rightarrow$F instability development.

 Indeed, because upstream of the speed peak  in the left lane the speed decreases,   L$\rightarrow$R lane-changing
 is realized [vehicle LR-2  in Fig.~\ref{S-F-onramp_traj}(b)]. This results in a time-headway increase for
  vehicle 11 [Fig.~\ref{S-F-onramp_traj_1}(d)]  and, therefore,
in overacceleration  of   vehicle 11 [$\lq\lq$11, overacceleration" in Fig.~\ref{S-F-onramp_traj_1}(c)].
Shortly afterward, however, vehicle RL-2 changes from the right lane to the left lane in front of vehicle 11 
    (Fig.~\ref{S-F-onramp_traj}). This R$\rightarrow$L lane-changing causes speed adaptation of vehicle 11
	[$\lq\lq$speed adaptation due to R$\rightarrow$L" in Figs.~\ref{S-F-onramp_traj_1}(c) and~\ref{S-F-onramp_traj_1}(d)].
	
	Thus,  there is a sequence of the local speed increase with
	the following speed decrease caused by speed adaptation.
	 As in the right lane (Sec.~\ref{Sequences_LC_ON_S}), the local speed increase  caused by overacceleration propagates
	upstream and it can become lower for some of the following vehicles moving
	upstream of vehicle 11
	[e.g., vehicles 12 and 14 in Fig.~\ref{S-F-onramp_traj_2}(b)]. Nevertheless, for other vehicles  
	overacceleration leads to the growth of the local speed increase [vehicle 13 in Fig.~\ref{S-F-onramp_traj_2}(b)].
	This occurs when due to L$\rightarrow$R lane-changing [vehicle LR-3 in Fig.~\ref{S-F-onramp_traj}(b)]
	time-headway increases considerably
	[vehicle 13 in Fig.~\ref{S-F-onramp_traj_2}(c)] leading to a greater overacceleration
	[$\lq\lq$13, overacceleration" in Figs.~\ref{S-F-onramp_traj_2}(b) and~\ref{S-F-onramp_traj_2}(c)].  
	On average due to overacceleration the local speed increase grows upstream 
	[vehicles 13 and 15 in Fig.~\ref{S-F-onramp_traj_2}(b)].
  Finally, 
the  S$\rightarrow$F instability is finished by the  S$\rightarrow$F transition in the left lane
[vehicle 16  in Fig.~\ref{S-F-onramp_traj_2}(b)].
 
\section{Discussion  \label{Dis}}

 \subsection{Microscopic  behaviors of vehicle motion as the cause of overacceleration     \label{OA_behavior_Sec}}

  The behavioral  origin of the discontinuous character
of   overacceleration applied in  Eq.~(\ref{a_OA}) has already been explained in~\cite{Kerner2023B}.
As we explain here, all other overacceleration mechanisms
as well as overacceleration cooperation found above
are caused by   microscopic   behaviors of vehicle motion  used in  model
(\ref{g_v_g_min1})--(\ref{g_prec_ACC}).

\subsubsection{Competition of speed adaptation   with safety acceleration  \label{SA-Safety-OA_S}}

In Figs.~\ref{No-overacceleration_traj}, \ref{Greater_q-in_traj}, 
	and~\ref{Greater_G_traj} we have shown that
  vehicle 1, which   must break while approaching    slower on-ramp vehicle on-1,
	accelerates later through safety acceleration. This reflects  the vehicle's desire to move in free flow. 
	However, the breaking of vehicle 1 forces 
	vehicles following vehicle 1 to decelerate; this speed adaptation
	is the opposite effect to safety acceleration.  

 When 
{\it for any vehicle speed} within the local speed decrease at the bottleneck
speed adaptation is on average stronger than safety acceleration,  
then safety acceleration cannot prevent congested traffic at the bottleneck. This is
independent of the    speed
 within the local speed decrease at the bottleneck (Fig.~\ref{LWR_Fig}). In this case, safety acceleration
does not lead to a nucleation character of traffic breakdown at the bottleneck  and, therefore,
  safety acceleration cannot be considered overacceleration (Sec.~\ref{1-lane-neglecting}).

  At other  traffic parameters
	(Secs.~\ref{Close_Safe_S} and~\ref{Greater_Syn_S}) speed adaptation is on average weaker than safety acceleration
at {\it high enough speeds only}, whereas at a low enough speed  
 speed adaptation becomes stronger than safety acceleration.
Then safety acceleration does prevent congested traffic at the bottleneck at high enough speeds within local speed decrease at the bottleneck: Safety acceleration
does   lead to the nucleation character of traffic breakdown at the bottleneck  (Figs.~\ref{Greater_q-in_induced} and~\ref{Greater_G_induced}). In these cases,
    in accordance with overacceleration definition,
safety acceleration is overacceleration.

 \subsubsection{Suppression of speed adaptation within indifferent zone through overacceleration \label{Suppression-AD-OA-Sec}}

We consider a vehicle that approaches the local speed decrease 
at the bottleneck while moving still upstream of the bottleneck  
within the indifferent zone $g_{\rm safe} \leq g \leq G$.
When condition $\alpha=0$ (\ref{alpha_0}) has been assumed (Sec.~\ref{SA-Safety-OA_S}), then 
the speed adaptation effect (\ref{g_v_g_min1_SA}) can be stronger than safety acceleration at any speed leading to
the absence of overacceleration (Sec.~\ref{1-lane-neglecting}).  

Contrarily, under opposite condition  $\alpha\neq 0$  due to overacceleration $a_{\rm OA}$ (\ref{a_OA}) 
  vehicle   
tries to escape from the local speed decrease at the bottleneck. This 
overacceleration  can suppress speed adaptation within the indifferent zone $g_{\rm safe} \leq g \leq G$
 as long as in (\ref{a_OA})
 speed  $v \geq v_{\rm syn}$. The suppression of speed adaptation through
overacceleration $a_{\rm OA}$ (\ref{a_OA}), (\ref{a_OA_gap}) causes
 the maintaining of free flow at the bottleneck within
a considerable larger flow-rate range [Figs.~\ref{2Z_Fig}(c) and~\ref{2Z_Fig}(d)] in comparison with the hypothetical case  
$\alpha=0$ (\ref{alpha_0}) [Fig.~\ref{2Z_Fig}(b)].

 \subsubsection{Overacceleration cooperation  in road lane}

Microscopic vehicle behaviors leading to overacceleration cooperation     in road lane (Sec.~\ref{Cooperation_Sec}) are as follows:
(i) Overacceleration $a_{\rm OA}$ (\ref{a_OA}), (\ref{a_OA_gap}) suppresses speed adaptation as
 explained in Sec.~\ref{Suppression-AD-OA-Sec}.
(ii) Independent of the effect of the suppression of speed adaptation,
there is mechanism of overacceleration caused by safety acceleration [$\lq\lq$1, overacceleration" in
Fig.~\ref{Greater_G_alpha_traj}].
(iii) Overacceleration caused by safety acceleration is changed to overacceleration $a_{\rm OA}$ just
after time headway   becomes longer than safe time-headway 
[$\lq\lq$1, change of overacceleration" in
Fig.~\ref{Greater_G_alpha_traj}]. Due to this overacceleration cooperation
 the maintaining of free flow at the bottleneck is realized within
a  larger flow-rate range [Fig.~\ref{2Z_Fig}(d)] in comparison with the   case  
of single overacceleration mechanism $a_{\rm OA}$ [Fig.~\ref{2Z_Fig}(c)].

  \subsubsection{Vehicle breaking capability as   reason for   critical speed}

	The suppression of speed adaptation through overacceleration $a_{\rm OA}$ can be realized only when in (\ref{a_OA}) 
  condition $v\geq v_{\rm syn}$ is satisfied. This can explain the existence of a critical speed
	for traffic breakdown  (Sec.~\ref{Veh_dyn_Sec}) as follows.  
		There is a finite breaking capability of a vehicle: The lower the minimum speed
		$v_{\rm min}$ within the local speed decrease at the bottleneck,
		the longer the vehicle breaking distance, therefore, the longer the width of the upstream front of
		the local speed decrease within which the vehicle decelerates.
		For this reason, the road location, at which
		condition $v<v_{\rm syn}$ has been just satisfied and, therefore, overacceleration $a_{\rm OA}=$ 0,  
		moves upstream, when $v_{\rm min}$ decreases.
		
		This effect is realized both
		on single-lane road [Figs.~\ref{Greater_G_spont_alpha}(b) and~\ref{Greater_G_spont_alpha_traj}(a)] and 
		two-lane road [Figs.~\ref{2-lane-alpha-1-spont_traj}(a), \ref{2-lane-alpha-1-spont_traj}(b),
		\ref{2-lane-alpha-1-spont_traj2}(a),  and~\ref{2-lane-alpha-1-spont_traj2}(b)].
		Thus, the suppression effect of speed adaptation through overacceleration becomes the weaker,
		the lower the minimum speed $v_{\rm min}$ 
		within the local speed decrease at the bottleneck. For this reason, there should be a critical  minimum speed within the local speed decrease  at which speed adaptation becomes on average stronger than overacceleration. This leads to
		traffic breakdown (F$\rightarrow$S transition).

\subsection{Generalization of deterministic   model for moving jam simulations \label{Gener_Sec}}

 In the   deterministic three-phase traffic flow model used for all simulations presented above
[Figs.~\ref{LWR_Fig}--\ref{S-F-onramp_traj_2}] we have used values of model parameters at which there is no vehicle overdeceleration and, therefore, no classical traffic flow instability (and no string instability) exists. For this reason,  moving jams occur in none of the simulations even when
the synchronized flow speed is very low. However, empirical data
have shown that moving jams do emerge in synchronized flow when the synchronized flow speed is low enough~\cite{KernerBook1}. Therefore, to simulate
moving jam emergence in  deterministic model 
(\ref{g_v_g_min1})--(\ref{g_prec_ACC}),
as already made in some stochastic three-phase traffic flow models (e.g.,~\cite{KKl,OA_Stoch}), 
we assume that dynamic model parameters (at least some of the model parameters)
     can change at a synchronized flow speed that is less than some characteristic speed
denoted by 
 $v_{\rm pinch}$: 
\begin{equation}
\mathcal{P}=\left\{
\begin{array}{ll}
\mathcal{P} \ {\rm at} \ v \geq v_{\rm pinch}
 \\
\mathcal{P}_{\rm pinch} \ {\rm at} \ v < v_{\rm pinch}, \\
\end{array}\right.
  \label{pinch-formula}
  \end{equation}
	where $\mathcal{P}$ denotes one of the 
 model parameters $\tau_{\rm G}$, $\tau_{\rm safe}$, $K_{1}$, $K_{2}$, $K_{3}$, $K^{(1)}_{4}$, 
 $K^{(2)}_{4}$, etc. used in  model (\ref{g_v_g_min1})--(\ref{g_prec_ACC}), whereas  
the subscript $pinch$ in  $\mathcal{P}_{\rm pinch}$ is used to
distinguish values of the same model parameters    for
  low speeds $v< v_{\rm pinch}$. We assume also that in Eq.~(\ref{a_OA})  condition
	\begin{equation}
 v_{\rm pinch}<v_{\rm syn}
  \label{pinch-syn-formula}
  \end{equation}
 is satisfied.
In the generalized model, rather than functions
(\ref{G_g_safe_simple}),
  we use   formulations for the speed-functions $G(v)$ and $g_{\rm safe}(v)$~\footnote{In general,
  synchronization space gap $G$ and   safe space gap  $g_{\rm safe}$ can also be formulated as functions of
different vehicle variables; for example, in~\cite{KKl} we have used
the  synchronization space gap $G$ as a function of both the speed $v$ and speed difference $\Delta v$.} in which at $v\rightarrow 0$ the space gap tends to some minimum gap $g_{\rm min}$ between vehicles~\footnote{The use of the minimum gap $g_{\rm min}$ is
	  well-known for many standard deterministic traffic flow models (see, e.g.,~\cite{Treiber-Kesting}). Note that
		the incorporation of the minimum gap $g_{\rm min}$ in the model 
		is physically equivalent to {\it slow-to-start-rule} used in cellular automation and other 
		stochastic traffic flow models~\cite{Barlovic,Takayasu}.
		Indeed, when a vehicle is in a standstill within a wide moving jam and the preceding vehicle begins to accelerate from the jam,
		then in accordance with (\ref{g_v_g_min1})--(\ref{K_Deltav2}), (\ref{g_safe-min}), (\ref{G-min})
		the vehicle waits to accelerate behind the preceding vehicle
		as long as the space gap is less than the minimum gap $g_{\rm min}$.}: Respectively,  
 in  (\ref{g_v_g_min1})--(\ref{g_prec_ACC}), rather than    $g_{\rm safe}=v\tau_{\rm safe}$ and   $G=v\tau_{\rm G}$
 (\ref{G_g_safe_simple})
[Figs.~\ref{LWR_Fig}--\ref{S-F-onramp_traj_2}], we get:
		\begin{equation}
g_{\rm safe}=\left\{
\begin{array}{ll}
v\tau_{\rm safe} \ {\rm at} \ v \geq v_{\rm pinch}
 \\
g_{\rm min}+v\left(\tau_{\rm safe, pinch}
-\tau_{\rm min}\right) \ {\rm at} \ v < v_{\rm pinch}, \\
\end{array}\right.
  \label{g_safe-min}
  \end{equation}
\begin{equation}
G=\left\{
\begin{array}{ll}
v\tau_{\rm G} \ {\rm at} \ v \geq v_{\rm pinch}
 \\
g_{\rm min}+v\left(\tau_{\rm G, pinch}
-\tau_{\rm min}\right) \ {\rm at} \ v < v_{\rm pinch}, \\
\end{array}\right.
  \label{G-min}  
  \end{equation}
		where $\tau_{\rm min}=g_{\rm min}/v_{\rm pinch}$,   
		$g_{\rm min}$  is a parameter.

\subsection{Empirical   traffic breakdown:  
Vehicle overacceleration versus vehicle overdeceleration  \label{OA-OV_Sec}}

In  real empirical traffic data, traffic breakdown at a bottleneck is an F$\rightarrow$S transition
that exhibits the empirical nucleation nature~\cite{KernerBook1}.  To study whether 
 vehicle overdeceleration caused by driver  overreaction in traffic flow consisting of human-driving vehicles (Sec.~\ref{Int}) effects
on the F$\rightarrow$S transition, we make the following
physical modeling.

\subsubsection{Simulations of spontaneous F$\rightarrow$S transition}  

\begin{figure*}
\begin{center}
\includegraphics[width = 13 cm]{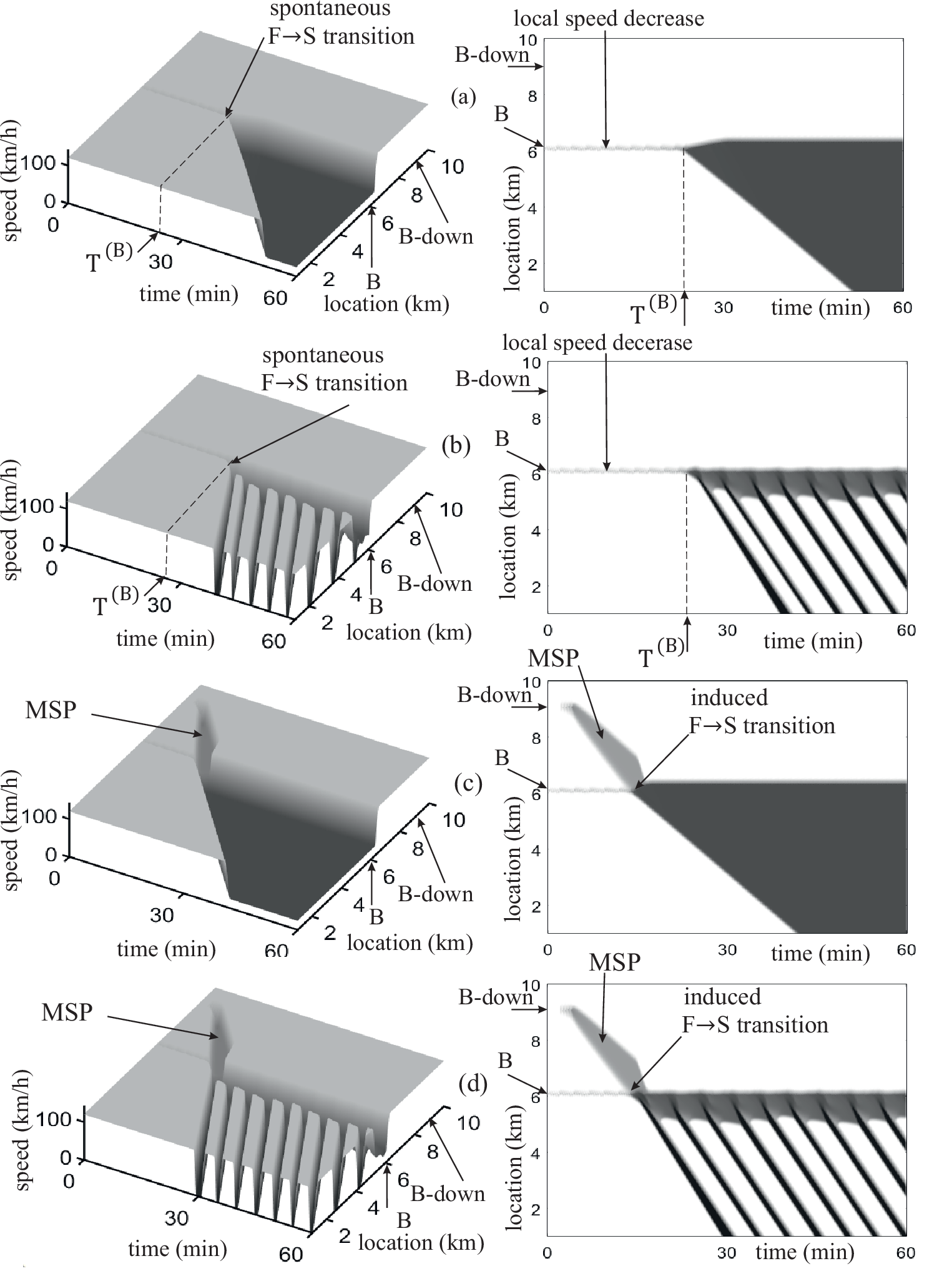}
\end{center}
\caption[]{Vehicle overacceleration versus vehicle overdeceleration: What is the cause of real (empirical) traffic breakdown 
at bottleneck? (a)--(d) Left subplots -- simulations of speed in space and time on single-lane road with two on-ramp bottlenecks denoted as $\lq\lq$B" and $\lq\lq$B-down" located 
at  $x_{\rm on}=$ 6 km and $x_{\rm on 2}=$ 9  km, respectively;
right subplots -- the same
vehicle speed data as those left, respectively, presented by regions with variable shades of gray [shades of gray
vary from white to black when the speed decreases from 120 km/h
(white) to 0 km/h (black)];
 simulations
at flow rates: $q_{\rm in}=$ 2250 vehicles/h, on-ramp inflow rates to bottleneck B $q_{\rm on}=$ 568.5 vehicles/h and
bottleneck B-down $q_{\rm on 2}=$ 0. In (a), (c), simulations of model 	 (\ref{g_v_g_min1})--(\ref{K_Deltav2}),  (\ref{G_g_safe_simple})
 with the same
model parameters  as those in
 Fig.~\ref{Greater_G_induced_alpha} in which no vehicle overdeceleration is realized. In (b), (d),
simulations of model 	 (\ref{g_v_g_min1})--(\ref{K_Deltav2}), (\ref{g_safe-min}), (\ref{G-min}), in which
vehicle overdeceleration is realized at  $v<v_{\rm pinch}$  with
 model parameters $g_{\rm min}=$ 5 m, $v_{\rm pinch}=$ 36 km/h, at $v<v_{\rm pinch}$ parameters
$K_{3, \rm pinch}= 0.1 \ s^{-2}$ and 
 $K^{(2)}_{4, \rm pinch}= 0.8 \ s^{-1}$; other model parameters   are the same as those in (a), (c).
In (c), (d),   MSP (moving  synchronized flow pattern) propagating upstream has been induced at downstream bottleneck B-down
at $t=$ 2 min through application of   on-ramp inflow impulse $\Delta q_{\rm on 2}=$
122 vehicles/h   of duration   2 min;  while reaching upstream bottleneck B, the MSP
induces F$\rightarrow$S transition (traffic breakdown) at upstream bottleneck B. 
}
\label{not-driver-overreaction}
\end{figure*}

In   model (\ref{g_v_g_min1})--(\ref{K_Deltav2}), (\ref{G_g_safe_simple}), in which no vehicle overdeceleration occurs,
spontaneous traffic breakdown (F$\rightarrow$S transition)
 is realized after a time delay $T^{\rm (B)}$ [Fig.~\ref{not-driver-overreaction}(a)] 
 (see Sec.~\ref{Physics_v_cr}). When we use  generalized model
(\ref{g_v_g_min1})--(\ref{K_Deltav2}), (\ref{g_safe-min}), (\ref{G-min})
of Sec.~\ref{Gener_Sec}, in which at speeds $v<v_{\rm pinch}$ vehicle overdeceleration does occur,
we find that after traffic breakdown has occurred wide moving jams appear almost immediately
[Fig.~\ref{not-driver-overreaction}(b)]; in this case, at $t>T^{\rm (B)}$ the spatiotemporal distribution of traffic congestion
is qualitatively the same as that well-known from 
 standard traffic flow models
(e.g.,~\cite{Bando_2,Treiber,KK1994,Krauss,Treiber-Kesting,Nagatani_R}), in which traffic breakdown is explained  by the classical traffic flow instability   resulting from vehicle overdeceleration.

Therefore, a question can arise: What is a qualitative difference between 
Fig.~\ref{not-driver-overreaction}(b) and related well-known results of the standard traffic flow models?
The qualitative difference is that traffic breakdown and its time delay    $T^{\rm (B)}$
do  not depend whether there is  vehicle overdeceleration    or not.
Indeed, we have found that
both the microscopic spatiotemporal development of a local speed decrease
in free flow at the bottleneck during  time interval $t<T^{\rm (B)}$ and  
time delay   $T^{\rm (B)}$ remain exactly the same in simulations presented
in Fig.~\ref{not-driver-overreaction}(a) when   vehicle overdeceleration does not exist  and
in Fig.~\ref{not-driver-overreaction}(b) when
vehicle overdeceleration does  exist:
Vehicle overdeceleration in  generalized model (\ref{g_v_g_min1})--(\ref{K_Deltav2}), (\ref{g_safe-min}), (\ref{G-min})
applied  in Fig.~\ref{not-driver-overreaction}(b)
 does not influence on the microscopic features of 
the development of the F$\rightarrow$S transition (traffic breakdown) at all.

Thus,  
the microscopic features of traffic breakdown do not depend on whether
vehicle  overdeceleration  is incorporated in the three-phase traffic flow model or not:
The physics of the F$\rightarrow$S transition is solely determined by vehicle overacceleration.
  The F$\rightarrow$S transition
leads to the emergence of  synchronized flow at the bottleneck.
Vehicle overdeceleration occurring in this synchronized flow  causes the classical traffic flow instability resulting in
the emergence of wide  moving jams
   (called as S$\rightarrow$J transitions): Wide moving jams result from 
a sequence of F$\rightarrow$S$\rightarrow$J transitions.

\subsubsection{Simulations of empirical induced  F$\rightarrow$S transition}

For a deeper understanding of the effect of vehicle overdeceleration on real (empirical) vehicular traffic,
we simulate empirical induced F$\rightarrow$S transition shown in Fig.~\ref{Hyp-OA}(a)
with either model (\ref{g_v_g_min1})--(\ref{K_Deltav2}), (\ref{G_g_safe_simple})
[Fig.~\ref{not-driver-overreaction}(c)] 
 or with   generalized model (\ref{g_v_g_min1})--(\ref{K_Deltav2}), (\ref{g_safe-min}), (\ref{G-min})
[Fig.~\ref{not-driver-overreaction}(d)].

 For physical modeling of  empirical data in Fig.~\ref{Hyp-OA}(a) we use
the fact that there is a long  time delay $T^{\rm (B)}\approx$ 24 min of the  F$\rightarrow$S transition
in both Figs.~\ref{not-driver-overreaction}(a) and~\ref{not-driver-overreaction}(b).  Therefore,
free flow at bottleneck B is in metastable state with respect to
the  F$\rightarrow$S transition at $t<T^{\rm (B)}$. 
With the use of a short on-ramp inflow impulse applied to downstream on-ramp
(B-down) we induce an MSP that reaches   upstream bottleneck B at $t<T^{\rm (B)}$.
The MSP induces
traffic breakdown at bottleneck B [Figs.~\ref{not-driver-overreaction}(c) and~\ref{not-driver-overreaction}(d)].  
Rather than on dynamics of the  F$\rightarrow$S transition, vehicle overdeceleration
  effects only on the congested pattern resulting from  the  F$\rightarrow$S transition: 
  When   model (\ref{g_v_g_min1})--(\ref{K_Deltav2}), (\ref{G_g_safe_simple}) is used, in which
no vehicle overdeceleration is realized, then either  spontaneous or induced F$\rightarrow$S transition
 causes the occurrence of the WSP in which no wide moving jams occur
 [Figs.~\ref{not-driver-overreaction}(a) and~\ref{not-driver-overreaction}(c)]; this is independent of 
how low synchronized flow speed is within the WSP.  In contrast, when    model
(\ref{g_v_g_min1})--(\ref{K_Deltav2}), (\ref{g_safe-min}), (\ref{G-min}) is used, in which
vehicle overdeceleration can occur, then
wide moving jams spontaneously emerge  in low-speed synchronized flow
 with the speed $v<v_{\rm pinch}$;
this slow synchronized flow, in turn, originally resulted from either a spontaneous or induced F$\rightarrow$S transition at bottleneck B 
 [Figs.~\ref{not-driver-overreaction}(b) and~\ref{not-driver-overreaction}(d)].   

We can conclude that the occurrence of the
 F$\rightarrow$S transition is solely determined by vehicle overacceleration:
 Independent of whether spontaneous [Figs.~\ref{not-driver-overreaction}(a)
and~\ref{not-driver-overreaction}(b)] or induced   
  F$\rightarrow$S transition [Figs.~\ref{not-driver-overreaction}(c) and~\ref{not-driver-overreaction}(d)] 
	is realized, vehicle overdeceleration
	does not effect on   features of the  F$\rightarrow$S transition at the bottleneck.

\subsection{A possible  confusion by 
  simulations of traffic congestion \label{Confusion-S}}
	
	\begin{figure}
\begin{center}
\includegraphics[width = 8 cm]{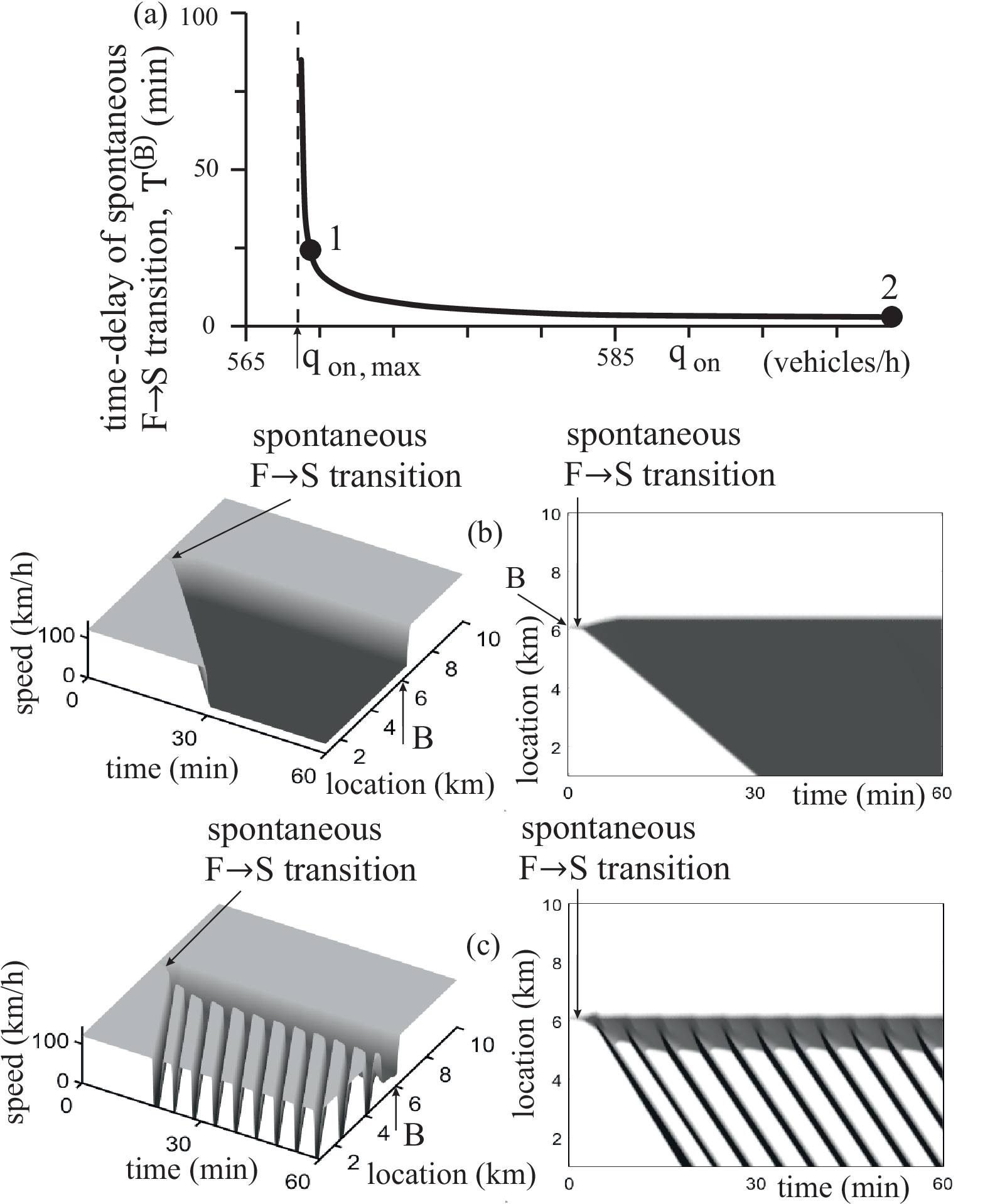}
\end{center}
\caption[]{Explanation of a possible  confusion by 
  simulations of traffic congestion.
Simulations of the same models as those in Figs.~\ref{not-driver-overreaction}(a) and~\ref{not-driver-overreaction}(b) 
 made on the same single-lane road
as well as the same flow rates 
  $q_{\rm in}=$ 2250 vehicles/h and $q_{\rm on 2}=$ 0.   
(a) Time-delay of spontaneous traffic breakdown (F$\rightarrow$S transition) $T^{\rm (B)}$
as function of on-ramp inflow rate $q_{\rm on}$;
	function $T^{\rm (B)}(q_{\rm on})$
		 does not depend on whether the model used in   Fig.~\ref{not-driver-overreaction}(a)
		or the model used in   Fig.~\ref{not-driver-overreaction}(b) have been applied; $q_{\rm on, \ max}=$ 568 vehicles/h.
       (b), (c) Left subplots -- simulations of speed in space and time at  
			$q_{\rm on}=$ 600 vehicles/h with the model used in   Fig.~\ref{not-driver-overreaction}(a) for (b)
			and with the model used in   Fig.~\ref{not-driver-overreaction}(b) for (c);   
right subplots -- the same
vehicle speed data as those left, respectively, presented by regions with variable shades of gray [shades of gray
vary from white to black when the speed decreases from 120 km/h
(white) to 0 km/h (black)]. In (a), point 1 is related to $q_{\rm on}=$ 568.5 vehicles/h used in
Fig.~\ref{not-driver-overreaction} for which $T^{\rm (B)}\approx$ 24 min, 
whereas point 2 is related to subplots (b) and (c) for which $T^{\rm (B)}\approx$ 3 min. 
}
\label{overreaction-OA}
\end{figure}
		
	To emphasize a possible  confusion by 
  simulations of traffic congestion, we consider Fig.~\ref{overreaction-OA}.
	The time delay of the spontaneous F$\rightarrow$S-transition $T^{\rm (B)}$ decreases rapidly as the on-ramp inflow rate 
$q_{\rm on}$, which is larger than $q_{\rm on, \ max}$, continues to increase
	[Fig.~\ref{overreaction-OA}(a)]. Therefore, at a large enough flow rate $q_{\rm on}$, time-delay $T^{\rm (B)}$ can be very short.
When model (\ref{g_v_g_min1})--(\ref{K_Deltav2}), (\ref{G_g_safe_simple}) is used, in which
 there is no vehicle overdeceleration,  the  result of traffic breakdown simulation
	[Fig.~\ref{overreaction-OA}(b)]
	is qualitatively different   from well-known simulation results
	of the standard traffic flow models in which the classical traffic flow instability in free flow at the bottleneck 
	is   the cause of traffic congestion (e.g.,~\cite{Bando_2,Treiber,KK1994,Krauss,Treiber-Kesting,Nagatani_R}). 
	
	Contrarily, when the classical traffic flow instability with wide moving jam emergence (S$\rightarrow$J transitions) is realized in synchronized flow
			[Fig.~\ref{overreaction-OA}(c)], then the  result of traffic breakdown  
			simulated with  generalized model (\ref{g_v_g_min1})--(\ref{K_Deltav2}), (\ref{g_safe-min}), (\ref{G-min}) is qualitatively the same as that simulated with
	one of the standard traffic flow models.
	This $\lq\lq$fraudulent" similarity is because there is almost no time-delay
	between the F$\rightarrow$S transition and S$\rightarrow$J transition within
	the sequence of the F$\rightarrow$S$\rightarrow$J transitions: 
	Moving jams occurs already in the emergent synchronized flow. 
	This can lead to a great confusion (and it does very often happen), when  a three-phase traffic flow model and
	a standard traffic flow model are compared.

 To solve this confusion, we should
 understand whether a traffic flow model 
can simulate the origin of the occurrence of congestion   in real (empirical) traffic or not. In other words,
the model should be able to simulate 
  the F$\rightarrow$S transition exhibiting the empirical nucleation nature at a bottleneck.
	This is because this is 
the basic empirical feature of   real  traffic breakdown    at the bottleneck.  
 As shown in
 books~\cite{KernerBook1,KernerBook2,KernerBook3,KernerBook4}, this
 basic empirical feature of   real  traffic breakdown (F$\rightarrow$S transition) can be explained by none
of the standard traffic flow models
 of Refs.~\cite{GM_Com1,GM_Com2,GM_Com3,KS,KS1,KS2,KS4,Newell1961,Newell,Gipps1981,Gipps1986,Wiedemann,Payne_1,Payne_2,Nagel_S,Bando_1,Bando,Bando_2,Bando_3,Nagatani_1,Nagatani_2,Treiber,Chen2012A,Chen2012B,Chen2014,Ashton,Drew,Gerlough,Gazis,Gartner,Barcelo,Elefteriadou,DaihengNi,Kessels,Treiber-Kesting,Schadschneider,Chowdhury,Helbing,Nagatani_R,Nagel}.
This should emphasize why in the three-phase traffic theory 
the physics of   real traffic breakdown is explained through vehicle overacceleration.
 
Microscopic physics of vehicle overacceleration disclosed in the paper
shows that over{\it acceleration} is a totally different behavior of microscopic vehicle motion 
  in comparison with   over{\it deceleration} caused by driver overreaction in traffic flow of human-driving vehicles.
This explains why features of traffic breakdown and highway capacity
simulated with   three-phase traffic flow models, as shown in
the books~\cite{KernerBook1,KernerBook2,KernerBook3,KernerBook4}, are qualitatively different from 
features of traffic breakdown and highway capacity simulated with 
  standard traffic flow models (e.g.,~\cite{GM_Com1,GM_Com2,GM_Com3,KS,KS1,KS2,KS4,Newell1961,Newell,Gipps1981,Gipps1986,Wiedemann,Payne_1,Payne_2,Nagel_S,Bando_1,Bando,Bando_2,Bando_3,Nagatani_1,Nagatani_2,Treiber,Chen2012A,Chen2012B,Chen2014,Ashton,Drew,Gerlough,Gazis,Gartner,Barcelo,Elefteriadou,DaihengNi,Kessels,Treiber-Kesting,Schadschneider,Chowdhury,Helbing,Nagatani_R,Nagel}).

\subsection{Conclusions \label{Con-S}}

1.  For a given flow rate in free flow  
upstream of an on-ramp bottleneck, the stronger the effect of vehicle overacceleration
 on  traffic flow, the larger  the maximum on-ramp flow rate   at which free flow   can
persist   at the bottleneck, i.e., the better traffic breakdown can be avoided.

2. In addition to known overacceleration in  road lane $a_{\rm OA}$ (\ref{a_OA})~\cite{Kerner2023B}, in vehicular traffic 
there can be  at least two mechanisms of vehicle overacceleration  in  road lane caused by safety acceleration at the bottleneck.

 3. Whether  an acceleration behavior becomes  overacceleration or  not depends on traffic characteristics and model parameters.
This conclusion is valid for both safety acceleration in  road lane and for  
vehicle acceleration through lane-changing.

 4. The microscopic effect of vehicle overacceleration on  traffic flow is determined by  
a spatiotemporal cooperation of  different   overacceleration mechanisms. The overacceleration cooperation     increases    the effect of  the maintaining of free flow at the bottleneck.

5. Both speed adaptation 
and    overacceleration in  road lane can effect qualitatively on  
   overacceleration mechanism
caused by lane-changing.

6. The S$\rightarrow$F instability on two-lane road without bottlenecks occurs through cooperation of different overacceleration mechanisms.

7. The S$\rightarrow$F instability on two-lane road with the bottleneck exhibits the following spatiotemporal dynamics.  

(i) In the right lane, the
S$\rightarrow$F instability is caused by overacceleration cooperation through
 R$\rightarrow$L lane-changing and overacceleration $a_{\rm OA}$ in road lane. Contrarily,  speed adaptation that occurs 
through merging of on-ramp vehicles together
with L$\rightarrow$R lane-changing tries   to prevent the S$\rightarrow$F instability development.

(ii) In the left lane, the
S$\rightarrow$F instability is caused by overacceleration cooperation through
 L$\rightarrow$R lane-changing and overacceleration $a_{\rm OA}$ in road lane. Contrarily,  speed adaptation 
through R$\rightarrow$L lane-changing tries   to prevent the S$\rightarrow$F instability development.

 8.  These    microscopic features of the effect of vehicle overacceleration on traffic flow
 (items 1--7 above) are related to traffic flow consisting of    human-driving and/or
automated-driving vehicles.

\end{document}